\documentclass[pra,reprint,superscriptaddress,longbibliography]{revtex4-1}

\usepackage{amsmath,amssymb}
\usepackage[euler]{textgreek}
\usepackage{subfigure}
\usepackage{psfrag}
\usepackage{graphicx}
\usepackage{color}
\usepackage[usenames,dvipsnames]{xcolor}
\usepackage{pifont}
\usepackage{fourier}
\usepackage{xr}
\usepackage{stmaryrd}
\usepackage{cancel}
\usepackage{upgreek}

\usepackage{fancyhdr}

\usepackage[process=auto]{pstool}
\usepackage[normalem]{ulem}
\definecolor{redg}{rgb}{1,0,0}
\definecolor{blueg}{rgb}{0.22,0.33,0.64}
\definecolor{greeng}{rgb}{0,0.63,0.29}
\definecolor{orangeg}{rgb}{0.96,0.47,0.13}

\DeclareMathAlphabet\mathbfcal{OMS}{cmsy}{b}{n}

\newcommand{\ve}[1]{\mathbf{#1}}
\newcommand{\vet}[1]{\mathbfcal{#1}}
\newcommand{\ves}[1]{\boldsymbol{#1}}

\newcommand{\tx}[1]{\text{#1}}
\newcommand{\te}[1]{\overline{\overline{#1}}}

\newcommand{\hatv}[1]{\hat{\mathbf{#1}}}

\begin{document}

\title{Electromagnetic Chirality}

\author{Christophe Caloz}
\email[]{christophe.caloz@polymtl.ca}
\affiliation{Polytechnique Montr\'{e}al, Montr\'{e}al, QC H3T-1J4, Canada}
\author{Ari Sihvola}
\affiliation{Aalto University, Espoo, FI-00076, Finland}

\begin{abstract}
This paper presents a first-principle and global perspective of electromagnetic chirality. It follows for this purpose a bottom-up construction, from the description of chiral particles or metaparticles (microscopic scale), through the electromagnetic theory of chiral media (macroscopic scale), to the establishment advanced properties and design principles of chiral materials and metamaterials. It preliminarily highlights the three fundamental concepts related to chirality -- mirror asymmetry, polarization rotation and magnetodielectric coupling -- and points out the nontrivial interdependencies existing between them. The first part (chiral particles) presents metamaterial as the most promising technology for chirality, compares two representative particles involving magnetoelectric coupling, namely the planar Omega particle and the twisted Omega or helix particle, and shows that only the latter is chiral, and finally links the response of microscopic particles to that of the medium formed by arranging them according to a subwavelength lattice structure. The second part (electromagnetic theory) infers from the previous microscopic study the chiral constitutive relations as a subset of the most general bianisotropic relations, derives parity conditions for the chiral parameters, computes the chiral eigenstates as circularly polarized waves, and finally shows that the circular birefringence of these states leads to polarization rotation. The third part (properties and design) introduces an explicit formulation of chirality based on spatial frequency dispersion or nonlocality, analyzes the temporal frequency dispersion or nonlocality of chiral media, and finally provides guidelines to design a practical chiral metamaterial.
\end{abstract}
%

\maketitle

\tableofcontents

\newpage

\section{Introduction}\label{sec:intro}

\begin{figure}[h]
    \centering
        \includegraphics[width=1\linewidth]{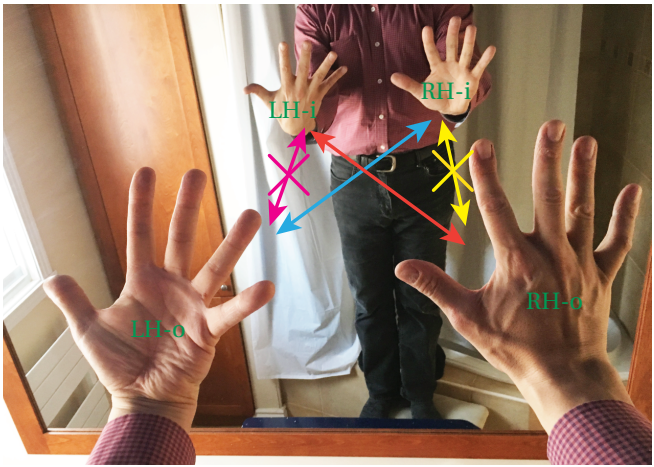}{
        \psfrag{R}[c][c][1]{\textcolor{ForestGreen}{RH-o}}
        \psfrag{L}[c][c][1]{\textcolor{ForestGreen}{LH-o}}
        \psfrag{A}[c][c][1]{\textcolor{ForestGreen}{RH-i}}
        \psfrag{B}[c][c][1]{\textcolor{ForestGreen}{LH-i}}}
        \caption{Human hands and their reflection in a mirror. RH-o and LH-o are the original right hand (RH) and left hand (LH), while RH-i and LH-i are their images in the mirror, with the correspondence RH-o$\leftrightarrow$LH-i and LH-o$\leftrightarrow$RH-i. The non-superimposability may be specifically understood as follows. RH-i is finger-to-finger aligned with RH-o, but it shows its palm whereas RH-o shows its back. Flipping RH-i brings about back-to-back translational symmetry, but loses finger-to-finger symmetry. So, RH-i and RH-o are fundamentally different, irrespectively to their orientation in space. The same naturally applies to the pair (LH-o,LH-i). {\it(Photograph: Rapha\"{e}l Caloz)}}
   \label{fig:mirror_picture}
\end{figure}

\begin{figure}[h]
    \centering
        \includegraphics[width=0.7\linewidth]{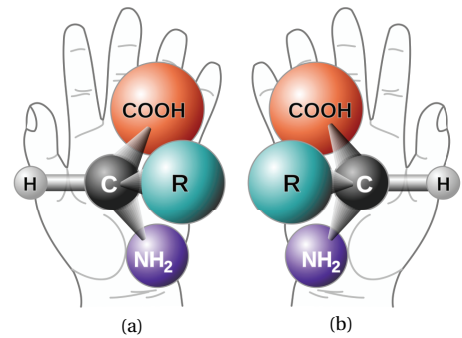}{
        \psfrag{a}[c][c][0.8]{(a)}
        \psfrag{b}[c][c][0.8]{(b)}}
        \caption{Chirality in a generic amino acid (--NH$_2$: amine, --COOH: caboxyl, --R: rest of the molecule). (a)~Left-handed (LH) enantiomer: with thumb along the C$\rightarrow$H axis, it takes the left hand for the fingers to point from COOH to NH$_2$ through R. (b)~Right-handed (RH) enantiomer: with thumb along the C$\rightarrow$H axis, it takes the right hand for the fingers to point from COOH to NH$_2$ through R. {\it(Picture: Wikimedia Commons)}}
   \label{fig:chir_handedness}
\end{figure}

The term \emph{chirality} comes from the Greek word \textchi\textepsilon$\acute{\text{\textiota}}$\textrho, which means \emph{hand}. It is the \emph{geometric property} according to which an object is \emph{mirror asymmetric} or, equivalently, different from its image in a mirror, irrespectively to orientation. The etymology and definition of chirality may be understood by considering Fig.~\ref{fig:mirror_picture}. The (mirror-plane) image of the right (resp. left) hand is \emph{not} superimposable with the right (resp. left) hand itself, but rather with the left (resp. right) hand. So, the human hands are chiral [sic], with right or left \emph{handedness}, and the right and left hands are called the enantiomers (from the Greek, $\acute{\text{\textepsilon}}$\textnu\textalpha\textnu\texttau\textiota{}: opposite,
\textmu$\acute{\text{\textepsilon}}$\textrho{}o$\mathrm{\varsigma}$: other) of each other. The fact that hands are certainly the most common chiral things that we deal with on a daily basis justifies the etymology. However, many other things are chiral, such as amino-acids -- see Fig.~\ref{fig:chir_handedness} -- and tris(bipyridine)ruthenium(II) chloride (red crystalline salt) in chemistry, DNA and sugars in biochemistry, sea snails and ammonite fossils in biology, screws and helical antennas in engineering, and fusilli pasta and twisted pastry in food!

A medium made of chiral molecules or particles is called a \emph{chiral medium}. Such a medium has the remarkable property of rotating the polarization of electromagnetic waves propagating through it. This phenomenon of \emph{polarization rotation}, also called \emph{optical activity}, was first observed more than 200 years ago with light passing through `translucent' substances by Arago in 1811~\cite{Arago_1812} and through quartz by Biot shortly later~\cite{Biot_1814}, and it was explained by Fresnel in terms of circular birefringence in 1821-22~\cite{Fresnel_1868}. It was further studied by Pasteur in salts of solutions of racemic mixtures of tartrates near the middle of the XIX$^\tx{th}$ century~\cite{Pasteur_1850}. In 1898, Bose reported the first microwave observation of chiral polarization rotation, in twisted jute (fiber produced by some plants) structures~\cite{Bose_1898}, and, about twenty years later, Lindman introduced wire spirals as more practical artificial chiral particles~\cite{Lindman_1914}. In the course of the XX$^\tx{th}$ century, it has been established that optical activity results from coupling between the electric and magnetic responses, or \emph{magnetoelectric coupling}, of chiral particles, and different related form of electromagnetic constitutive relations have been proposed~\cite{Condon_1937,Tellegen_1948,Post_1962,Kong_1972,Fedorov_1976}. Towards the end of the century appeared the first textbooks on chiral and biisotropic media~\cite{Lakhtakia_1989,Lindell_EWCBM_1994}.

Until the turn of the current century, chiral media had been mostly restricted to theoretical electromagnetic studies~\cite{Lakhtakia_1989,Lindell_EWCBM_1994,Kong_EWT_2008}. The advent of modern \emph{metamaterials} (e.g.~\cite{Capolino_2009}) and, even more, \emph{metasurfaces} (e.g.~\cite{Achouri_NP_2018}) has dramatically changed the situation, and chiral media, along with their bianisotropic extension~\cite{Kong_EWT_2008}, have now become a practical reality that is poised to revolutionize microwave, terahertz and photonics technologies.

Chirality involves a number of concepts that are sometimes misunderstood and confused. These concepts include Pasteur and Tellegen biisotropy~\cite{Lindell_EWCBM_1994}, biisotropy and bianisotropy~\cite{Kong_EWT_2008}, circular birefringence and circular dichroism~\cite{Born_Wolf_1999}, reciprocal and nonreciprocal polarization rotation~\cite{Caloz_PRAp_10_2018}, temporal and spatial electromagnetic symmetry~\cite{Jackson_1998}, and temporal and spatial dispersion or nonlocality~\cite{Jackson_1998,Landau_1984}. This paper presents a global, intuitive and yet rigorous, first-principle perspective of chiral media and metamaterials that is intended to dissipate misunderstandings and confusions, and hence help further research in the field.

The rest of the paper is organized as follows. After making a global comment on the three fundamental aspects of chirality -- mirror asymmetry, polarization rotation and magnetodielectric coupling -- and their interrelations in Sec.~\ref{sec:prelim_com}, and setting global assumptions in Sec.~\ref{sec:glob_assump}, the document follows a bottom-up development in three main parts. The first part, composed of Secs.~\ref{sec:chir_mat_meta} to~\ref{sec:rel_ma_mi}, deals with the microscopic, particle aspect of chirality; specifically, it presents metamaterial technology as the most promising approach to electromagnetic chirality (Sec.~\ref{sec:chir_mat_meta}), analyses two representative metaparticles involving magnetoelectric coupling -- the achiral planar Omega particle and the volume-twisted Omega or helix particle (Sec.~\ref{sec:two_part}) -- and links the microscopic metaparticle response to the macroscopic response of the corresponding metamaterial (Sec.~\ref{sec:rel_ma_mi}). The second part, composed of Secs.~\ref{sec:macr_const_rel} to~\ref{sec:chir_eig_states}, covers the electromagnetic theory of chiral media; specifically, it infers from the previous microscopic study the chiral constitutive relations as a subset of the bianisotropic relations (Sec.~\ref{sec:macr_const_rel}), derives parity conditions for the chiral parameters (Sec.~\ref{sec:spat_sym}), computes the chiral eigenstates as circularly-polarized waves (Sec.~\ref{sec:chir_eig_states}), and finally shows that the circular birefringence of these states leads to polarization rotation (Sec.~\ref{sec:pol_rot}). The third part, composed of Secs.~\ref{sec:spat_nl} to~\ref{sec:des_princ}, introduces an explicit
formulation of chirality based on spatial frequency dispersion or nonlocality (Sec.~\ref{sec:spat_nl}), studies the temporal frequency dispersion or nonlocality of chiral media, and finally provides guidelines to (Sec.~\ref{sec:temp_disp_design}), and elaborates a synthesis procedure based on iterative full-wave analysis for the design of chiral metamaterials (Sec.~\ref{sec:des_princ}). Finally, conclusions are given in Sec.~\ref{sec:concl}.

\section{Preliminary Comment: the Chiral Trinity}\label{sec:prelim_com}
Section~\ref{sec:intro} has mentioned that chirality is intimately related to the concepts of mirror asymmetry, polarization rotation and magnetoelectric coupling. Mirror asymmetry is today's geometrical definition of chirality, polarization rotation is the fundamental chiral effect on light observed two centuries ago by Arago and Biot, and magnetoelectric coupling has been found to be a fundamental electromagnetic feature of chiral media in the course of the past century.

\begin{figure}[h]
    \centering
        \includegraphics[width=\linewidth]{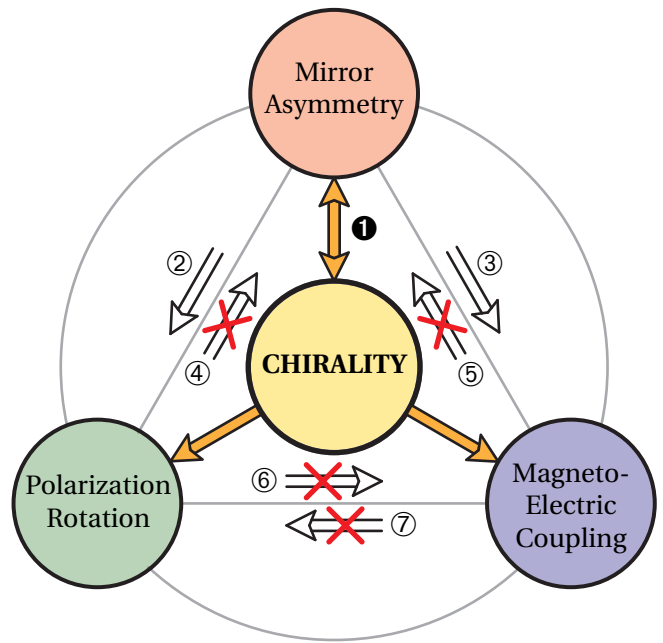}{
        \psfrag{X}[c][c][1.1]{\bf CHIRALITY}
        \psfrag{A}[c][c][1.1]{\begin{minipage}{2.5cm}\centering Mirror \\ Asymmetry \end{minipage}}
        \psfrag{B}[c][c][1.1]{\begin{minipage}{2.5cm}\centering Polarization \\ Rotation \end{minipage}}
        \psfrag{C}[c][c][1.1]{\begin{minipage}{2.5cm}\centering Magneto- \\ Electric \\ Coupling \end{minipage}}
        \psfrag{1}[c][c][1.3]{\ding{202}}
        \psfrag{2}[c][c][1.3]{\ding{193}}
        \psfrag{3}[c][c][1.3]{\ding{194}}
        \psfrag{4}[c][c][1.3]{\ding{195}}
        \psfrag{5}[c][c][1.3]{\ding{196}}
        \psfrag{6}[c][c][1.3]{\ding{197}}
        \psfrag{7}[c][c][1.3]{\ding{198}}
        }
        \vspace{-1mm}
        \caption{The chiral trinity with implication relations between the three fundamentally related concepts of mirror asymmetry, polarization rotation and magnetoelectric coupling. The arrows indicate implications and the barred arrows indicate nonimplications.}
   \label{fig:trilogy}
\end{figure}

However, these concepts are not systematically interdependent, and it is of crucial importance to distinguish the implication relations existing between them. These relations are represented in Fig.~\ref{fig:trilogy}, which will be demonstrated throughout the paper. They are the following:
\begin{itemize}
  \item Mirror asymmetry is a necessary and sufficient conditions for chirality (\ding{202}). This requires no proof since the former is the definition of the latter, and the two may thus be considered as merged together in Fig.~\ref{fig:trilogy}.
  \item Mirror asymmetry implies both polarization rotation (\ding{193}) and magnetoelectric coupling (\ding{194}), but neither polarization rotation (\ding{195}) nor electromagnetic coupling (\ding{196}) implies mirror asymmetry.
  \item Polarization rotation does not imply magnetoelectric coupling (\ding{197}), and magnetolectric coupling does not imply polarization rotation (\ding{198}).
\end{itemize}

\section{Global Assumptions}\label{sec:glob_assump}

The following assumptions hold throughout the paper:

\begin{enumerate}
  \item \label{it:LTI} All the media are \emph{linear} and \emph{time-invariant} (LTI).
  \item \label{it:HT_SS} They are excited by waves with \emph{harmonic time dependence}, and \emph{steady-state conditions} are assumed.
  \item \label{it:Gen_ST_field} Due to \ref{it:LTI}) and~\ref{it:HT_SS}), the electromagnetic responses of the media, and hence all the fields involved, have the same time dependence, and may thus be generally written versus space ($\ve{r}$) and time ($t$) in the elliptical polarization form
      \begin{equation}\label{eq:time_harm_dep}
        \vet{F}(\ve{r},t)
        =\ve{F}_1\cos[\omega t-\phi(\ve{r})]+\ve{F}_2\sin[\omega t-\phi(\ve{r})],
      \end{equation}
      where $\ve{F}_1$ and $\ve{F}_2$ are real perpendicular vectors, with circular polarization if $|\ve{F}_1|=|\ve{F}_2|$, and where $\omega$ is the temporal angular frequency ($\omega=2\pi f$, $f:$ frequency) and $\phi(\ve{r})$ is the spatial phase.
  \item As implicitly assumed in~\ref{it:Gen_ST_field}), scalar and vector quantities are denoted by regular and bold characters, respectively, while tensors are denoted by a double overline, as for instance $\te{\chi}$.
  \item Given~\ref{it:Gen_ST_field}), we use the customary \emph{phasor notation}, which conveniently allows to drop the time dependence in most calculations. The phasor corresponding to~\eqref{eq:time_harm_dep} is the auxiliary complex vector
      \begin{equation}\label{eq:phasor}
        \ve{F}(\ve{r})
        =\left(\ve{F}_1+i\ve{F}_2\right)e^{i\phi(\ve{r})}.
      \end{equation}
  \item The field~\eqref{eq:time_harm_dep} is then retrieved from~\eqref{eq:phasor} via the operation
      \begin{equation}\label{eq:wt_conv}
        \vet{F}(\ve{r},t)=\tx{Re}\left\{\ve{F}(\ve{r})e^{-i\omega t}\right\},
      \end{equation}
      where the physical field and its phasor are distinguished by calligraphic and regular characters, respectively.
\item The complex harmonic time dependence $e^{-i\omega t}$ in~\eqref{eq:wt_conv} corresponds to the convention that is generally adopted in the physics community~\cite{Kong_EWT_2008,Jackson_1998}. The engineering community rather uses the equivalent convention $e^{+j\omega t}$\cite{Harrington_THEF_1961,Pozar_ME_2011}, where $j=-i$. We choose here the former convention because it is more common in the literature on complex media.
\item If the medium is \emph{isotropic}, it is convenient to select a coordinate system that coincides with the direction of propagation, $\hatv{k}$. Here, assuming $\hatv{k}=\hatv{z}$, we therefore choose $\hatv{r}=\hatv{z}$. The corresponding phasor has the plane-wave form
    \begin{equation}\label{eq:PW_space}
      \ve{F}(z)
      =\left(\ve{F}_1+i\ve{F}_2\right)e^{\pm i\beta z},
    \end{equation}
    which is related to~\eqref{eq:phasor} by $\phi(\ve{r})=\phi(z)=\pm\beta z$ ($\beta=k=2\pi/\lambda$, $\lambda$: wavelength). The corresponding complex spacetime function is $e^{i(\pm\beta z-\omega t)}$, and the phase velocity is found by monitoring a wave point of fixed phase -- i.e., $\partial(\pm\beta z-\omega t)/\partial t=\pm\beta\partial z/\partial t-\omega=0$ -- as $v_\tx{p}=\partial z/\partial t=\pm\omega/\beta\gtrless 0$ ($\beta>0$), indicating that the positive and negative signs in~\eqref{eq:PW_space} correspond to wave propagation in the $+z$ (forward) and $-z$ (backward) directions, respectively. Note that the choice of the plane-wave form in~\eqref{eq:PW_space} is not restrictive since any wave in an LTI medium can be decomposed in a spectrum of plane waves from Fourier theory~\cite{Clemmow_2013}.
\item The sourceless time-harmonic Maxwell-Faraday and Maxwell-Amp\`{e}re equations for the chosen $e^{-i\omega t}$ time dependence are
    \begin{subequations}\label{eq:Maxwell_curl}
      \begin{align}
        \nabla\times\ve{E}&=i\omega\ve{B}, \\
        \nabla\times\ve{H}&=-i\omega\ve{D},
      \end{align}
    \end{subequations}
    where $\ve{E}$ (V/m), $\ve{H}$ (A/m), $\ve{D}$ (C/m$^2=$~As/m$^2$) and $\ve{B}$ (Wb/m$^2=$~Vs/m$^2$) are the usual electric field, magnetic field, electric flux density (or displacement field), and magnetic flux density (or magnetic induction field), respectively, and $\ve{J}$ (A/m$^2$) is the electric current density.
\end{enumerate}

\section{Metamaterial Implementation}\label{sec:chir_mat_meta}

\subsection{Motivation and Definition}\label{sec:meta_motiv_def}

Although chiral molecules and substances, such as amino acids and sugars~\cite{Lindell_EWCBM_1994,Sihvola_1999}, are abundant in nature, they are generally not amenable to electromagnetic applications, due to their  chemical instability, high loss and restricted spectrum. Artificial chirality, in the form of metamaterials, is a much more promising avenue in this regard, since one can engineer chiral metamaterials with high robustness, low loss, arbitrary operation frequency, and tailorable overall properties.

\begin{figure}[h]
    \centering
        \includegraphics[width=0.9\linewidth]{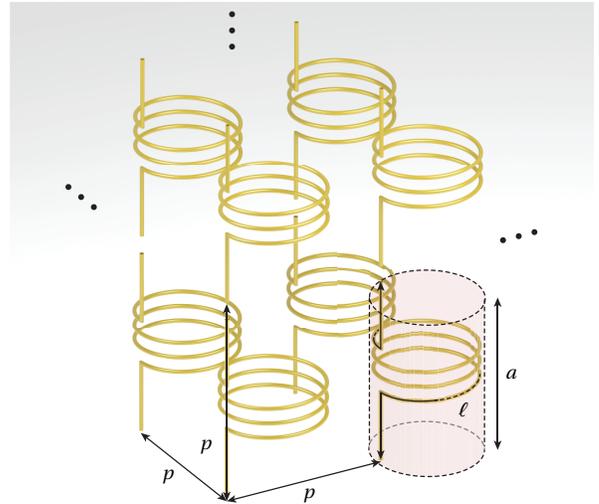}{
        \psfrag{a}[c][c][0.8]{$a$}
        \psfrag{p}[c][c][0.8]{$p$}
        \psfrag{l}[c][c][0.8]{$\ell$}
        }
        \caption{Example of a 3D chiral metamaterial with important dimensional parameters. Here the lattice is periodic with a cubic unit cell of dimension $p$, and it is made of multiturn-helix metaparticles of external size $a$ and unfolded (resonant, $\lambda_\tx{res}/2$) length $\ell$, with $a\ll\ell$ [Eq.~\eqref{eq:mtm_dims}].}
   \label{fig:metam_ex}
\end{figure}

A metamaterial is a medium constituted of a 1D, 2D or 3D subwavelength-lattice array of scattering particles -- or `metaparticles' -- whose key properties are due more to the geometry and orientation of these particles than to the molecular-scale nature of the materials that compose them. Figure~\ref{fig:metam_ex} shows an example of a 3D chiral metamaterial constituted of multiturn-helix-shaped metaparticles.

\subsection{Dimensional Constraints}\label{sec:dim_constr}
The subwavelength-lattice array condition is necessary for the structure to be \emph{homogeneizable}, and hence really operate as a medium, without spurious diffraction and with well-defined constitutive parameters. Denoting the lattice feature (or period in the most common case of a periodic -- crystal-like -- structure) $p$, and the size of the metaparticle $a$, one must thus satisfy the relation
\begin{equation}\label{eq:subl_latt}
|k|a\leqslant|k|p\ll 2\pi
\quad\tx{or}\quad
a\leqslant{}p\ll\lambda.
\end{equation}
At the same time, to interact with an incoming wave, and hence transform that wave according to specifications, the metaparticle must be operated close to its resonance, which occurs at the frequency where its \emph{resonant size}, $\ell$, is half the wavelength, i.e.,
\begin{equation}\label{eq:reson_part}
\ell=\lambda_\tx{res}/2.
\end{equation}
Note that in the case of a metamaterial, one invariably uses this halfwavelength (or first or lowest) resonance ($m=1$ in $\ell=m\lambda_\tx{res}/2$), because higher resonances would imply larger metaparticle electric sizes, which opposes the fundamental homogeneizable medium requirement~\eqref{eq:subl_latt}.

The conditions~\eqref{eq:subl_latt} and~\eqref{eq:reson_part} are clearly antagonistic, since $a$ and $\ell$ are both related to the size of the metaparticle. Fortunately, the `external size', which we define as the size of the smallest box fully containing the particle, i.e., here $a$, can be made substantially smaller than the resonant size, $\ell$, by folding an initially simple (e.g. straight or single-looped) structure of dimension $\ell$ upon itself in the three directions of space, as illustrated in Fig.~\ref{fig:metam_ex}, and by leveraging reactive (inductive and capacitive) loading, which ultimately leads to the viable and typical metamaterial regime
\begin{equation}\label{eq:mtm_dims}
a\leqslant{}p\ll\ell\approx\lambda_\tx{res}/2<\lambda_{\tx{res},0},
\end{equation}
where $\lambda$ represents the wavelength of the wave in the (possible) medium that embeds or surrounds (e.g. supporting substrate) the particle and $\lambda_0$ represents the wavelength of the wave in free space ($\lambda<\lambda_0$). Figure~\ref{fig:metam_ex} illustrates Eq.~\eqref{eq:subl_latt}, with $\lambda=\lambda_0$ if the particles stand in free-space, and $\lambda<\lambda_0$ if they include material loads or are supported by a dielectric matrix frame. Typical metamaterials involve parameters in the order of $a\in[\lambda_0/15-\lambda_0/4]$ and $p=\in[\lambda_0/12-\lambda_0/4]$~\cite{Capolino_2009}.

\subsection{Metaparticle Selection}\label{sec:metpart_sel}
As regular materials owe their macroscopic properties to their constitutive atoms and molecules, metamaterials owe their macroscopic properties to their metaparticles. Since they are deeply subwavelength, these metaparticles are restricted to \emph{dipolar responses}~\footnote{In general, a multipole expansion is required for modeling the scattering from an object~\cite{Jackson_1998}. However, the multipoles of order larger than the dipoles have a negligible effect when the object is deeply subwavelength, as is the case for a metaparticle, according to~\eqref{eq:mtm_dims}.}, characterized by the electric dipole moment $\ve{p}_\tx{e}$ (Asm) and by the magnetic dipole moments $\ve{p}_\tx{m}$ (Vsm), which are respectively defined as~\cite{Jackson_1998,Landau_1984}
\begin{subequations}\label{eq:pm_dip_mmt}
\begin{equation}
\ve{p}_\tx{e}=\int_V\ve{r}'\rho(\ve{r}')d\ve{r}',\label{eq:m_dip_mm}
\end{equation}
\begin{equation}
\ve{p}_\tx{m}=\mu_0\ve{m}=\frac{\mu_0}{2}\int_V\ve{r}'\times\ve{J}(\ve{r}')d\ve{r}'\label{eq:m_dip_mmt},
\end{equation}
\end{subequations}
where $\rho(\ve{r})$ and $\ve{J}(\ve{r})$ are the spatial distributions of electric charge density and current density, respectively, and where the volume integration corresponds to the structure of the particle. Note that we have here redefined the usual magnetic dipole moment $\ve{m}$~\cite{Jackson_1998} as $\ve{p}_\tx{m}=\mu_0\ve{m}$ for symmetry in the forthcoming chiral relations. In a simple artificial-dielectric metamaterial, the electric dipolar response ($\ve{p}_\tx{e}$) is exclusively due to the electric excitation ($\ve{E}$), and we denote it here $\ve{p}_\tx{ee}$, while the magnetic dipolar response ($\ve{p}_\tx{m}$) is exclusively due to the magnetic excitation ($\ve{H}$), and we denote it here $\ve{p}_\tx{mm}$. The best particle for ($\ve{p}_\tx{ee}$) is a straight conducting wire or a straight dielectric rod, according to Maxwell-Faraday law, in the microwave and optical regimes, respectively, while the best particle for ($\ve{p}_\tx{mm}$) is a looped conducting wire or a looped dielectric rod, according to Maxwell-Amp\`{e}re law. A combined $\ve{p}_\tx{ee}$--$\ve{p}_\tx{mm}$ particle leads then generally to a Lorentz-dispersive composite positive/negative-index metamaterial~\cite{Caloz_2005}, with negative index~\cite{Shelby_2001} below the electric and magnetic plasma frequencies and positive index above~\cite{Capolino_2009} (see Sec.~\ref{sec:temp_disp_design}).

\begin{figure}[h]
    \centering
        \includegraphics[width=\linewidth]{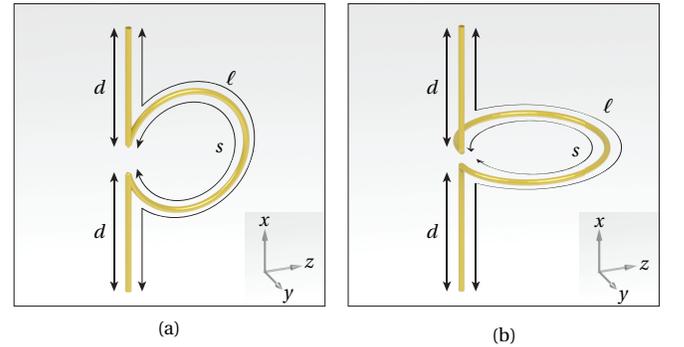}{
        \psfrag{a}[c][c][0.8]{(a)}
        \psfrag{b}[c][c][0.8]{(b)}
        \psfrag{x}[c][c][0.8]{$x$}
        \psfrag{y}[c][c][0.8]{$y$}
        \psfrag{z}[c][c][0.8]{$z$}
        \psfrag{d}[c][c][0.8]{$d$}
        \psfrag{s}[c][c][0.8]{$s$}
        \psfrag{l}[c][c][0.8]{$\ell$}
        }
        \vspace{-8mm}
        \caption{Two metaparticles with merged straight and looped sections. (a)~Planar Omega particle. (b)~Twisted Omega or helix particle (single-turn version of the helices in Fig.~\ref{fig:metam_ex}).}
   \label{fig:two_omega_part}
\end{figure}

However, as mentioned in Sec.~\ref{sec:intro} and as will be seen later, chirality is fundamentally related to a magnetoelectric response within the chiral particle. The most natural strategy to realize such a coupled response is to structurally merge the aforementioned straight and looped elements into a `single-block' particle, so that conduction or displacement current continuity in the resulting block adds $\ve{p}_\tx{em}$ to $\ve{p}_\tx{ee}$ and $\ve{p}_\tx{me}$ to $\ve{p}_\tx{mm}$. Such a single-block straight-looped metaparticle could look like the particles that are shown in Fig.~\ref{fig:two_omega_part}, with straight and looped sections of respective lengths $2d$ and $s$, summing up to the unfolded length $\ell$ and interrelated from~\eqref{eq:reson_part} as
\begin{subequations}\label{eq:antag_constr}
\begin{equation}
\ell=2d+s=\frac{\lambda_\tx{res}}{2},
\end{equation}
i.e.,
\begin{equation}\label{eq:antag_constr_ds}
s=\frac{\lambda_\tx{res}}{2}-2d
\quad\tx{or}\quad
d=\frac{\lambda_\tx{res}}{4}-\frac{s}{2}.
\end{equation}
\end{subequations}
Equation~\eqref{eq:antag_constr_ds} reveals that the straight- and looped-section lengths are \emph{antagonistic} to each other, an observation that will be seen in~\ref{sec:des_princ} and~\ref{sec:des_princ} to be of great importance in the response and design of chiral metamaterials.

\subsection{Metaparticle Polarizabilities}\label{sec:mtp_pol}

\begin{figure}[h]
    \centering
        \includegraphics[width=0.75\linewidth]{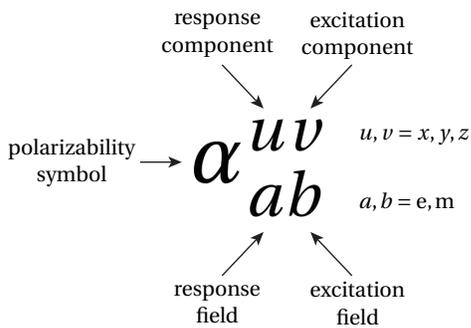}{
        \psfrag{x}[c][c][0.9]{\begin{minipage}{2cm}\centering polarizability \\ symbol\end{minipage}}
        \psfrag{u}[c][c][0.9]{\begin{minipage}{2cm}\centering response \\ component \end{minipage}}
        \psfrag{v}[c][c][0.9]{\begin{minipage}{2cm}\centering excitation \\ component \end{minipage}}
        \psfrag{a}[c][c][0.9]{\begin{minipage}{2cm}\centering response \\ field \end{minipage}}
        \psfrag{b}[c][c][0.9]{\begin{minipage}{2cm}\centering excitation \\ field \end{minipage}}
        \psfrag{y}[l][l][0.9]{$u,v=x,y,z$}
        \psfrag{z}[l][l][0.9]{$a,b=\tx{e},\tx{m}$}
        }
        \vspace{-2mm}
        \caption{Notation for the components of the polarizability dyadic tensors in Eq.~\eqref{eq:polariz_tens}, and susceptibility dyadic tensors to appear farther, in a Cartesian coordinate system. Here, the excitation fields are considered to be $\ve{E}$ and $\ve{H}$, and the responses are the vectorial dipole moments $\ve{p}_\tx{e}$ and $\ve{p}_\tx{m}$ corresponding respectively to the response fields $\ve{D}$ and $\ve{B}$ in the medium formed by these moments (see Sec.~\ref{sec:rel_ma_mi}). The polarizability may be most efficiently read out as ``$u$-directed $a$ response due to $v$-directed $b$ excitation.'' For instance, $\alpha_\tx{em}^{xy}$ is the polarizability component corresponding to the $x$-directed electric response due to a $y$-directed magnetic excitation, or to the polarization $p^x_{\tx{e},H_y}$.}
   \label{fig:alpha_notation}
\end{figure}

As the atoms and molecules in regular materials, the metaparticles in a metamaterial may be conveniently characterized in terms of \emph{polarizabilities}~\cite{Jackson_1998,Ishimaru_1990}. In the case of a general metamaterial, involving anisotropy and magnetoelectric coupling, such characterization may be expressed in terms of the electric dipole moments and magnetic dipole moments induced by the electric and magnetic fields as
\begin{equation}\label{eq:polariz_tens}
\begin{pmatrix}
\ve{p}_\tx{ee} & \ve{p}_\tx{em} \\
\ve{p}_\tx{me} & \ve{p}_\ve{mm}
\end{pmatrix}
=
\begin{pmatrix}
\te{\alpha}_\tx{ee}\cdot\ve{E}_\tx{loc} & \te{\alpha}_\tx{em}\cdot\ve{H}_\tx{loc} \\
\te{\alpha}_\tx{me}\cdot\ve{E}_\tx{loc} & \te{\alpha}_\tx{mm}\cdot\ve{H}_\tx{loc}
\end{pmatrix}
\end{equation}
where $\te{\alpha}_\tx{ee}$, $\te{\alpha}_\tx{em}$, $\te{\alpha}_\tx{me}$ and $\te{\alpha}_\tx{mm}$ are the electric-to-electric, magnetic-to-electric, electric-to-magnetic and magnetic-to-magnetic $3\times{}3$ coupling dyadic tensors, respectively, which are measured in Asm$^2$/V, sm$^2$, sm$^2$ and Vsm$^2$/A (see Appendix~\ref{app:units_const_par}), and where $\ve{E}_\tx{loc}$ and $\ve{H}_\tx{loc}$ are the local excitation fields. Figure~\ref{fig:alpha_notation} details the notation used in this paper for the components of the polarizability tensors in a Cartesian coordinate system. The local excitation fields are the difference between the global excitation fields and the fields produced by the polarization of the neighboring particles in a dense medium~\cite{Jackson_1998,Sihvola_1999}, and reduce to the excitation fields, $\ve{E}$ and $\ve{H}$, in a sufficiently dilute medium. The paper assumes that the dilute-medium approximation is valid, until the final design guidelines in Sec.~\ref{sec:des_princ}, which does not represents a severe restriction in terms of qualitative description.

\section{Two Metaparticle Study Cases}\label{sec:two_part}

\subsection{Mirror-Symmetry Test}\label{sec:mir_sym_test}

\begin{figure}[h]
    \centering
        \includegraphics[width=\linewidth]{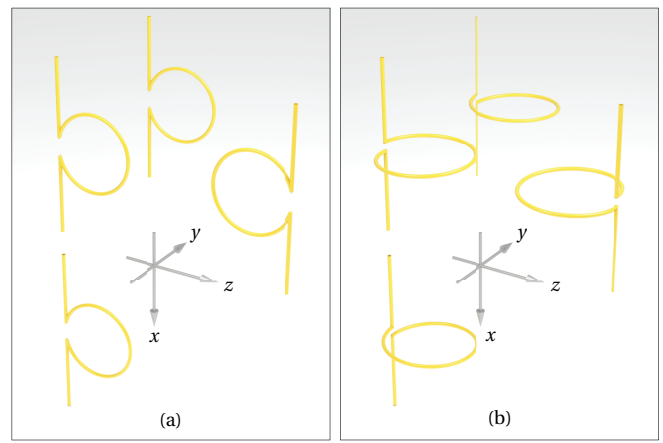}{
        \psfrag{a}[c][c][0.8]{(a)}
        \psfrag{b}[c][c][0.8]{(b)}
        \psfrag{x}[c][c][0.8]{$x$}
        \psfrag{y}[c][c][0.8]{$y$}
        \psfrag{z}[c][c][0.8]{$z$}}
        \vspace{-6mm}
        \caption{Mirror reflections of the two particles in Fig.~\ref{fig:two_omega_part}. (a)~Planar Omega particle [Fig.~\ref{fig:two_omega_part}(a)] (lying here in the $xz$ plane), which is mirror-symmetric and hence not chiral. (b)~Twisted Omega or helix particle [Fig.~\ref{fig:two_omega_part}(b)] (extending in the 3 directions space), which is mirror-asymmetric, and hence chiral. The original particle in the left top is RH, whereas its three images are LH.}
   \label{fig:omega_mirror}
\end{figure}

According to the rationale in Sec.~\ref{sec:metpart_sel}, the planar and twisted straight/looped-section metaparticles, which are respectively shown in Figs.~\ref{fig:two_omega_part}(a) and~\ref{fig:two_omega_part}(b), are both potential candidates for chiral particles, since they both involve magnetoelectric coupling, as we shall verify in Secs.~\ref{sec:MaxAmp_straight_omega} and~\ref{sec:MaxAmp_twisted_omega}. However, according to the definition of Sec.~\ref{sec:intro}, such coupling is only a necessary condition for chirality, the absolute (necessary and sufficient condition) criterion being mirror asymmetry (Sec.~\ref{sec:prelim_com}). Let us then apply the mirror test, depicted in Fig.~\ref{fig:omega_mirror}, to the two metaparticles.

Let us start with the planar Omega particle [Fig.~\ref{fig:two_omega_part}(a)], tested in Fig.~\ref{fig:omega_mirror}(a). Upon reflection in the $x$, $y$ and $z$ directions, this particle transforms into images that are exactly identical to itself. The $z$ direction image is flipped in space, but can be flipped back, without any structural change, to perfectly superimpose with the original particle, and is hence indeed also identical to it. So, the particle is identical to any of its mirror images. Therefore, it is not chiral or, equivalently, has no handedness. It should therefore not induce any polarization rotation, as shall be verified in Sec.~\ref{sec:MaxAmp_straight_omega}.

How about the twisted Omega or helix particle [Fig.~\ref{fig:two_omega_part}(b)], tested Fig.~\ref{fig:omega_mirror}(b)? This particle differs from its planar counterpart only by the 90$^\circ$ twist of the loop section with respect to the straight section. However, this volume twist plays a determinant role in the mirror test: the three images are now different from the original particle; they are LH whereas the original one is RH. So, the twisted Omega particle, as the human hand (Fig.~\ref{fig:mirror_picture}), is chiral, and we shall see in Sec.~\ref{sec:MaxAmp_twisted_omega} that it possesses the consequently expected polarization rotation property~\footnote{\label{fn:odd_refl}If the $z$-mirrored helix particle were next $y$-mirrored, it would naturally flip handedness again, and hence retrieve the original RH handedness. However, such a double-reflection operation ($z\rightarrow-z$ followed by $y\rightarrow-y$) does not correspond to a reflection in the $xy$ direction but to a rotation of 180$^\circ$ about the $x$ axis. In contrast, yet an additional reflection along $x$ would yield again, as a single reflection, the reverse (LH) handedness. This is because a triple reflection  along $x$, $y$ and $z$ ($z\rightarrow-z$, $y\rightarrow-y$, $x\rightarrow-x$) is equivalent to a single reflection in the $xyz$ direction ($\ve{r}\rightarrow-\ve{r}$).}.

Despite the fact that the planar Omega particle is not chiral, we shall still analyze it in the sequel of this section, as its comparison with the twisted Omega or helix particle is instructive for a strong understanding of chirality.

\subsection{Volume Necessary (but Insufficient) Condition}\label{sec:vol}

Comparing Fig.~\ref{fig:two_omega_part}(b) with Fig.~\ref{fig:two_omega_part}(a) shows that the RH helix particle is obtained by twisting the planar particle about the $z$ axis in the clockwise direction, as a key in a lock to open the door, whereas the LH helix particle is obtained by twisting the planar particle about the $z$ axis in the counterclockwise direction, as to close the door. So, the volume twist has imparted handedness, and hence chirality, to the particle.

Such `handedness-ization' would not have been possible without transforming the initially planar structure into a volume one. Indeed, a volume-less, planar structure, such as the planar Omega particle, looks indeed identical from its two sides, and it is only the existence of nonplanarity or volume, as in the helix particle or the hand (Fig.~\ref{fig:mirror_picture})~\footnote{If the hand had no thickness, it would reduce to its mere projection, and hence have neither a palm nor a back. It would therefore look identical from both sides, and have thus no handedness!}, that allows handedness. So, a particle must necessarily include a volume, or thickness, or depth, to be chiral. Moreover, the smallest dimension of this volume must be a significant fraction of the wavelength for a significant chiral effect, since otherwise either the wire section or the loop section would be too small.

However, three-dimensionality is only a necessary condition -- and not a sufficient condition! -- for chirality. The simplest example demonstrating the latter insufficiency is that of a spherical particle. Also, transforming the planar Omega particle into a volume particle by adding to it untwisted looped sections about the $x$ axis (e.g. in the $xz$  plane) still does not make the particle different from its mirror image and hence chiral.

\subsection{Planar Omega Particle (Achiral)}\label{sec:MaxAmp_straight_omega}
\begin{figure}[h]
    \centering
        \includegraphics[width=\linewidth]{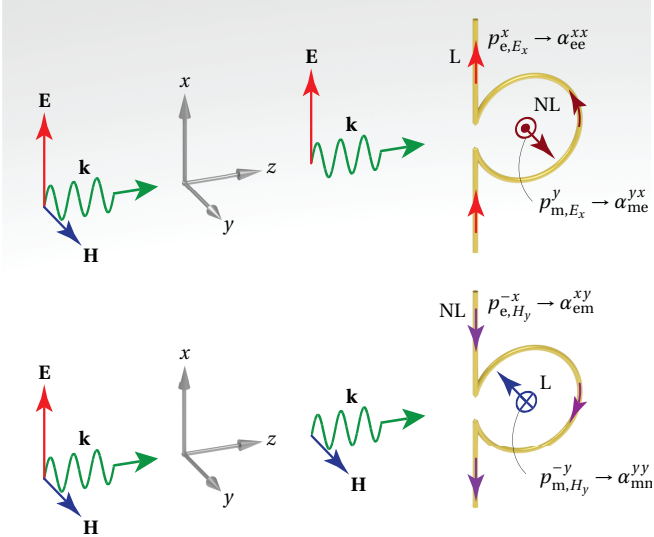}{
        \psfrag{x}[c][c][0.8]{$x$}
        \psfrag{y}[c][c][0.8]{$y$}
        \psfrag{z}[c][c][0.8]{$z$}
        \psfrag{L}[c][c][0.8]{L}
        \psfrag{N}[c][c][0.8]{NL}
        \psfrag{E}[c][c][0.8]{$\ve{E}$}
        \psfrag{H}[c][c][0.8]{$\ve{H}$}
        \psfrag{k}[c][c][0.8]{$\ve{k}$}
        \psfrag{p}[l][l][0.8]{$p^x_{\tx{e},{E_x}}\rightarrow\alpha_\tx{ee}^{xx}$}
        \psfrag{m}[l][l][0.8]{$p^y_{\tx{m},{E_x}}\rightarrow\alpha_\tx{me}^{yx}$}
        \psfrag{q}[l][l][0.8]{$p^{-y}_{\tx{m},{H_y}}\rightarrow\alpha_\tx{mm}^{yy}$}        \psfrag{r}[l][l][0.8]{$p^{-x}_{\tx{e},{H_y}}\rightarrow\alpha_\tx{em}^{xy}$}
        }
        \caption{Electromagnetic response of the planar Omega particle (achiral) to plane-wave excitation. The top and bottom represent the response to the electric and magnetic parts, respectively, of the electromagnetic field excitation. Only the polarization case corresponding to nonzero polarizabilities is shown (polarization $E_x$--$H_y$). Using~\eqref{eq:antag_constr}, one may evaluate here the length of each of the straight sections, $d$, and the circumference of the loop, $s$ to $d=\lambda/10$ and $s=3\lambda/10$.}
   \label{fig:straight_omega}
\end{figure}

The $\ve{p}_\tx{ee,em}$ and $\ve{p}_\tx{me,mm}$ dipolar responses of the planar Omega particle, which is now known to be achiral, are depicted in Fig.~\ref{fig:straight_omega}, where the particle lies in the $xz$ plane with its straight section directed along $x$. Let us separately examine the responses to the electric part (top of the figure) and magnetic part (bottom of the figure) of the electromagnetic field excitation to separately determine the electric-response and magnetic-response polarizability pairs $(\te{\alpha}_\tx{ee},\te{\alpha}_\tx{em}$) and ($\te{\alpha}_\tx{me},\te{\alpha}_\tx{mm}$).

An $x$-directed electric field excitation induces an $x$-directed electric dipole moment $p_{\tx{e},{E_x}}^x$ in the straight section of the particle, which corresponds to the electric-to-electric polarizability $\alpha_\tx{ee}^{xx}$. The current associated with this dipole moment must then flow in the loop section, due to current continuity, and this occurs in the same (upward in the figure) direction given the subwavelength, $\lambda/2$-dimension of the unfolded particle [Eq.~\eqref{eq:mtm_dims}]. This looped current gives rise to the $y$-directed magnetic dipole moment $p_{\tx{m},{E_x}}^y$, which corresponds to the electric-to-magnetic polarizability $\alpha_\tx{me}^{yx}$. Due to symmetry and due to the subwavelength size of the loop, scattering from the $z$-oriented parts of this (looped) current cancels out, and hence does not produce any $\alpha_\tx{ee}^{zx}$; in contrast, scattering from its $x$-oriented parts does not cancel out, due to the asymmetry induced by the gap, and therefore slightly contribute to $\alpha_\tx{ee}^{xx}$.

The response to the magnetic field is found by a symmetric reasoning. A $y$-directed magnetic field excitation induces a $y$-directed magnetic dipole moment $p_{\tx{m},{H_y}}^{-y}$ in the direction opposing the incident field (Lenz law) in the looped section of the particle, which corresponds to the magnetic-to-magnetic polarizability $\alpha_\tx{mm}^{yy}$. Due to the same symmetry reason as before, the associated looped current produces a small $\alpha_\tx{em}^{xy}$ response but no $\alpha_\tx{em}^{zy}$ response. The looped current must then flow along the straight sections, due to current continuity, and this occurs in the same (downward in the figure) direction given the subwavelength, $\lambda/2$-dimension of the unfolded particle. This straight current gives rise to the $x$-directed electric dipole moment $p_{\tx{e},{H_y}}^{-x}$, which corresponds to the main part of the magnetic-to-electric polarizability $\alpha_\tx{em}^{xy}$.

Note that the polarizabilities $\alpha_\tx{me}^{yx}$ and $\alpha_\tx{em}^{xy}$ must have the same magnitude since they involve the same geometrical parts. However, they are oppositely directed. Therefore,
\begin{equation}\label{eq:alpha_meyx_emxy}
\alpha_\tx{me}^{yx}=-\alpha_\tx{em}^{xy}.
\end{equation}
One may easily verify that the $E_x$--$H_y$ polarization considered in Fig.~\ref{fig:straight_omega} is the only one that yields nonzero polarizabilities, assuming that the conductor forming the particle has a deeply-subwavelength, and hence negligible, diameter~\footnote{If the diameter measures a substantial fraction of the wavelength, small responses would also exist in the other directions, leading to extra tensorial components. One could then approximate the different sections by straight and looped ellipsoids, and specifically prolate spheroids with large (needle-type) axis ratio. For instance, if the aspect ratio is $10:1$, the polarizability of a straight needle reads $\alpha=V\epsilon_0(\epsilon_\tx{r}-1)/[1+K(\epsilon_\tx{r}-1)]=V\epsilon_0/K$, assuming that it is perfectly conducting ($\epsilon_\tx{r}\rightarrow i\infty$), where $V$ is the volume of the needle, $\epsilon_0$ the free-space permittivity, and $K$ the depolarization factor in the direction considered. For an $x$-oriented needle, the depolarization factors are found to be $K_x=0.02$ and $K_y=K_z=0.49$~\cite{Sihvola_1999}. Hence there exists polarizability components perpendicular to the needle axis, but they are $0.49/0.02=24.5$ smaller than that along the axis, and may thus be neglected. So, for sections with a length-to-diameter ratio of more than 10, such effects can be safely ignored.}. The global susceptibility tensors of the planar Omega particle in Fig.~\ref{fig:straight_omega} are then~\cite{Saadoun_1992}
\begin{equation}\label{eq:planar_Omeg_tens}
\begin{pmatrix}
\te{\alpha}_\tx{ee} & \te{\alpha}_\tx{em} \\
\te{\alpha}_\tx{me} & \te{\alpha}_\tx{mm}
\end{pmatrix}
=
\begin{pmatrix}
\begin{pmatrix}
\alpha_\tx{ee}^{xx} & 0 & 0 \\
0 & 0 & 0 \\
0 & 0 & 0
\end{pmatrix}
\begin{pmatrix}
0 & \alpha_\tx{em}^{xy} & 0 \\
0 & 0 & 0 \\
0 & 0 & 0
\end{pmatrix} \\
\begin{pmatrix}
0 & 0 & 0 \\
-\alpha_\tx{em}^{xy} & 0 & 0 \\
0 & 0 & 0
\end{pmatrix}
\begin{pmatrix}
0 & 0 & 0 \\
0 & \alpha_\tx{mm}^{yy} & 0 \\
0 & 0 & 0
\end{pmatrix}
\end{pmatrix}.
\end{equation}
The tensor~\eqref{eq:planar_Omeg_tens} can be modified by transforming the `monoatomic' metaparticle of Fig.~\ref{fig:straight_omega}(a) into a `biiatomic' metaparticle or `triatomic' metaparticle, obtained by adding copies of the initial particle (here in the $xz$ or $zx$ plane) in the other two or three planes. This can be done in a diversity of ways. In each plane, the planar Omega particle can take one out of 4 distinct orientations: the straight section can be directed along the two perpendicular directions of the plane and for each of these orientations the looped section may point to two opposite directions. This may be best seen by using proper labeling. For instance, the particle of Fig.~\ref{fig:straight_omega}(a) may be labelled $(zx,x,+z)$, indicating that it is lying in the $zx$ plane, with straight section in the $x$ direction and looped section pointing towards the $+z$ direction, and the same plane supports also the three other orientations $(zx,x,-z)$, $(zx,z,+x)$ and $(zx,z,-x)$. There are then $4^1=4$ possibilities for a monoatomic particle, $4^2=16$ possibilities for a biatomic particle, and $4^3=64$ possibilities for a triatomic particle.

The triatomic particle corresponding to the cyclic permutations of the particle in Fig.~\ref{fig:straight_omega}(a), i.e., [$(zx,x,+z)$,$(xy,y,+x)$,$(yz,z,+y)$] can be easily found, by the same permutations, to correspond to the metaparticle tensors
\begin{equation}\label{eq:planar_Omeg_tens_3}
\begin{pmatrix}
\te{\alpha}_\tx{ee} & \te{\alpha}_\tx{em} \\
\te{\alpha}_\tx{me} & \te{\alpha}_\tx{mm}
\end{pmatrix}
=
\begin{pmatrix}
\alpha_\tx{ee}\te{I} & \alpha_\tx{em}\te{I}_\tx{P} \\
\alpha_\tx{me}\te{I}_\tx{P} & \alpha_\tx{mm}\te{I}
\end{pmatrix},
\end{equation}
where $\te{I}$ and $\te{I}_\tx{P}$ are the symmetric and permutated unit tensors $\te{I}=\hatv{x}\hatv{x}+\hatv{y}\hatv{y}+\hatv{z}\hatv{z}$ and $\te{I}_\tx{P}=\hatv{x}\hatv{y}+\hatv{y}\hatv{z}+\hatv{z}\hatv{x}$, respectively. In such a metaparticle, the direct (ee and mm) tensors have reduced to scalar, but the cross (em and me) tensors have not, so that the overall response is still anisotropic.

The most `symmetric' nontrivial medium that can be obtained with the the planar Omega particle is in fact the hexatomic medium with the particles $(zx,x,+z)$ and $(yz,y,+z)$ providing $\hatv{z}\times\te{I}$, $(xy,y,+x)$ and $(zx,z,+x)$ providing $\hatv{x}\times\te{I}$, and $(xy,x,+y)$ and $(yz,z,+y)$ providing $\hatv{y}\times\te{I}$. The corresponding polarizability tensors may be written
\begin{equation}\label{eq:planar_Omeg_tens_6}
\begin{pmatrix}
\te{\alpha}_\tx{ee} & \te{\alpha}_\tx{em} \\
\te{\alpha}_\tx{me} & \te{\alpha}_\tx{mm}
\end{pmatrix}
=
\begin{pmatrix}
\alpha_\tx{ee}\te{I} & \alpha_\tx{em}\te{I}_\tx{A} \\
\alpha_\tx{me}\te{I}_\tx{A} & \alpha_\tx{mm}\te{I}
\end{pmatrix},
\end{equation}
where $\te{I}_\tx{A}$ is the (6-component) antisymmetric tensor $\te{I}_\tx{A}=\ve{r}\times\te{I}$. However, such a metaparticle involves 2 particles per plane, which might be at odds with with the dimensional constraint~\eqref{eq:mtm_dims}.

Note that allowing more than one planar Omega particle per plane may also wash out the magnetoelectric coupling effects of the resulting multiatomic metaparticle. For instance, if the $zy$ plane were allowed to support the $(zx,x,-z)$ particle in addition to the $(zx,x,+z)$ particle in Fig.~\ref{fig:straight_omega}, then the responses $\alpha_\tx{me}^{yx}$ and $\alpha_\tx{em}^{xy}$ would disappear due to cancelation of the scattering from the two particles. The hexaomic metaparticle obtained by adding the two complementary cancelling pairs would naturally lead to a magnetoelectic-less structure.

How about polarization rotation? We have seen above that, for the polarization selected in Fig.~\ref{fig:straight_omega}, the electric response of the metaparticle is fully parallel the electric excitation ($\ve{p}_\tx{ee}\|\ve{E}$ and $\ve{p}_\tx{em}\|\ve{E}$) and the magnetic response is fully parallel to its magnetic excitation ($\ve{p}_\tx{me}\|\ve{H}$ and $\ve{p}_\tx{mm}\|\ve{H}$), despite coupling, as may be easily checked in~\eqref{eq:planar_Omeg_tens}, ~\eqref{eq:planar_Omeg_tens_3} and~\eqref{eq:planar_Omeg_tens_6}. This means that the particle does induce any rotation for this polarization. It only induces a phase shift, say $\phi_x$, corresponding to its interaction with the wave. But the situation is a little more subtle. Consider now the polarization rotated by 90$^\circ$ (i.e. $\ve{E}\|\hatv{y}$ and $\ve{H}\|\hatv{y}$). In this case, the particle does essentially not interact with the wave and is hence invisible to it, so that it does not induce any phase shift, or $\phi_y=0$. The particle is hence \emph{birefringent}, which indeed rotates the obliquely polarized incident wave. However, as will be seen in Sec.~\ref{sec:nonlocal_omega}, this birefringence is fundamentally different from chiral birefringence, as can readily be realized by the fact that it does not occur for all the incident polarizations.

In summary, we have found that the planar Omega particle in Fig.~\ref{fig:straight_omega} has the following properties:
\begin{enumerate}
  \item It indeed involves magnetoelectric coupling, as predicted in Sec.~\ref{sec:metpart_sel}, since $\te{\alpha}_\tx{em},\te{\alpha}_\tx{me}\neq 0$, which is the reason it was called pseudo-chiral in~\cite{Saadoun_1992}, and $\te{\alpha}_\tx{me}=-\te{\alpha}_\tx{em}$;
  \item Despite such coupling, it is achiral, by definition, since it is identical to its mirror image, as shown in Sec.~\ref{sec:mir_sym_test} (proof of \ding{196} in Fig.~\ref{fig:trilogy});
  \item It cannot reduce to an isotropic particle upon adding copies in the other two directions of space.
  \item As expected from its achiral nature (Sec.~\ref{sec:intro}) and despite its magnetoelectric coupling, it does not involve systematic polarization rotation, i.e., it is not completely gyrotropic (proof of \ding{198} in Fig.~\ref{fig:trilogy}).
\end{enumerate}

\subsection{Twisted Omega or Helix Particle (Chiral)}\label{sec:MaxAmp_twisted_omega}
\begin{figure}[h]
    \centering
        \includegraphics[width=\linewidth]{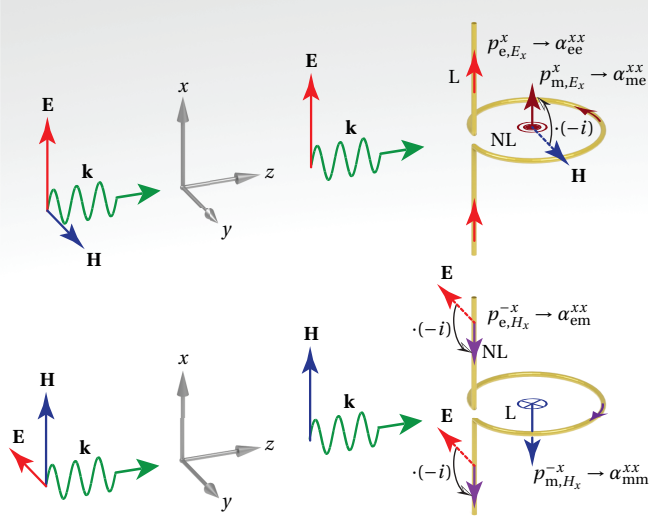}{
        \psfrag{i}[c][c][0.8]{$\cdot(-i)$}
        \psfrag{x}[c][c][0.8]{$x$}
        \psfrag{y}[c][c][0.8]{$y$}
        \psfrag{z}[c][c][0.8]{$z$}
        \psfrag{L}[c][c][0.8]{L}
        \psfrag{N}[c][c][0.8]{NL}
        \psfrag{E}[c][c][0.8]{$\ve{E}$}
        \psfrag{H}[c][c][0.8]{$\ve{H}$}
        \psfrag{k}[c][c][0.8]{$\ve{k}$}
        \psfrag{p}[l][l][0.8]{$p^x_{\tx{e},{E_x}}\rightarrow\alpha_\tx{ee}^{xx}$}
        \psfrag{m}[l][l][0.8]{$p^x_{\tx{m},{E_x}}\rightarrow\alpha_\tx{me}^{xx}$}
        \psfrag{q}[l][l][0.8]{$p^{-x}_{\tx{m},{H_x}}\rightarrow\alpha_\tx{mm}^{xx}$}        \psfrag{r}[l][l][0.8]{$p^{-x}_{\tx{e},{H_x}}\rightarrow\alpha_\tx{em}^{xx}$}
        }
        \caption{Electromagnetic response of the twisted Omega or helix (here RH) particle (chiral) to plane-wave excitation. The top and bottom represent the responses to the electric and magnetic parts, respectively, of the electromagnetic field excitation. Only the polarization cases corresponding to nonzero polarizabilities are shown (polarizations $E_x$--$H_y$ and $-E_y$--$H_x$). The dimensional comment in the caption of Fig.~\ref{fig:straight_omega} also holds here.}
   \label{fig:twisted_omega}
\end{figure}

The $\ve{p}_\tx{ee,em}$ and $\ve{p}_\tx{me,mm}$ dipolar responses of the twisted Omega or helix particle, which is now known to be chiral, are depicted in Fig.~\ref{fig:twisted_omega}, where the straight section is directed along $x$, while the loop section lies in the $yz$ plane. Let us again separately examine the responses to the electric part (top of the figure) and magnetic part (bottom of the figure) of the electromagnetic field excitation to separately determine the electric-response and magnetic-response polarizability pairs $(\te{\alpha}_\tx{ee},\te{\alpha}_\tx{em}$) and ($\te{\alpha}_\tx{me},\te{\alpha}_\tx{mm}$).

An $x$-directed electric field excitation induces again an $x$-directed electric dipole moment $p_{\tx{e},{E_x}}^x$ on the straight section of the particle, corresponding to $\alpha_\tx{ee}^{xx}$, and the current associated with this dipole moment again flows in the loop section from current continuity. However, given the $90^\circ$ twist of the loop, this current now gives rise to the $x$-directed magnetic dipole moment $p_{\tx{m},E_x}^x$, which corresponds to $\alpha_\tx{me}^{xx}$. Note in passing the if the angle of the loop twist were not exactly $90^\circ$ -- e.g. $60^\circ$ -- then the induced magnetic moment would be tilted, which would introduce parasitic off-axis contributions to the response. Moreover, an $x$-directed magnetic field excitation (different form the $y$-directed one at the bottom of Fig.~\ref{fig:straight_omega}) induces an $x$-directed magnetic dipole moment $p_{\tx{m},H_x}^{-x}$ in the looped section of the particle, corresponding to $\alpha_\tx{mm}^{xx}$, plus, from current continuity, the $x$-directed electric dipole moment $p_{\tx{e},{H_y}}^{-x}$, corresponding to $\alpha_\tx{em}^{xx}$. Again, the cross polarizations are opposite, i.e.,
\begin{equation}\label{eq:alphamexxemxx}
\alpha_\tx{me}^{xx}=-\alpha_\tx{em}^{xx}.
\end{equation}
In addition, the looped current produces the responses $\alpha_\tx{ee}^{yx}$ and $\alpha_\tx{em}^{yx}$ (without $\alpha_\tx{ee}^{zx}$ and $\alpha_\tx{em}^{zx}$) due to the gap asymmetry; however, these responses are negligibly small as the $yz$ plane projection of the gap is very small and may even reach zero, and we therefore henceforth consider them negligible. These observations indicate that a mirror asymmetry, corresponding to a volume twist, necessarily induces magnetoelectric coupling (proof of \ding{194} in Fig.~\ref{fig:trilogy}).

\begin{figure}[h]
    \centering
        \includegraphics[width=0.6\linewidth]{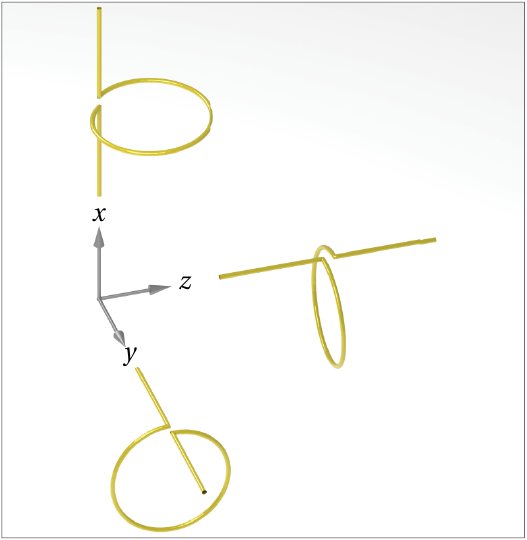}{
        \psfrag{x}[c][c][0.8]{$x$}
        \psfrag{y}[c][c][0.8]{$y$}
        \psfrag{z}[c][c][0.8]{$z$}}
        \vspace{-3mm}
        \caption{`Triatomic' metaparticle from by the combination of helical metaparticles (Fig.~\ref{fig:twisted_omega}) oriented in the three directions of space.}
   \label{fig:twisted_omegas_3_orientations}
\end{figure}

So, the four main susceptibilities identified above correspond to the dyadic component $xx$, and it is easy to verify that all the other susceptibility components are negligible. The helix metaparticle in Fig.~\ref{fig:twisted_omega} has thus the polarizability tensors
\begin{equation}\label{eq:twisted_Omeg_tens}
\begin{pmatrix}
\te{\alpha}_\tx{ee} & \te{\alpha}_\tx{em} \\
\te{\alpha}_\tx{me} & \te{\alpha}_\tx{mm}
\end{pmatrix}
=
\begin{pmatrix}
\begin{pmatrix}
\alpha_\tx{ee}^{xx} & 0 & 0 \\
0 & 0 & 0 \\
0 & 0 & 0
\end{pmatrix}
\begin{pmatrix}
\alpha_\tx{em}^{xx} & 0 & 0 \\
0 & 0 & 0 \\
0 & 0 & 0
\end{pmatrix} \\
\begin{pmatrix}
-\alpha_\tx{em}^{xx} & 0 & 0 \\
0 & 0 & 0 \\
0 & 0 & 0
\end{pmatrix}
\begin{pmatrix}
\alpha_\tx{mm}^{xx} & 0 & 0 \\
0 & 0 & 0 \\
0 & 0 & 0
\end{pmatrix}
\end{pmatrix},
\end{equation}
and the `triatomic' metaparticle formed by adding copies of that particle in the $y$ and $z$ directions, as shown in Fig.~\ref{fig:twisted_omegas_3_orientations}, has the polarizability tensors
\begin{equation}\label{eq:twisted_Omeg_tens_3_or}
\begin{pmatrix}
\te{\alpha}_\tx{ee} & \te{\alpha}_\tx{em} \\
\te{\alpha}_\tx{me} & \te{\alpha}_\tx{mm}
\end{pmatrix}
=
\begin{pmatrix}
\alpha_\tx{ee}\te{I} & \alpha_\tx{em}\te{I} \\
-\alpha_\tx{em}\te{I} & \alpha_\tx{mm}\te{I}
\end{pmatrix},
\end{equation}
which have this time reduced to scalars. Remarkably, and in contrast with the 64~possibilities of the straight Omega particle, this arrangement the helix particle is unique~\footnote{This, of course, assumes only one type of handedness, either RH or LH. If we allow both LH and RH particles, we have $2^3=8$ possible combinations (2 handednesses in each of the 3 directions of space). However, 6 of these combinations lead to different signs in different directions, which breaks biisotropy. The arrangement of Fig.~\ref{fig:twisted_omegas_3_orientations} -- and its LH counterpart whose $\chi$ has the opposite sign -- are the only ones yielding purely scalar polarizabilities, or biisotropy (see Secs.~\ref{sec:biisot_chir} and~\ref{sec:gen_biis_med}).}! In this sense, the helix particle is much more symmetric and fundamental than the planar Omega particle.

Given the particle chirality, we expect to have here polarization rotation. Indeed, we have, along with $\ve{p}_\tx{ee}\|\ve{E}$ and $\ve{p}_\tx{mm}\|\ve{H}$, that $\ve{p}_\tx{me}\perp\ve{H}$ and $\ve{p}_\tx{em}\perp\ve{E}$, indicating that part of the magnetic field and electric field have been rotated by 90$^\circ$, which is a clear indication the expected gyrotropy. Specifically, the initially $+y$-directed field $\ve{H}$ (top of the figure) has rotated towards the $+x$ direction into $p_{\tx{m},E_x}^x$, while the initially $-y$-directed field $\ve{E}$ (bottom of the figure) has rotated towards the $-x$ direction into $p_{\tx{e},{H_x}}^{-x}$. The electromagnetic field has therefore rotated about the $z$ axis in the direction corresponding to the left-hand with the thumb pointing in the propagation direction ($+z$). In other words, the field phasors associated with the cross coupling terms have rotated by the angle of $-\pi/2$ ($y$ to $x$ direction) or, equivalently, have been multiplied by the factor $e^{-i\pi/2}=-i$, and therefore the cross coupling polarizabilities are in quadrature with their direct coupling counterparts. A LH particle naturally leads to rotation in the opposite direction.

In contrast to the planar Omega particle (Fig.~\ref{fig:straight_omega}), the twisted Omega or helix particle involves two different polarization states in its nonzero polarizabilities. Specifically, in Fig.~\ref{fig:twisted_omega}, the polarization $(E_x,H_y)$ involves $\alpha_\tx{ee}^{xx}$ and $\alpha_\tx{me}^{xx}$, while the polarization $(-E_y,H_x)$ involves $\alpha_\tx{mm}^{xx}$ and $\alpha_\tx{em}^{xx}$. In the case of a circularly polarized exciting wave, these two states are separated by the time interval $T/4$ ($T=\omega/(2\pi)$: time period) of the harmonic wave, which correspond to a factor $e^{-i\pi/2}=-i$ in phasor notation. These observations indicate mirror asymmetry necessarily induces polarization rotation (proof of \ding{193} in Fig.~\ref{fig:trilogy}).

In summary, we have found that the twisted Omega or helix particle in Fig.~\ref{fig:straight_omega} has the following properties:
\begin{enumerate}
  \item It indeed involves magnetoelectric coupling, as predicted in Sec.~\ref{sec:metpart_sel}, since $\te{\alpha}_\tx{em},\te{\alpha}_\tx{me}\neq 0$, and $\te{\alpha}_\tx{me}=-\te{\alpha}_\tx{em}$;
  \item It is chiral, by definition, since it is different to its mirror image, as shown in Sec.~\ref{sec:mir_sym_test}, as allowed by its volume configuration (Sec.~\ref{sec:vol});
  \item It reduces to an isotropic particle upon adding copies in the other two directions of space, since the response is then purely diagonal.
  \item As expected from its chiral nature (Sec.~\ref{sec:intro}), it induces polarization rotation, i.e., it is gyrotropic.
  \item \label{list:polar_quadr} The polarizability pairs ($\alpha_\tx{ee}^{xx},\alpha_\tx{me}^{xx}$) and ($\alpha_\tx{mm}^{xx},\alpha_\tx{em}^{xx}$) are in a quadrature relationship.
\end{enumerate}

Analytical formulas for the values of the polarizabilities of the helix particle in terms of its geometrical parameters are given in~\cite{Tretyakov_1996}. Alternative chiral particles may be easily inferred from the observations made in this section. First, a chiral particle must necessarily have a volume, according to Sec.~\ref{sec:vol}, which disqualifies all purely planar particles. Even twisted planar particles, such as all kinds of gammadion crosses, are achiral. Such structures have a sort of handedness, but this is a weak form of handedness, since it changes with the side from which they are observed, so that they are identical to their mirror image.  However, chiral particles may take alternative shapes, such their multiturn-helix, or spring, version in Fig.~\ref{fig:metam_ex}, their square-loop version of the helix in Fig.~\ref{fig:two_omega_part}, or structures looking like that the amino acids represented in Fig.~\ref{fig:chir_handedness}.

\section{Microscopic-to-Macroscopic Scales Conversion}\label{sec:rel_ma_mi}
Section~\ref{sec:mtp_pol} has shown that metaparticles can be characterized by polarizability tensors, according to the relation~\eqref{eq:polariz_tens}, and Secs~\ref{sec:MaxAmp_straight_omega} and~\ref{sec:MaxAmp_twisted_omega} have shown how to determine the structure of these tensors for the planar Omega particle and for the twisted Omega or helix particle, respectively. Arranging copies of such particles in the three dimensions of space, either randomly or according to a crystal-like lattice, and under the dimensional constraints outlined in Sec.~\ref{sec:dim_constr}, forms a metamaterial, as described in Sec.~\ref{sec:meta_motiv_def}. This section outlines the conversion from the \emph{microscopic} scale of the metaparticles to the \emph{macroscopic} scale of the corresponding metamaterial.

As previously mentioned, assume here dilute metamaterials, i.e. metamaterials with relatively low metaparticle density, and hence negligible inter-particle coupling. This assumption is often not valid, particularly in metamaterials leveraging tight inter-particle coupling for broad operational bandwidth~\cite{Caloz_2005}. However, it is acceptable to understand the essence of chirality, and it may be relaxed by including interaction tensors~\cite{Collin_1990} and overcome by full-wave analysis (see Sec.~\ref{sec:des_princ}).

Under the aforementioned assumption of negligible inter-particle interaction, the microscopic electric and magnetic dipole moments $\ve{p}_\tx{ee,em}$ and $\ve{p}_\tx{me,mm}$ can be simply averaged in space and orientation to provide the corresponding macroscopic electric and magnetic polarization densities, $\ve{P}_\tx{e}$ (As/m$^2$) and $\ve{P}_\tx{m}=\mu_0\ve{M}$ (Vs/m$^2$), where $\ve{M}$ is the usual magnetization~\cite{Jackson_1998,Sihvola_1999,Ishimaru_1990}, from which the medium properties follow. This leads here to the relations
\begin{subequations}\label{eq:pol}
\begin{equation}
\begin{split}
\ve{P}_\tx{e}&=N\langle\ve{p}_\tx{e}\rangle
=N\big(\langle\ve{p}_\tx{ee}\rangle+\langle\ve{p}_\tx{em}\rangle\big) \\
&=N\left(\langle\te{\alpha}_\tx{ee}\cdot\ve{E}\rangle
+\langle\te{\alpha}_\tx{em}\cdot\ve{H}\rangle\right) \\
&=N\left(\langle\te{\alpha}_\tx{ee}\rangle\cdot\ve{E}
+\langle\te{\alpha}_\tx{em}\rangle\cdot\ve{H}\right),
\end{split}
\end{equation}
\begin{equation}
\begin{split}
\ve{P}_\tx{m}&=N\langle\ve{p}_\tx{m}\rangle
=N\big(\langle\ve{p}_\tx{em}\rangle+\langle\ve{p}_\tx{mm}\rangle\big) \\
&=N\left(\langle\te{\alpha}_\tx{me}\cdot\ve{E}\rangle
+\langle\te{\alpha}_\tx{mm}\cdot\ve{H}\rangle\right) \\
&=N\left(\langle\te{\alpha}_\tx{me}\rangle\cdot\ve{E}
+\langle\te{\alpha}_\tx{mm}\rangle\cdot\ve{H}\right),
\end{split}
\end{equation}
\end{subequations}
where the $\langle\cdot\rangle$ symbol represents the averaging operation, $N$ (1/m$^3$) denotes the particle density, and the other quantities were defined in Secs.~\ref{sec:metpart_sel} and~\ref{sec:mtp_pol}. Consistently with the negligible inter-particle coupling assumption, we have dropped the subscript `loc' that appeared in~\eqref{eq:polariz_tens} (see Sec.~\ref{sec:mtp_pol}) in the second equalities, and considered the metaparticles to be aligned according to a lattice with well-defined coordinate system in the third equalities.

The particle average densities of polarizabilities in~\eqref{eq:pol} are related with the corresponding medium \emph{susceptibilities} $\te{\chi}_{ab}$ by
\begin{subequations}\label{eq:rel_al_chi}
\begin{equation}
N\,\langle\te{\alpha}_\tx{ab}\rangle=c_{ab}\te{\chi}_{ab},
\end{equation}
with the normalizing factors
\begin{equation}
c_\tx{ee}=\epsilon_0,\quad
c_\tx{em}=c_\tx{em}=\sqrt{\epsilon_0\mu_0},\quad
c_\tx{mm}=\mu_0,
\end{equation}
\end{subequations}
where $\epsilon_0=8.854\cdot 10^{-12}$~As/Vm and $\mu_0=4\pi\cdot 10^{-7}$~Vs/Am are the free-space permittivity and permeability, respectively~\cite{Jackson_1998,Sihvola_1999,Ishimaru_1990}. Substituting~\eqref{eq:rel_al_chi} into~\eqref{eq:pol} transforms these relations into
\begin{subequations}\label{eq:bianis_polar}
\begin{align}
\ve{P}_\tx{e}
&=\epsilon_0\te{\chi}_\tx{ee}\cdot\ve{E}
+\sqrt{\epsilon_0\mu_0}\,\te{\chi}_\tx{em}\cdot\ve{H}, \\
\ve{P}_\tx{m}
&=\sqrt{\epsilon_0\mu_0}\,\te{\chi}_\tx{me}\cdot\ve{E}
+\mu_0\te{\chi}_\tx{mm}\cdot\ve{H},
\end{align}
\end{subequations}
where $\te{\chi}_\tx{ee}$, $\te{\chi}_\tx{em}$, $\te{\chi}_\tx{me}$ and $\te{\chi}_\tx{mm}$ are the (unitless) electric-to-electric, magnetic-to-electric, electric-to-magnetic and magnetic-to-magnetic susceptibility dyadic tensors, respectively, whose notation follows the conventions in Fig.~\ref{fig:alpha_notation}.

\section{Constitutive Relations}\label{sec:macr_const_rel}

\subsection{General Bianisotropic Relations}
The behavior of an electromagnetic medium can generally be expressed by the relations~\cite{Kong_EWT_2008,Jackson_1998,Ishimaru_1990}
\begin{subequations}\label{eq:gen_const_rel}
\begin{align}\label{eq:gen_const_rel_D}
\ve{D}&=\epsilon_0\ve{E}+\ve{P}_\tx{e}, \\
\label{eq:gen_const_rel_B}
\ve{B}&=\mu_0\ve{H}+\ve{P}_\tx{m},
\end{align}
\end{subequations}
where $\epsilon_0\ve{E}$ and $\mu_0\ve{H}$ respectively represent the electric and magnetic responses of free space, which corresponds to the spacings between the molecules or metaparticles, while the $\ve{P}_\tx{e}$ and $\ve{P}_\tx{e}$ represents the response of the particle forming the medium and given by~\eqref{eq:bianis_polar}. Equations~\eqref{eq:gen_const_rel} correspond to the most usual electromagnetics convention of $[\ve{D};\ve{B}]$ being expressed in terms of $[\ve{E};\ve{H}]$, where $\ve{E}$ and $\ve{H}$ are considered as the excitations while $\ve{B}$ and $\ve{D}$ are considered as the medium responses~\cite{Jackson_1998,Ishimaru_1990}~\footnote{The book~\cite{Kong_EWT_2008} also includes the alternative formulations $[\ve{D};\ve{H}]$ versus $[\ve{E};\ve{B}]$, based on the motivation that $\ve{E}$ and $\ve{B}$ are the fields that are directly measurable experimentally via the Lorentz force ($\ve{F}=q\left(\ve{E}+\ve{v}\times\ve{B}\right)$; $q$: charge, $\ve{v}$: charge velocity), and $[\ve{E};\ve{H}]$ versus $[\ve{D};\ve{B}]$, based on the motivation that $\ve{D}$ and $\ve{B}$ form with the spectral spatial wavevector $\ve{k}$ an electromagnetic basis that does not dependent on the nature of the medium.}.

We now relate the field responses to the medium susceptibilities by inserting~\eqref{eq:bianis_polar} into~\eqref{eq:gen_const_rel}, which yields
\begin{subequations}\label{eq:bianis_decomp}
\begin{align}
\ve{D}
&=\epsilon_0\left(\te{I}+\te{\chi}_\tx{ee}\right)\cdot\ve{E}
+\sqrt{\epsilon_0\mu_0}\,\te{\chi}_\tx{em}\cdot\ve{H}, \\
\ve{B}
&=\sqrt{\epsilon_0\mu_0}\,\te{\chi}_\tx{me}\cdot\ve{E}
+\mu_0\left(\te{I}+\te{\chi}_\tx{mm}\right)\cdot\ve{H}.
\end{align}
\end{subequations}
Defining
\begin{equation}\label{eq:exzm_from_xhi}
\begin{pmatrix}
\te{\epsilon} & \te{\xi} \\
\te{\zeta} & \te{\mu}
\end{pmatrix}
=
\begin{pmatrix}
\epsilon_0\left(\te{I}+\te{\chi}_\tx{ee}\right) & \sqrt{\epsilon_0\mu_0}\,\te{\chi}_\tx{em}\\
\sqrt{\epsilon_0\mu_0}\,\te{\chi}_\tx{me}& \mu_0\left(\te{I}+\te{\chi}_\tx{mm}\right)
\end{pmatrix}
\end{equation}
transforms~\eqref{eq:bianis_decomp} into the conventional \emph{bianisotropic}~\footnote{In the term `bianisotropic', coined by Kong in~\cite{Cheng_1968}, `bi' refers to the fact that each (electric and magnetic) response depends on both the electric excitation and the magnetic excitation, while `anisotropic' refers to the fact the responses are not parallel to their excitations or, equivalently, are characterized by tensorial parameters. From~\eqref{eq:rel_al_chi}, it is clear that this term applies to the polarizabilities as well as to the susceptibilities.} relations~\cite{Kong_1972,Lindell_EWCBM_1994,Kong_EWT_2008}
\begin{subequations}\label{eq:bianis_const_rel}
\begin{align}\label{eq:bianis_const_rel_D}
\ve{D}&=\te{\epsilon}\cdot\ve{E}+\te{\xi}\cdot\ve{H}, \\
\label{eq:bianis_const_rel_B}
\ve{B}&=\te{\zeta}\cdot\ve{E}+\te{\mu}\cdot\ve{H},
\end{align}
\end{subequations}
where $\te{\epsilon}$, $\te{\mu}$, $\te{\xi}$ and $\te{\zeta}$ are the permittivity, permeability, magnetic-to-electric coupling and electric-to-magnetic coupling dyadic tensors, respectively, which are measured in As/Vm, Vs/Am and s/m (see Appendix~\ref{app:units_const_par}). The four medium tensors are generally of dimension $3\times 3$, and involve thus 36~complex parameters overall.

The coupled equations~\eqref{eq:bianis_const_rel} -- where $\ve{D}$ and $\ve{B}$ are interdependent through the coupling tensors $\te{\xi}$ and $\te{\zeta}$ -- represent the most general explicit constitutive relations for an LTI medium. Linear time-invariance [Assumption~\ref{it:LTI}) in Sec.~\ref{sec:glob_assump}] is indeed a necessary condition for such relations to hold, since nonlinearity would involve powers of $\ve{E}$ and $\ve{H}$~\cite{Boyd_2008}, while time variance would require replacing the scalar products by convolution products~\cite{Kalluri_ETVCM_2010}.

As we have seen in Sec.~\ref{sec:MaxAmp_straight_omega}, the straight Omega particle [Fig.~\ref{fig:two_omega_part}(a) or Fig.~\ref{fig:straight_omega}(a)], whether unique or combined with copies of itself in different directions of space, is always characterized by non-reducible bianisotropic relations [Eqs.~\eqref{eq:planar_Omeg_tens}, \eqref{eq:planar_Omeg_tens_3} and~\eqref{eq:planar_Omeg_tens_6}]. Therefore, they correspond to the most general relations~\eqref{eq:bianis_const_rel}, whose tensorial parameters are obtained from~\eqref{eq:planar_Omeg_tens}, \eqref{eq:planar_Omeg_tens_3} or~\eqref{eq:planar_Omeg_tens_6} via~\eqref{eq:rel_al_chi}.

\subsection{Biisotropic Chiral Media}\label{sec:biisot_chir}
In the particular case where the tensors $\te{\epsilon}$, $\te{\mu}$, $\te{\xi}$ and $\te{\zeta}$ in~\eqref{eq:bianis_const_rel} reduce to $\epsilon\te{I}$, $\mu\te{I}$, $\xi\te{I}$ and $\zeta\te{I}$, and hence ultimately to scalars, the medium is called \emph{biisotropic}. This is for instance the case of the medium formed by the triatomic helix particle shown in Fig.~\ref{fig:twisted_omegas_3_orientations} and characterized by the polarizability tensor~\eqref{eq:twisted_Omeg_tens_3_or}. Equations~\eqref{eq:bianis_const_rel} reduce then to
\begin{subequations}\label{eq:biis_const_rel}
\begin{align}\label{eq:biis_const_rel_D}
\ve{D}&=\epsilon\ve{E}+\xi\ve{H}, \\
\label{eq:biis_const_rel_B}
\ve{B}&=\zeta\ve{E}+\mu\ve{H},
\end{align}
\end{subequations}
and a biisotropic medium is thus characterized only by the 4~complex parameters $\epsilon$, $\xi$, $\zeta$ and $\mu$.

We have established in Sec.~\ref{sec:MaxAmp_twisted_omega} that the cross coupling parameters of a chiral structure were opposite to each other [Eq.~\eqref{eq:alphamexxemxx}] and in quadrature with the direct coupling parameters (Fig.~\ref{fig:twisted_omega}). In terms of the constitutive parameters in~\eqref{eq:biis_const_rel}, this translates into
\begin{equation}\label{eq:xizetaepsmu_rel}
\zeta=-\xi
\quad\tx{and}\quad
\angle\zeta,\angle\xi=\angle\epsilon\pm\pi/2=\angle\mu\pm\pi/2,
\end{equation}
where the last equality expresses the fact the the electric and magnetic responses of a plane wave in a simple biisotropic medium -- such as free space -- are in phase, as seen by inserting~\eqref{eq:PW_space} into~\eqref{eq:Maxwell_curl}. In order to make sure to select the correct signs in~\eqref{eq:xizetaepsmu_rel}, let us examine the behavior of the RH helix particle described in Fig.~\ref{fig:twisted_omega} in some more details.

\begin{figure}[h]
    \centering
        \includegraphics[width=\linewidth]{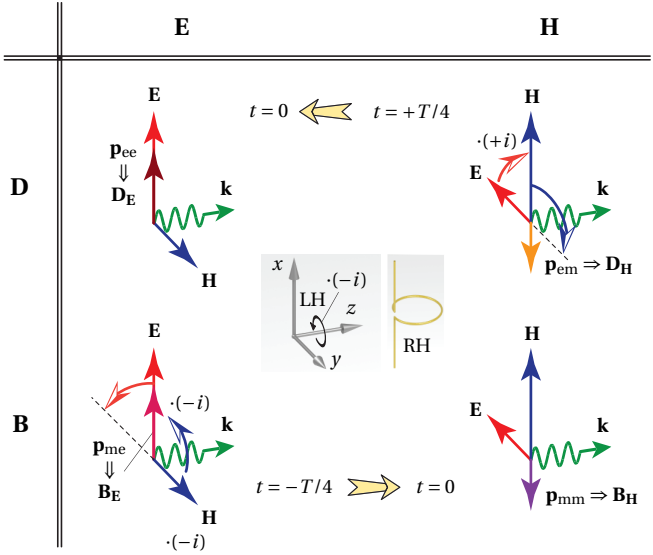}{
        \psfrag{A}[c][c][1]{$\ve{E}$}
        \psfrag{Y}[c][c][1]{$\ve{H}$}
        \psfrag{C}[c][c][1]{$\ve{D}$}
        \psfrag{Z}[c][c][1]{$\ve{B}$}
        \psfrag{R}[c][c][0.8]{RH}
        \psfrag{L}[c][c][0.8]{LH}
        \psfrag{1}[c][c][0.8]{$t=0$}
        \psfrag{2}[c][c][0.8]{$t=+T/4$}
        \psfrag{3}[c][c][0.8]{$t=-T/4$}
        \psfrag{i}[c][c][0.8]{$\cdot(-i)$}
        \psfrag{l}[c][c][0.8]{$\cdot(+i)$}
        \psfrag{m}[c][c][0.8]{$\cdot(-i)$}
        \psfrag{x}[c][c][0.8]{$x$}
        \psfrag{y}[c][c][0.8]{$y$}
        \psfrag{z}[c][c][0.8]{$z$}
        \psfrag{E}[c][c][0.8]{$\ve{E}$}
        \psfrag{H}[c][c][0.8]{$\ve{H}$}
        \psfrag{k}[c][c][0.8]{$\ve{k}$}
        \psfrag{D}[c][c][0.8]{$\ve{D}$}
        \psfrag{B}[c][c][0.8]{$\ve{B}$}
        \psfrag{a}[c][c][0.8]{\begin{minipage}{1cm}\centering$\ve{p}_\tx{ee}$ \\ $\Downarrow$ \\ $\ve{D}_\ve{E}$\end{minipage}}
        \psfrag{b}[c][c][0.8]{\begin{minipage}{1cm}\centering$\ve{p}_\tx{me}$ \\ $\Downarrow$ \\ $\ve{B}_\ve{E}$\end{minipage}}
        \psfrag{c}[l][l][0.8]{$\ve{p}_\tx{em}\Rightarrow\ve{D}_\ve{H}$}
        \psfrag{d}[l][l][0.8]{$\ve{p}_\tx{mm}\Rightarrow\ve{B}_\ve{H}$}
        }
        \vspace{-5mm}
        \caption{Electric responses ($\ve{D}$) and magnetic responses ($\ve{B}$) of the RH helix particle in Figs.~\ref{fig:two_omega_part}(a) to the electric excitation ($\ve{E}$) and magnetic excitation ($\ve{H}$). The direct responses $\ve{D}_\ve{E}$ and $\ve{B}_\ve{H}$ are synchronized (e.g. reference time $t=0$), whereas the cross responses $\ve{D}_\ve{H}$ and $\ve{B}_\ve{E}$ are respectively advanced ($t=+T/4$) and delayed ($t=-T/4$) with respect to them.}
   \label{fig:sign_of_i}
\end{figure}

For this purpose, Fig.~\ref{fig:sign_of_i} isolates and compares the electric and magnetic responses to the electric and magnetic fields. The figure first recalls the result established in conjunction with Fig.~\ref{fig:twisted_omega} that the RH helix particle rotates the field polarization in the transverse plane in the direction corresponding to the left hand with the thumb pointing in the propagation direction, or in the $y$ to $x$ direction, corresponding to phasor multiplication by the factor $(-i)$, or yet to a delay $T/4=(2\pi/\omega)/4=\pi/(2\omega)$. Comparing the responses $\ve{D}_\ve{E}$ and $\ve{D}_\ve{H}$ shows then that the latter is $\pi/2$ beyond (or $T/4$ later) in the mentioned direction than the former; it must therefore be rotated backward, in the positive ($x$ to $y$) direction, for matching (or synchronization), or multiplied by the factor $+i$. Similarly comparing the responses $\ve{B}_\ve{H}$ and $\ve{B}_\ve{E}$ shows that the latter is $\pi/2$ behind (or $T/4$ earlier) in the mentioned direction than the former; it must therefore be rotated forward, in the negative ($y$ to $z$) direction, for matching (or synchronization), or multiplied by the factor $-i$. Opposite signs would naturally be obtained in the case of a LH helix particle.

We have thus, with respect to the references of $\epsilon$ and $\mu$, that $\xi=+i\chi$ and $\zeta=-i\chi$. This reformulates the chiral constitutive relations~\eqref{eq:biis_const_rel} as
\begin{subequations}\label{eq:chir_const_rel}
\begin{align}\label{eq:chir_const_rel_D}
\ve{D}&=\epsilon\ve{E}+i\chi\ve{H}, \\
\label{eq:chir_const_rel_B}
\ve{B}&=-i\chi\ve{E}+\mu\ve{H},
\end{align}
\end{subequations}
where $\chi$ is called the chiral parameter, which be shown in Sec.~\ref{sec:loss} to be purely real in the lossless case and complex in the presence of loss~\cite{Lindell_EWCBM_1994,Kong_EWT_2008}. A chiral medium, characterized by the relations~\eqref{eq:chir_const_rel}, is also called a \emph{Pasteur medium}. Inserting~\eqref{eq:chir_const_rel} into~\eqref{eq:Maxwell_curl} provides then the following explicit form of Maxwell equations for a chiral or Pasteur medium:
\begin{subequations}\label{eq:Maxwell_curl_mod}
\begin{align}
\nabla\times\ve{E}
&=i\omega\left(-i\chi\ve{E}+\mu\ve{H}\right)
=\omega\left(\chi\ve{E}+i\mu\ve{H}\right), \\
\nabla\times\ve{H}
&=-i\omega\left(\epsilon\ve{E}+i\chi\ve{H}\right)
=\omega\left(-i\epsilon\ve{E}+\chi\ve{H}\right).
\end{align}
\end{subequations}

Note the Greek symbol $\chi$ is now associated with both the susceptibility, from~\eqref{eq:bianis_polar}, and with the chirality factor, from~\eqref{eq:chir_const_rel}. However, the former will always appear with the double subscript ee, em, me or mm, whereas the latter is always on its own. So, this coincidence should not pose any notational ambiguity.

The \emph{reciprocity} of a chiral medium, which is imposed by the assumed absence of an external force or of nonlinearity combined with spatial asymmetry~\cite{Caloz_2005}, is expressed in the constitutive relations~\eqref{eq:chir_const_rel} by the fact that $\te{\xi}=i\chi\te{I}=-\te{\zeta}$ satisfies the reciprocity condition
\begin{equation}\label{eq:recipr_cond}
\te{\xi}=-\te{\zeta}^T,
\end{equation}
where the subscript $T$ denotes the transpose operation, and which results from the Lorentz nonreciprocity relations~\cite{Kong_EWT_2008,Caloz_PRAp_10_2018}.

\begin{figure}[h]
    \centering
        \includegraphics[width=\linewidth]{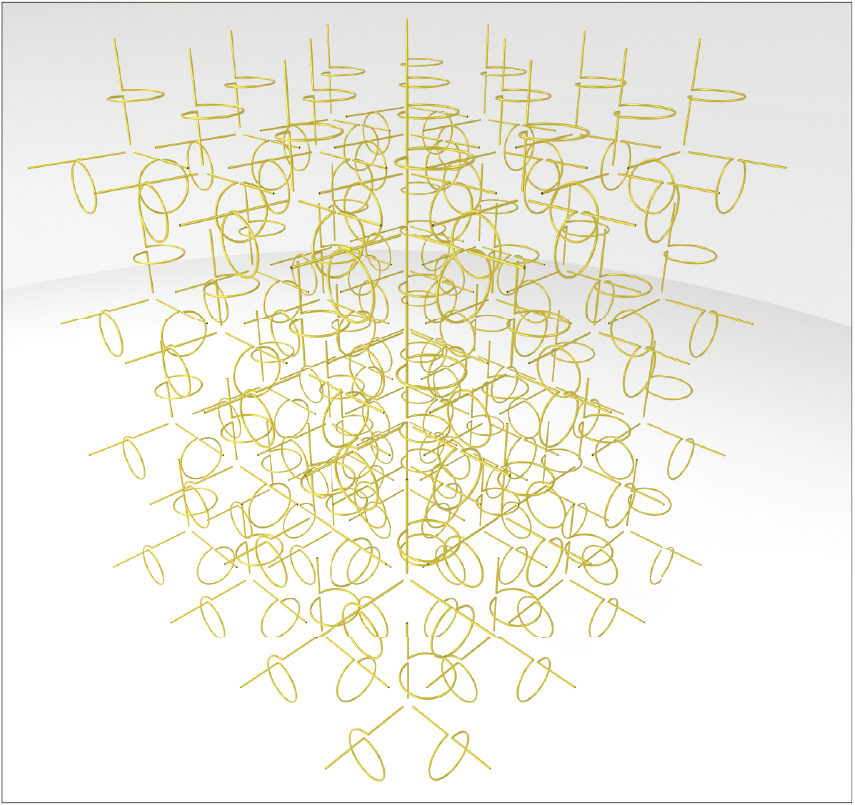}{}
        \vspace{-3mm}
        \caption{Chiral metamaterial made of a 3D array of twisted Omega or helix metaparticles [Fig.~\ref{fig:two_omega_part}(b)] oriented in the $x$, $y$ and $z$ directions of space [Fig.~\ref{fig:twisted_omegas_3_orientations}].}
   \label{fig:chiral_metamaterial}
\end{figure}

As we have seen in Sec.~\ref{sec:MaxAmp_twisted_omega}, the metamaterial constituted by the triatomic helix metaparticle shown in Fig.~\ref{fig:twisted_omegas_3_orientations} is isotropic [Eq.~\eqref{eq:twisted_Omeg_tens_3_or}]. Therefore, it indeed corresponds to the biisotropic chiral relations~\eqref{eq:chir_const_rel}. Such a medium is represented in Fig.~\ref{fig:chiral_metamaterial}. However, the monoatomic helix particle in Fig.~\ref{fig:two_omega_part}, described by the anisotropic polarizability tensors~\eqref{eq:twisted_Omeg_tens}, is also chiral and gyrotropic, as shown in Fig.~\ref{fig:twisted_omega}. This indicates that the biisotropic formulation~\eqref{eq:chir_const_rel} is excessively restrictive. However, it is what is generally called chirality (actually meaning biisotropic chirality) in the electromagnetic community, and we shall not dispute this term here.

\subsection{Generalized Biisotropic Media}\label{sec:gen_biis_med}

The chiral relations~\eqref{eq:chir_const_rel} do not not represent the only form of biisotropy. The magnetoelectric responses can also be in phase with their responses. In this case, we can write $\xi=\tau$ and $\zeta=\tau$, which particularizes~\eqref{eq:biis_const_rel} to
\begin{subequations}\label{eq:Tellegen_rel}
\begin{align}
\ve{D}&=\epsilon\ve{E}+\tau\ve{H}, \\
\ve{B}&=\tau\ve{E}+\mu\ve{H},
\end{align}
\end{subequations}
where $\tau$ will also be shown in Sec.~\ref{sec:loss} to be purely real in the lossless case and complex in the presence of loss. These equations~\eqref{eq:Tellegen_rel} are the constitutive relations of a \emph{Tellegen medium}. Such medium is nonreciprocal since $\te{\xi}=\tau\te{I}=\te{\zeta}=\te{\zeta}^T${\kern-0.6em} violates the reciprocity condition~\eqref{eq:recipr_cond}. Therefore, it must involve an external force, specifically permanent or induced electric and magnetic dipole moments parallel or antiparallel to each other~\cite{Lindell_EWCBM_1994}.

Thus, the coupling parameters $\xi$ and $\zeta$ of a biisotropic medium [Eq.~\eqref{eq:biis_const_rel}] may generally be written as~\cite{Lindell_EWCBM_1994,Kong_EWT_2008}
\begin{subequations}\label{eq:biis_xi_ze_split}
\begin{align}
\xi&=\tau+i\chi, \\
\zeta&=\tau-i\chi.
\end{align}
\end{subequations}
Inserting~\eqref{eq:biis_xi_ze_split} into~\eqref{eq:biis_const_rel} provides the alternative general biisotropic constitutive relations
\begin{subequations}\label{eq:biis_const_rel_split}
\begin{align}
\ve{D}&=\epsilon\ve{E}+(\tau+i\chi)\ve{H}, \\
\ve{B}&=(\tau-i\chi)\ve{E}+\mu\ve{H},
\end{align}
\end{subequations}
which correspond to a chiral of Pasteur medium if $\tau=0$ and $\chi\neq 0$ and to a Tellegen medium if $\tau\neq 0$ and $\chi=0$.

It is useful to relate the three parameters $\epsilon$, $\mu$ and $\chi$ in~\eqref{eq:chir_const_rel} to the corresponding susceptibilities, which are then obtainable from the metaparticle polarizabilities via~\eqref{eq:rel_al_chi}. These relations may be found by
reducing the tensors to scalars in~\eqref{eq:exzm_from_xhi} and using~\eqref{eq:biis_xi_ze_split} with $\tau=0$, i.e., $\xi=i\chi$ and $\zeta=-i\chi$, which yields
\begin{equation}\label{eq:exzm_from_xhi_chir}
\begin{pmatrix}
\epsilon & \chi \\
\chi & \mu
\end{pmatrix}
=
\begin{pmatrix}
\epsilon_0\left(1+\chi_\tx{ee}\right) & -i\sqrt{\epsilon_0\mu_0}\,\chi_\tx{em}\\
i\sqrt{\epsilon_0\mu_0}\chi_\tx{me} & \mu_0\left(1+\chi_\tx{mm}\right)
\end{pmatrix},
\end{equation}
implying that $\chi_\tx{me}=-\chi_\tx{em}$, consistently with the previously found results for the twisted Omega or helix particle (Sec.~\ref{sec:MaxAmp_twisted_omega}).

\subsection{Loss}\label{sec:loss}
According to the Poynting theorem, a medium is lossless if the divergence of the divergence of its Poynting vector in it is always purely imaginary~\cite{Jackson_1998,Ishimaru_1990}, i.e.,
\begin{equation}\label{eq:Re_div_0}
\tx{Re}\left\{\nabla\cdot\ve{S}\right\}=0,
\quad\tx{with}\quad
\ve{S}=\ve{E}\times\ve{H}^*,
\end{equation}
where the `*' superscripts denotes the complex conjugate operation. The divergence in~\eqref{eq:Re_div_0} can be expanded as follows:
\begin{equation}\label{eq:loss_der}
\begin{split}
\nabla\cdot\ve{S}
&=\frac{1}{2}\nabla\cdot\left(\ve{E}\times\ve{H}^*\right) \\
&=\frac{1}{2}\left[\ve{H}^*\cdot\left(\nabla\times\ve{E}\right)-\ve{E}\cdot\left(\nabla\times\ve{H}^*\right)\right] \\
&=\frac{1}{2}\left\{\omega\left[-i\epsilon^*\left|\ve{E}\right|^2+i\mu\left|\ve{H}\right|^2
+\left(\chi-\chi^*\right)\ve{E}\cdot\ve{H}^*\right]\right\}
\end{split}
\end{equation}
where we have applied the identity $\nabla\cdot(\ve{a}\times\ve{b})=\ve{b}\cdot(\nabla\times\ve{a})-\ve{a}\cdot(\nabla\times\ve{b})$ in the second equality, and replaced $\nabla\times\ve{E}$ and $\nabla\times\ve{H}$ by their Maxwell expressions~\eqref{eq:Maxwell_curl_mod}. Nullifying the real part of the last expression in~\eqref{eq:loss_der} according to~\eqref{eq:Re_div_0} finally yields the lossless conditions
\begin{subequations}\label{eq:lossless_cond}
\begin{align}\label{eq:lossless_cond_eps}
\tx{Re}\left\{-i\epsilon^*\right\}&=0
\quad\tx{or}\quad
\epsilon\in\mathbb{R}, \\
\label{eq:lossless_cond_mu}
\tx{Re}\left\{i\mu^*\right\}&=0
\quad\tx{or}\quad
\mu\in\mathbb{R}, \\
\label{eq:lossless_cond_chi}
\chi-\chi^*&=0.
\end{align}
\end{subequations}
The demonstration of these relations may go as follows. First consider a point of space and phase of the harmonic cycle where $\ve{H}=0$. At such a point, Eq.~\eqref{eq:loss_der} implies~\eqref{eq:lossless_cond_eps}, and that relation must always be true since the medium is assumed to be time invariant. Then consider another point and phase where $\ve{E}=0$, which now implies~\eqref{eq:lossless_cond_mu}. This leaves out only the third term in the last expression of~\eqref{eq:loss_der}. The factor $\chi-\chi^*$ in this term is purely imaginary, but it still must vanish for if it would not, then $\ve{E}\cdot\ve{H}^*$, which is generally complex and thus includes an imaginary part, would otherwise assign a real part to the overall term in violation to the the zero real requirement. Note in passing that it is easy to verify that in a Tellegen medium $\tau$ is also purely real in the lossless case and complex in the lossy case.

In the presence of loss, the parameters $\epsilon$, $\mu$ and $\chi$ may be complex, i.e.,
\begin{equation}
\epsilon=\epsilon'+i\epsilon'',\quad
\mu=\mu'+i\mu'',\quad
\chi=\chi'+i\chi'',
\end{equation}
where the primed and unprimed quantities are real numbers denoting the real and imaginary parts. Therefore, the RH and LH eigenwavenumbers, which will be derived in Sec.~\ref{sec:mod_sol} as $\beta^+_\pm=\omega(\sqrt{\epsilon\mu}\pm\chi)$ [Eq.~\eqref{eq:beta_4sols}], generally do not have  only different real parts but also different imaginary parts, and then different absorptions. This difference in absorption for the RH and LH waves results in transforming the lossless circularly polarized states into elliptical polarized states, as will become clear in Sec.~\ref{sec:circ_pol}. This phenomenon is referred to as \emph{circular dichroism}~\cite{Fasman_2013}.

Finally, note that the imaginary parts of ($\epsilon,\chi_\tx{ee}$) and ($\mu,\chi_\tx{mm}$) are purely positive in a lossy (or dissipative) medium with the assumed $e^{-i\omega t}$ harmonic time dependence (Appendix~\ref{sec:loss_imag_sign}). In contrast, the passive lossy condition on ($\chi,\chi_\tx{em,me}$) is less restrictive~\cite{Lindell_MEFA_1992}; for $\chi$, this condition is found from $\tx{Im}\{\beta^+_\pm\}>0$ as $\tx{Im}\{\chi\}<\tx{Im}\{\sqrt{\epsilon\mu}\}$.

\section{Parity Conditions}\label{sec:spat_sym}

The operation of mirror reflection, which substantiates the definition of chirality (Sec.~\ref{sec:intro}), is equivalent to the operation of \emph{space reversal}~\cite{Jackson_1998}. Space reversal is defined via the operator ${\cal S}$ (called the parity operator -- ${\cal P}$ -- in quantum mechanics) as
\begin{subequations}\label{eq:space_rev_r}
\begin{equation}
{\cal S}\left\{\ve{r}\right\}=\ve{r}'=-\ve{r}
\quad\tx{or}\quad
{\cal S}:\ve{r}\mapsto\ve{r}'=-\ve{r}
\end{equation}
when trivially applied to the space variable, $\ve{r}$, and generally, when applied to any other physical quantity $\ves{\Psi}(\ve{r})$, as
\begin{equation}
{\cal S}\left\{\ves{\Psi}(\ve{r})\right\}
=\ves{\Psi}'(\ve{r}')=\ves{\Psi}(-\ve{r})
=\pm\ves{\Psi}(-\ve{r}),
\end{equation}
\end{subequations}
where positive and negative signs correspond to even parity and odd parity, respectively.
This section recalls the fundamental rules of electromagnetic space-reversal symmetries~\cite{Jackson_1998}, and deduces from them useful parity properties for the electromagnetic constitutive parameters introduced in Sec.~\ref{sec:macr_const_rel}. Note that the space reversal operation~\eqref{eq:space_rev_r} corresponds to mirror reflection if it is applied once or an odd number of times, since ${\cal S}^m(\ve{r})=(-1)^m\ve{r}$ only if $m$ is an odd integer~\footnote{\label{fn:mirror_parity}We have ${\cal S}^1(\ve{r})={\cal S}(\ve{r})=-\ve{r}$ by definition [Eq.~\eqref{eq:space_rev_r}] (mirror symmetry), ${\cal S}^2(\ve{r})={\cal S}[{\cal S}(\ve{r})]={\cal S}(-\ve{r})=\ve{r}$ (return to original spatial frame), ${\cal S}^3(\ve{r})={\cal S}[{\cal S}^2(\ve{r})]={\cal S}(\ve{r})=-\ve{r}$ (mirror symmetry), etc., which is rather trivial. The situation is more subtle when rotations in different directions are involved, as in Sec.~\ref{sec:mir_sym_test}.}.

According to~\eqref{eq:space_rev_r}, $\ve{r}$ is odd under space reversal, or space-reversal odd or antisymmetric, by definition. It follows that the vectorial spatial derivative operator $\nabla=\nabla_\ve{r}$, involving only spatial quantities of single multiplicity, is also space-reversal odd. In contrast, the charge and charge density, $q$ and $\rho$, are space-reversal even, since reversing the coordinate system leaves charges unaffected. The space-reversal symmetry properties of $\nabla$ and $\rho$ imply then, via Gauss law ($\nabla\cdot\ve{D}=\rho$) that $\ve{D}$ is space-reversal odd, which also implies space-reversal odd parity for $\ve{E}$, considering for instance the simple case of vacuum. Combining this last result with Maxwell-Faraday law ($\nabla\times\ve{E} =i\omega\ve{B}$) indicates then that $\ve{B}$ is space-reversal even, and so is then also $\ve{H}$. We have thus
\begin{subequations}\label{eq:field_sr_par}
\begin{align}
[\ve{D},\ve{E}](-\ve{r})&=-[\ve{D},\ve{E}](\ve{r}), \\
[\ve{B},\ve{H}](-\ve{r})&=+[\ve{B},\ve{H}](\ve{r}).
\end{align}
\end{subequations}
Considering these field space-reversal parities in the constitutive relations in Sec.~\ref{sec:macr_const_rel} reveals then that the parity properties of the different constitutive parameters: $\te{\epsilon},\epsilon$ and $\te{\mu},\mu$ are even, while $\te{\xi},\xi$, $\te{\zeta},\zeta$ and $\tau,\chi$ or odd. Specifically, enforcing parity compatibility in~\eqref{eq:chir_const_rel} reveals that in a chiral medium $\epsilon$ and $\mu$ are even functions of space while $\chi$ is an odd function of space, or
\begin{align}\label{eq:par_cond}
\epsilon(-\ve{r})&=+\epsilon(\ve{r}), \\
\mu(-\ve{r})&=+\mu(\ve{r}), \\
\chi(-\ve{r})&=-\chi(\ve{r}).
\end{align}
These relations may serve as a precious sanity check in the modeling of chiral materials. Incidentally, note that the Tellegen parameter [Eq.~\ref{eq:Tellegen_rel}] is odd, as $\chi$, i.e., $\tau(-\ve{r})=-\tau(\ve{r})$.

\section{Chiral Eigenstates}\label{sec:chir_eig_states}

\subsection{Modal Solutions}\label{sec:mod_sol}
Consider a $+z$-propagating plane-wave excitation. Such a wave has the phasor form~\eqref{eq:PW_space}, corresponding in a Cartesian coordinate system to the fields
\begin{subequations}\label{eq:fwd_EH}
\begin{align}
\ve{E}^+(z)&=(\hatv{x}E_x+\hatv{y}E_y)e^{+i\beta^+ z}, \\
\ve{H}^+(z)&=(\hatv{x}H_x+\hatv{y}H_y)e^{+i\beta^+ z},
\end{align}
\end{subequations}
where the $+$ superscript has been introduced to emphasize $+z$ propagation. Substituting~\eqref{eq:fwd_EH} into the coupled system of equations~\eqref{eq:Maxwell_curl_mod} yields the modal wavenumbers and modal fields~(see Appendix~\ref{app:der_chir_eig})
\begin{subequations}\label{eq:eigen_states}
\begin{equation}\label{eq:beta_4sols}
\beta_\pm^+
=\omega\left(\sqrt{\epsilon\mu}\pm\chi\right),
\end{equation}
\begin{equation}\label{eq:E_4sols}
\ve{E}_\pm^+(z)
=E_0\left(\hatv{x}\pm i\hatv{y}\right)e^{+i\beta_\pm^+ z},
\end{equation}
\begin{equation}\label{eq:H_4sols}
\ve{H}_\pm^+(z)
=\frac{\hatv{z}\times\ve{E}_\pm^+(z)}{\eta}
=\frac{E_0}{\eta}\left(\hatv{y}\mp i\hatv{x}\right)e^{+i\beta_\pm^+ z}
=\mp\frac{i}{\eta}\ve{E}_\pm^+(z),
\end{equation}
\end{subequations}
\noindent where $E_0$ is assumed to be real (choice of proper phase origin), and $\eta=\sqrt{\mu/\epsilon}$ is the medium intrinsic impedance. In these relations, the $\pm$ signs correspond to RH/LH handedness, respectively, according to the observations in Sec.~\ref{sec:MaxAmp_twisted_omega}.

\subsection{Circular Polarization}\label{sec:circ_pol}
The physical waves corresponding to the chiral modes~\eqref{eq:eigen_states} are found from~\eqref{eq:wt_conv} as
\begin{equation}\label{eq:inst_field}
\begin{split}
\vet{E}_\pm^+(z,t)
&=\tx{Re}\left\{\ve{E}_\pm^+(z)e^{-i\omega t}\right\} \\
&=\text{Re}\left\{E_0\left(\hatv{x}\pm i\hatv{y}\right)e^{-i(\omega t-\beta_\pm^+ z)}\right\} \\
&=E_0\text{Re}\left\{\left(\hatv{x}\pm i\hatv{y}\right)\left[\cos(\omega t-\beta_\pm^+ z)-i\sin(\omega t-\beta_\pm^+ z)\right]\right\} \\
&=E_0\left[\hatv{x}\cos(\omega t-\beta_\pm^+ z)\pm\hatv{y}\sin(\omega t-\beta_\pm^+ z)\right].
\end{split}
\end{equation}

\begin{figure}[h]
    \centering
        \includegraphics[width=\linewidth]{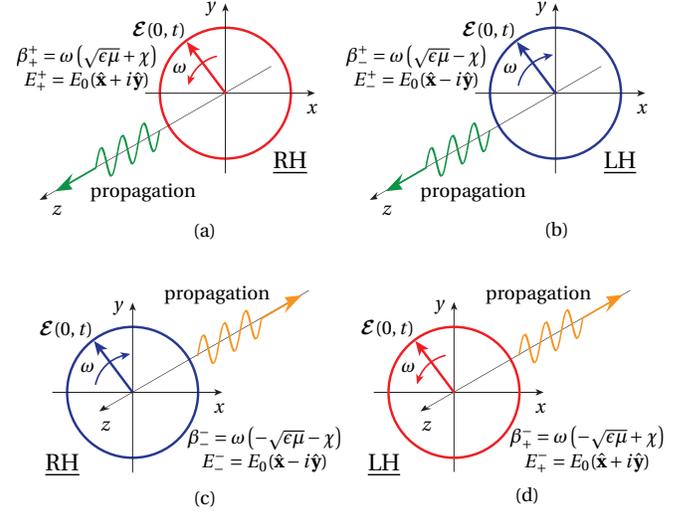}{
        \psfrag{a}[c][c][0.8]{(a)}
        \psfrag{b}[c][c][0.8]{(b)}
        \psfrag{c}[c][c][0.8]{(c)}
        \psfrag{d}[c][c][0.8]{(d)}
        \psfrag{x}[c][c][0.8]{$x$}
        \psfrag{y}[c][c][0.8]{$y$}
        \psfrag{z}[c][c][0.8]{$z$}
        \psfrag{w}[c][c][0.8]{$\omega$}
        \psfrag{v}[c][c][0.8]{$\vet{E}(0,t)$}
        \psfrag{p}[c][c][0.8]{propagation}
        \psfrag{E}[c][c][1]{\underline{RH}}
        \psfrag{F}[c][c][1]{\underline{LH}}
        \psfrag{G}[c][c][1]{\underline{RH}}
        \psfrag{H}[c][c][1]{\underline{LH}}
        \psfrag{A}[c][c][0.8]{
        \begin{minipage}{2.5cm}\centering
        $\beta^+_+=\omega\left(\sqrt{\epsilon\mu}+\chi\right)$ \\
        $E^+_+=E_0(\hatv{x}+i\hatv{y})$
        \end{minipage}}
        \psfrag{B}[c][c][0.8]{
        \begin{minipage}{2.5cm}\centering
        $\beta^+_-=\omega\left(\sqrt{\epsilon\mu}-\chi\right)$ \\
        $E^+_-=E_0(\hatv{x}-i\hatv{y})$
        \end{minipage}}
        \psfrag{C}[c][c][0.8]{
        \begin{minipage}{2.8cm}\centering
        $\beta^-_-=\omega\left(-\sqrt{\epsilon\mu}-\chi\right)$ \\
        $E^-_-=E_0(\hatv{x}-i\hatv{y})$
        \end{minipage}}
        \psfrag{D}[c][c][0.8]{
        \begin{minipage}{2.8cm}\centering
        $\beta^-_+=\omega\left(-\sqrt{\epsilon\mu}+\chi\right)$ \\
        $E^-_+=E_0(\hatv{x}+i\hatv{y})$
        \end{minipage}}
        }
        \vspace{-4mm}
        \caption{Circular polarization (CP) of the chiral modes in the plane $z=0$ [Eqs.~\eqref{eq:eigen_states} and~\eqref{eq:eigen_states_m}]. (a)~FWD-prop. ($+$) and $(\hatv{x}+i\hatv{y})$ phasor ($+$), leading to RH-CP. (b)~FWD-prop. ($+$) and $(\hatv{x}-i\hatv{y})$ phasor ($-$), leading to LH-CP. (c)~BWD-prop. ($-$) and $(\hatv{x}-i\hatv{y})$ ($-$) phasor, leading to RH-CP. (d)~BWD-prop. ($-$) and $(\hatv{x}+i\hatv{y})$ ($+$) phasor, leading to LH-CP.}
   \label{fig:circ_pol}
\end{figure}

Figures~\ref{fig:circ_pol}(a) and~(b) show the temporal evolution of the electric field vector $\vet{E}^+_+(z,t)$  in a given plane plane transverse to the propagation direction (here $z=0$). The tip of $\vet{E}^+_+(0,t)$ traces a circle following the fingers of the right hand with the thumb pointing in the direction of wave propagation, as shown in Fig.~\ref{fig:circ_pol}(a); it is thus right-handed (RH) circularly polarized~\footnote{This corresponds to the polarization handedness convention generally used in the electrical engineering community (e.g.~\cite{Kong_EWT_2008,Pozar_ME_2011,Balanis_2012,Ulaby_2014}), for which all the results following the time-harmonic convention $e^{-i\omega t}$ (used in this paper) apply upon substituting $-i\rightarrow j$ for transformation to the time-harmonic $e^{+j\omega t}$ convention (e.g., Eq.~\eqref{eq:inst_field}: the resulting opposite signs in both the complex vector and the complex phase eventually cancel out to yield the same real result.). The physics and optics community generally follows a convention leading to the opposite handedness, referring to clock direction rather than handedness, and considering that when the field rotation is counterclockwise (resp. clockwise) with the observer facing the oncoming wave, the wave is left-handed (resp. right-handed) (e.g.~\cite{Jackson_1998,Saleh_Teich_FP_2007}). This is all quite confusing, but important to note!}. In contrast, the tip of the vector $\vet{E}^+_-(z,t)$ traces a circle following the fingers of the left hand with the thumb pointing in the direction of wave propagation, as shown in Fig.~\ref{fig:circ_pol}(b); it is thus left-handed (LH) circularly polarized. Thus, the modes of a chiral medium are \emph{RH and LH circularly polarized} waves.

It should be emphasized that the notion of circular polarization refers here to the temporal variation of the field, and not to its spatial evolution, which is opposite. This easily seen by setting $t=0$ (instead of $z=0$) in~\eqref{eq:inst_field}. The handedness of the spatial spiral formed by the tip of the field in space is always opposite to the temporal handedness, as properly illustrated in~\cite{Ulaby_2014}.

The fact that the medium presents different wavenumbers (or refractive indices) to RH-CP waves ($\beta^+_+$) and LH-CP waves ($\beta^+_-$) is referred to as \emph{circular birefringence}, and is the fundamental cause of polarization rotation, as will be seen in Sec.~\ref{sec:pol_rot}.

\subsection{Opposite Propagation Direction}

Reversing the direction of wave propagation from $+z$ to $-z$ implies reversing the sign of the argument of the spatial exponentials -- and corresponding superscripts -- in~\eqref{eq:fwd_EH}, without altering the complex vector $\left(\hatv{x}\pm i\hatv{y}\right)$. This operation transforms~\eqref{eq:eigen_states} to~\footnote{Here, we chose to reverse the sign of $\beta$ while leaving the sign of $z$ unchanged. Doing the opposite would transform~\eqref{eq:beta_4sols_m} to the equivalent result $\beta_\pm^-=\omega\left(\sqrt{\epsilon\mu}\mp\chi\right)$ in the reversed coordinate system, where the reversal of the sign of $\chi$ compared to~\eqref{eq:beta_4sols} is a manifestation of this space-reversal odd symmetry of chirality [Eq.~\eqref{eq:beta_4sols}].}
\begin{subequations}\label{eq:eigen_states_m}
\begin{equation}\label{eq:beta_4sols_m}
\beta_\pm^-
=-\omega\left(\sqrt{\epsilon\mu}\mp\chi\right)
=\omega\left(-\sqrt{\epsilon\mu}\pm\chi\right),
\end{equation}
\begin{equation}\label{eq:E_4sols_m}
\ve{E}_\pm^-(z)
=E_0\left(\hatv{x}\pm i\hatv{y}\right)e^{+i\beta_\pm^- z},
\end{equation}
\begin{equation}\label{eq:H_4sols_m}
\ve{H}_\pm^-(z)
=-\frac{\hatv{z}\times\ve{E}_\pm^-(z)}{\eta}
=\pm\frac{i}{\eta}\ve{E}_\pm^-(z),
\end{equation}
\end{subequations}
which results in reversing the circular-polarization handedness, as shown in Figs.~\ref{fig:circ_pol}(c) and ~\ref{fig:circ_pol}(d), since the thumb is now pointing in the opposite direction whereas the signs of the complex vector have not changed.

One must be very careful to avoid confusion in determining the (temporal) wave handedness. In~\eqref{eq:eigen_states} and~\eqref{eq:eigen_states_m}, the superscripts $\pm$ unambiguously indicate the direction of wave propagation, while the subscripts $\pm$ unambiguously indicate the sign of the complex vector $\hat{x}\pm i\hatv{y}$. But neither the superscripts nor the subscripts indicate handedness. For instance, as shown in Fig.~\ref{fig:circ_pol}, the vector $\left(\hatv{x}+i\hatv{y}\right)$ is associated with right-handedness in positive propagation [Fig.~\ref{fig:circ_pol}(a)], but left-handedness in backward propagation [Fig.~\ref{fig:circ_pol}(d)].

\subsection{Modal Isotropy and Eigenstate Perpective}\label{sec:mod_isotr_eig_persp}
Now that we have found the modes of a chiral medium, let us see how these modes turn out in the constitutive relations.

Substituting~\eqref{eq:E_4sols} and \eqref{eq:H_4sols} into~\eqref{eq:chir_const_rel} yields for the forward propagation direction
\begin{subequations}\label{eq:eig_const_rel}
\begin{align}\label{eq:eig_const_rel_DE}
\ve{D}_\pm^+
&=\epsilon\ve{E}_\pm^++i\chi\ve{H}_\pm^+
=\left(\epsilon\pm\frac{\chi}{\eta}\right)\ve{E}_\pm^+
=\epsilon_\pm^+\ve{E}_\pm^+, \\
\label{eq:eig_const_rel_BH}
\ve{B}_\pm^+
&=-i\chi\ve{E}_\pm^++\mu\ve{H}_\pm^+
=\left(\mu\pm\chi\eta\right)\ve{H}_\pm^+
=\mu_\pm^\pm\ve{H}_\pm^+,
\end{align}
\end{subequations}
while substituting~\eqref{eq:E_4sols_m}-\eqref{eq:H_4sols_m} into~\eqref{eq:chir_const_rel} yields for the backward propagation direction
\begin{subequations}\label{eq:eig_const_rel_m}
\begin{align}\label{eq:eig_const_rel_DE_m}
\ve{D}_\pm^-
&=\epsilon\ve{E}_\pm^-+i\chi\ve{H}_\pm^-
=\left(\epsilon\mp\frac{\chi}{\eta}\right)\ve{E}_\pm^-
=\epsilon_\pm^-\ve{E}_\pm^-, \\
\label{eq:eig_const_rel_BH_m}
\ve{B}_\pm^-
&=-i\chi\ve{E}_\pm^-+\mu\ve{H}_\pm^-
=\left(\mu\mp\chi\eta\right)\ve{H}_\pm^-
=\mu_\pm^-\ve{H}_\pm^-.
\end{align}
\end{subequations}

Equations~\eqref{eq:eig_const_rel} and~\eqref{eq:eig_const_rel_m} reveal that $\ve{D}_\pm^\pm\|\ve{E}_\pm^\pm$ and $\ve{B}_\pm^\pm\|\ve{H}_\pm^\pm$, which means that a chiral medium, although generally biisotropic [Eq.~\eqref{eq:chir_const_rel}], reduces to a \emph{monoisotropic} ($\chi_\pm^\pm=\zeta_\pm^\pm=0$) medium for its modes. Equations~\eqref{eq:eig_const_rel} also mean that a chiral medium preserves the nature and handedness of the exciting wave, since $\ve{D}_\pm^\pm$ (resp. $\ve{B}_\pm^\pm$) differ from $\ve{E}_\pm^\pm$ (resp. $\ve{H}_\pm^\pm$) only by a scalar complex coefficient, which only alters the phase of the wave.

Moreover, note that Eqs.~\eqref{eq:eig_const_rel} have the eigenform \mbox{${\cal L}\ves{\Psi}=\lambda\ves{\Psi}$}, where ${\cal L}$ is an implicit operator mapping $\ve{E}_\pm^\pm$ to $\ve{D}_\pm^\pm$ for~\eqref{eq:eig_const_rel_DE} and $\ve{H}_\pm^\pm$ to $\ve{B}_\pm^\pm$ for~\eqref{eq:eig_const_rel_BH}. The circularly-polarized modes~\eqref{eq:eigen_states} represent thus the \emph{eigenstates} of a chiral medium, with eigenvalues $\epsilon_\pm^\pm=\pm(\epsilon\pm\chi/\eta)$ for~\eqref{eq:eig_const_rel_DE} and eigenvalues $\mu_\pm^\pm=\pm(\mu\pm\chi\eta)$ for~\eqref{eq:eig_const_rel_BH}.

\section{Polarization Rotation or Gyrotropy}\label{sec:pol_rot}
After finding that a chiral medium preserves the nature of circularly polarized waves and only changes their phase, an obvious question is: how does such a medium affect linearly polarized (LP) waves?

\begin{figure}[h]
    \centering
        \includegraphics[width=\linewidth]{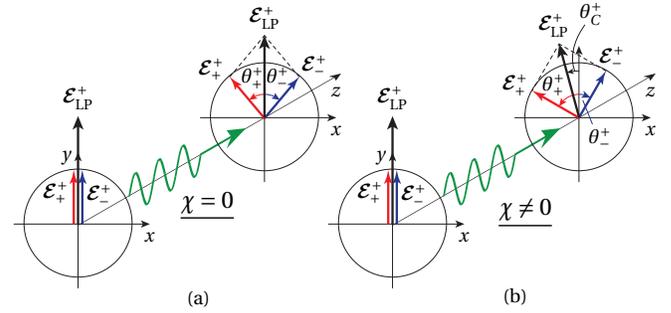}{
        \psfrag{a}[c][c][0.8]{(a)}
        \psfrag{b}[c][c][0.8]{(b)}
        \psfrag{x}[c][c][0.8]{$x$}
        \psfrag{y}[c][c][0.8]{$y$}
        \psfrag{z}[c][c][0.8]{$z$}
        \psfrag{l}[c][c][0.8]{$\vet{E}^+_+$}
        \psfrag{n}[c][c][0.8]{$\vet{E}^+_-$}
        \psfrag{e}[c][c][0.8]{$\vet{E}^+_\tx{LP}$}
        \psfrag{q}[c][c][0.8]{$\theta^+_+$}
        \psfrag{r}[c][c][0.8]{$\theta^+_-$}
        \psfrag{t}[c][c][0.8]{$\theta^+_{C}$}
        \psfrag{1}[c][c][1]{\underline{$\chi=0$}}
        \psfrag{2}[c][c][1]{\underline{$\chi\neq0$}}
        }
        \vspace{-4mm}
        \caption{Chiral rotation of an LP wave (figure inspired by~\cite{Dionne_LLJ_2005}). (a)~Monoisotropic medium ($\chi=0$), where the RH/LH-CP phasors rotate by the same amounts $\theta^+_\pm=\pm\omega\sqrt{\epsilon\mu}z/2$, resulting in no LP rotation. (b)~Chiral medium ($\chi\neq 0$), where the RH/LH-CP phasors rotate by the different amounts $\theta^+_\pm=\omega(\sqrt{\epsilon\mu}\pm\chi)z/2$, resulting in a chiral rotation angle of $\theta^+_\tx{C}=\theta^+_+-\theta^+_-=\omega\chi z$. Note that this graph also holds for Faraday rotation~\cite{Dionne_LLJ_2005} upon replacing the electric phasor and quantities by their magnetic counterparts, and $\chi$ by the magnetization saturation $M_\tx{s}$~\cite{Pozar_ME_2011,Lax_1962}.}
   \label{fig:cir_dichroism}
\end{figure}

The most natural way to address this question is to decompose a test LP wave into the medium (RH-CP and LH-CP) eigenstates, since these states form a simple, isotropic basis with well-defined eigenwavenumbers. Consider for instance the $\hatv{x}$-polarized $+z$-propagating plane wave in the plane $z=0$, which reads, according to~\eqref{eq:PW_space},
\begin{equation}\label{eq:x_PW}
\ve{E}_\tx{LP}^+(z=0)=\hatv{x}E_0.
\end{equation}
The chiral medium resolves such a wave into its RH-CP and a LH-CP components as
\begin{equation}\label{eq:x_PW}
\ve{E}_\tx{LP}^+(z=0)
=\frac{E_0}{2}\left(\hatv{x}+i\hatv{y}\right)+\frac{E_0}{2}\left(\hatv{x}-i\hatv{y}\right).
\end{equation}
The propagation along the medium is then found by assigning to the RH-CP and LH-CP waves composing the LP wave their respective eigenwavenumbers, $\beta_+^+$ and $\beta_-^+$, given by~\eqref{eq:beta_4sols}, which results in
\begin{equation}\label{eq:ELP_CP_dec}
\begin{split}
\ve{E}_\tx{LP}^+(z)
&=\left[\frac{E_0}{2}\left(\hatv{x}+i\hatv{y}\right)e^{i\beta_+^+ z}
+\frac{E_0}{2}\left(\hatv{x}-i\hatv{y}\right)e^{i\beta_-^+ z}\right].
\end{split}
\end{equation}
where multiplication by $e^{-i\omega t}$ shows that the two CP-wave vectors composing $\ve{E}_\tx{LP}^+(z)$ rotate at different spatial rates, since $\beta_+^+\neq\beta_-^+$, resulting in a net rotation of the (linear) polarization, or \emph{gyrotropy}, as illustrated in Fig.~\ref{fig:cir_dichroism}. This is the \emph{circular birefringence} explanation of Fresnel~\cite{Fresnel_1868}, where different medium properties are seen by the two CP waves.

This may be mathematically shown by grouping the terms of same polarization in~\eqref{eq:ELP_CP_dec}, and next factoring out the resulting exponential sums and difference to form sines and cosines, i.e.,
\begin{subequations}
\begin{equation}\label{eq:chir_rot}
\begin{split}
\ve{E}_\tx{LP}^+(z)
=&E_0\left[\hatv{x}\frac{e^{i\beta_+^+ z}+e^{i\beta_-^+ z}}{2}
+i\hatv{y}\frac{e^{i\beta_+^+ z}-e^{i\beta_-^+ z}}{2}\right] \\
=&e^{i\big(\frac{\beta_+^++\beta_-^+}{2}\big)z}E_0\left\{
\hatv{x}\left[\frac{e^{i\big(\frac{\beta_+^+-\beta_-^+}{2}\big)z}
+e^{-i\big(\frac{\beta_+^+-\beta_-^+}{2}\big)z}}{2}\right]\right. \\
&\qquad\qquad\qquad\left.+i^2\hatv{y}\left[\frac{e^{i\big(\frac{\beta_+^+-\beta_-^+}{2}\big)z}
-e^{-i\big(\frac{\beta_+^+-\beta_-^+}{2}\big)z}}{2i}\right]
\right\} \\
=&e^{i{\beta^+_\tx{e}}z}E_0\left\{
\hatv{x}\cos\left(\beta^+_{\tx{o}}z\right)
-\hatv{y}\sin\left(\beta^+_{\tx{o}}z\right)
\right\},
\end{split}
\end{equation}
where
\begin{equation}
\beta^+_{\tx{e}}=\frac{\beta_+^++\beta_-^+}{2}=\omega\sqrt{\epsilon\mu},
\quad
\beta^+_{\tx{o}}=\frac{\beta_+^+-\beta_-^+}{2}=\omega\chi.
\end{equation}
\end{subequations}
The last expression of~\eqref{eq:chir_rot} reveals that the chiral medium rotates the polarization of the LP wave  in space as it propagates a distance $z$, by the angle
\begin{subequations}\label{eq:chiral_rot}
\begin{equation}\label{eq:chiral_rot_ang}
\begin{split}
\theta_\tx{C}(z;\beta^+_\tx{o})
&=\tan^{-1}\left(\frac{E_y}{E_x}\right)
=\tan^{-1}\left[\frac{-\sin(\beta^+_\tx{o}z)}{\cos(\beta^+_\tx{o}z)}\right] \\
&=-\tan^{-1}\left[\tan(\beta^+_\tx{o}z)\right]
=-\beta^+_\tx{o}z
=-\omega\chi z
\end{split}
\end{equation}
and with the phase shift
\begin{equation}\label{eq:chiral_rot_pha}
\phi_\tx{C}(z;\beta_\tx{e}^+)
=\beta^+_\tx{e}z
=\omega\sqrt{\epsilon\mu}z.
\end{equation}
\end{subequations}
So a chiral medium preserves the LP nature of an LP wave, but rotates its polarization as it propagates, by an amount that is proportional to the \emph{difference} of the RH-CP and LH-CP wavenumbers, or refractive indices $n^\pm_\pm=\beta^\pm_\pm/k_0=c\beta^\pm_\pm/\omega$, and with a phase shift that is proportional to the \emph{sum} of the same RH-CP and LH-CP wavenumbers.

\begin{figure}[h]
    \centering
        \includegraphics[width=\linewidth]{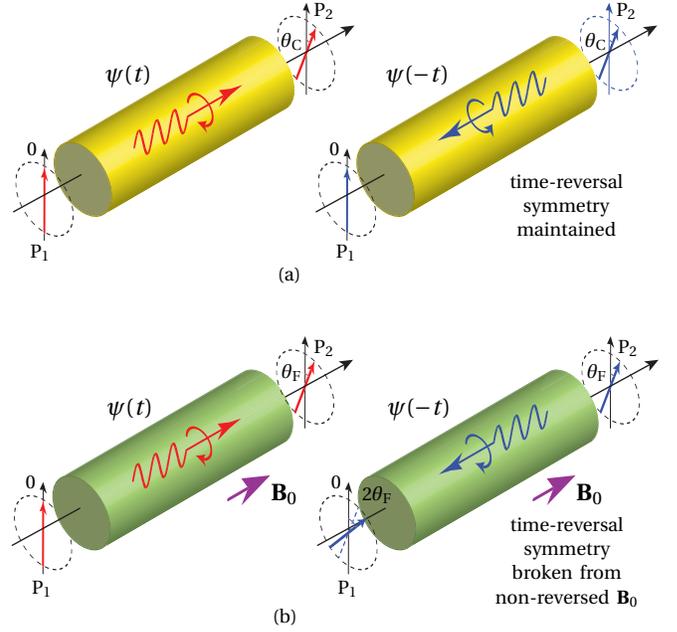}{
        \psfrag{A}[c][c][0.8]{(a)}
        \psfrag{B}[c][c][0.8]{(b)}
        \psfrag{K}[c][c][0.8]{$\ve{J}(\uparrow)$}
        \psfrag{L}[c][c][0.8]{$\ve{E}(\nnearrow)$}
        \psfrag{M}[c][c][0.8]{$\ve{E}'(\uparrow)$}
        \psfrag{N}[c][c][0.8]{$\ve{J}'(\nnearrow)$}
        \psfrag{0}[c][c][0.8]{0}
        \psfrag{1}[c][c][0.8]{P$_1$}
        \psfrag{2}[c][c][0.8]{P$_2$}
        \psfrag{H}[c][c]{$\ve{B}_0$}
        \psfrag{X}[l][l]{$\psi(t)$}
        \psfrag{Y}[l][l]{$\psi(-t)$}
        \psfrag{p}[c][c][0.8]{$\theta_\tx{C}$}
        \psfrag{r}[c][c][0.8]{$\theta_\tx{F}$}
        \psfrag{q}[c][c][0.8]{$2\theta_\tx{F}$}
        \psfrag{a}[c][c][0.8]{\begin{minipage}{4cm}\centering
            time-reversal \\ symmetry \\ maintained \end{minipage}}
        \psfrag{b}[c][c][0.8]{\begin{minipage}{3cm}\centering
            time-reversal \\ symmetry \\ broken from \\ non-reversed $\ve{B}_0$\end{minipage}}
        }
        \caption{Comparison of chirality gyrotropy and Faraday gyrotropy using direct-time (left) and time-reversed (right) experiments. (a)~Chiral medium (reciprocal), where $\theta_\tx{C}(z)=\omega\chi z$. (b)~Faraday medium (nonreciprocal), where $\mu$ and $\kappa$ form the permeability tensor as $\theta_\tx{F}(z)=\omega(\sqrt{\epsilon(\nu+\kappa)}-\sqrt{\epsilon(\nu-\kappa)})z/2$ with $\te{\mu}=[\nu,-i\kappa,0;i\kappa,\nu,0;0,0,\mu_0]$~\cite{Pozar_ME_2011,Lax_1962}.}
   \label{fig:gyro_rec_med}
\end{figure}

This spatial rotation effect for an LP wave seems essentially identical to Faraday rotation~\cite{Faraday_1933} in magnetized plasmas~\cite{Stix_1992} and magnetized ferrites~\cite{Lax_1962} or transistor-loaded magnetless metaferrites~\cite{Kodera_AWPL_2018}, despite the fact that none of these media involve chiral particles (proof of \ding{195} in Fig.~\ref{fig:trilogy}) and are purely monoanisotropic (non magnetelectric coupling), with anisotropy $\te{\epsilon}$ for plasmas and anisotropy $\te{\mu}$ for ferrites or metaferrites (proof $\rightarrow$ of \ding{197} in Fig.~\ref{fig:trilogy}). However, there is a fundamental difference between chiral rotation and Faraday rotation~\cite{Faraday_1933}: the former is reciprocal, as shown from the medium reciprocity condition in Sec.~\ref{sec:macr_const_rel}, whereas the latter breaks that reciprocity condition, and is hence nonreciprocal~\cite{Caloz_PRAp_10_2018}. The two types of gyrotropy are illustrated in Fig.~\ref{fig:gyro_rec_med}.

The reciprocity of chiral rotation is readily apparent in~\eqref{eq:chiral_rot}. Since reversing the direction of wave propagation is equivalent to reversing the sign of the argument of the spatial phase, i.e., of either $\beta$ or $z$, consider reversing the sign of $\beta_\tx{o}$  while maintaining the coordinate system fixed in~\eqref{eq:chiral_rot}. This results in
\begin{equation}\label{eq:C_rot}
\theta_\tx{C}(z;-\beta^+_\tx{o})
=-(-\beta^+_\tx{o})z
=\beta^+_\tx{o}z
=-\theta_\tx{C}(z;\beta^+_\tx{o}),
\end{equation}
which shows that that if the LP wave has accumulated a rotation angle of $\theta$ as propagating in the $+z$ direction, this angle is is undone as the wave returns, propagating in the $-z$ direction, so that the initial polarization state is retrieved, as shown in Fig.~\ref{fig:gyro_rec_med}(a). Such a medium is therefore time-reversal symmetric, and hence reciprocal, as it should be due to the absence of external force and asymmetric nonlinearity~\cite{Caloz_PRAp_10_2018}.

At the microscopic scale of the helix chiral particle [Fig.~\ref{fig:two_omega_part}(b)], reciprocity may be understood by reversing the direction of propagation from $+z$ to $-z$ in Fig.~\ref{fig:twisted_omega}, which corresponds to reversing the direction of either $\ve{E}$ or $\ve{H}$ so as to preserve the handedness of the triad $(\ve{E},\ve{H},\ve{k})$. First, consider the top of Fig.~\ref{fig:twisted_omega}, with $\ve{E}$ still pointing towards $+x$, and hence $\ve{H}$ pointing now towards $-y$; since the induced dipolar moments are unchanged, $\ve{H}$ rotates from the $-y$ direction to the $+x$ direction. Similarly, consider now the bottom of Fig.~\ref{fig:twisted_omega}, with $\ve{H}$ still pointing towards $+x$, and hence $\ve{E}$ pointing now towards $+y$; given the again unchanged dipolar moments, $\ve{E}$ rotates from the $+y$ direction to the $-x$ direction. The electromagnetic field has therefore rotated about the $z$ axis in the direction corresponding to the right-hand with the thumb pointing in the propagation direction ($-z$), which is the opposite rotation direction than that found for $+z$ propagation. Note that with the chosen convention for rotation handedness, depicted in Fig.~\ref{fig:circ_pol}, a RH helices (as in Fig.~\ref{fig:twisted_omega}) induce LH CP (and RH spatial spirals), and LH helices induce RH CP (and LH spatial spirals), which corresponds to~\eqref{eq:C_rot} and the rotation rewinding in Fig.~\ref{fig:gyro_rec_med}.

In contrast to chiral rotation, in Faraday rotation, the angle acquired by the wave propagating along the $+z$ direction, $\theta$, keeps accumulating in the same absolute direction, dictated by the fixed bias field $\ve{B}_0$, when propagating back in the $-z$ direction. This leads to a total rotation angle of $2\theta$ for the round trip, as shown in Fig.~\ref{fig:gyro_rec_med}(b). So, the initial state of the system is not retrieved upon time reversal, and the system is nonreciprocal~\cite{Caloz_PRAp_10_2018}, which allows the realization of devices such as isolators, circulators and nonreciprocal phase shifters.

\section{Spatial Dispersion (or Spatial Nonlocality)}\label{sec:spat_nl}
\subsection{Definition and Identification}
Spatial nonlocality, for a medium, means that its response at a given point of space does not depend on the excitation only at that point but also in the vicinity of it~\cite{Jackson_1998,Landau_1984}.

The planar Omega and helix particles are clearly nonlocal, with local and nonlocal responses indicated by the respective labels `L' and `NL' in Figs.~\ref{fig:straight_omega} and~\ref{fig:twisted_omega}. Consider first the planar Omega particle. The direct responses $p_{\tx{e},{E_x}}^x$, corresponding to $\alpha_\tx{ee}^{xx}$, and $p_{\tx{m},{H_y}}^{-x}$, corresponding to $\alpha_\tx{mm}^{xy}$ are produced exactly at the points of excitation, namely at the straight sections and at the looped section, respectively; they are thus local. In contrast, the cross responses $p_{\tx{m},E_x}^y$ and $p_{\tx{e},{H_y}}^{-x}$, respectively corresponding to $\alpha_\tx{me}^{yx}$ and $\alpha_\tx{em}^{xy}$, are produced off the points of excitation, namely at the straight sections and at the looped section, respectively; they are thus nonlocal. The medium is thus globally nonlocal. The nonlocality of the twisted Omega or helix particle may be understood from the same current continuity thought experiment, with nonlocal polarizabilities $\alpha_\tx{me}^{xx}$ and $\alpha_\tx{em}^{xx}$.

\subsection{Spatial-Spectral Constitutive Relations}\label{sec:speat_spec_const_rel}
Mathematically, a medium is nonlocal if its electric and magnetic responses do not depend only on the respective electric and magnetic excitations, but also on their spatial derivatives. To see how the chiral constitutive relations are nonlocal, we therefore set to find out how they involve spatial derivatives of the excitations. For this purpose, we first solve~\eqref{eq:Maxwell_curl_mod} for $\ve{E}$ and for $\ve{H}$ in terms of $\nabla\times\ve{E}$ and $\nabla\times\ve{H}$, which yields
\begin{subequations}\label{eq:EH_vs_curls}
\begin{align}\label{eq:E_vs_curls}
\ve{E}
&=\frac{1}{\omega\gamma}\left(i\mu\nabla\times\ve{H}-\chi\nabla\times\ve{E}\right), \\
\label{eq:H_vs_curls}
\ve{H}
&=-\frac{1}{\omega\gamma}\left(i\epsilon\nabla\times\ve{E}+\chi\nabla\times\ve{H}\right),
\end{align}
where
\begin{equation}\label{eq:gam}
\gamma
=\epsilon\mu-\chi^2.
\end{equation}
\end{subequations}
Next, we substitute these relations into the expressions of~\eqref{eq:chir_const_rel} that are associated with chirality, i.e., having the coefficient $\chi$. Specifically, inserting thus~\eqref{eq:H_vs_curls} into~\eqref{eq:chir_const_rel_D} and \eqref{eq:E_vs_curls} into~\eqref{eq:chir_const_rel_B} provides the alternative constitutive relations
\begin{subequations}\label{eq:const_rel_nl}
\begin{align}\label{eq:const_rel_nl_D}
\ve{D}
&=\underbrace{\epsilon\ve{E}}_{\tx{local}}
+\underbrace{\frac{\chi}{\omega\gamma}\left(\epsilon\nabla\times\ve{E}-i\chi\nabla\times\ve{H}\right)}_{\tx{nonlocal}}, \\
\label{eq:const_rel_nl_B}
\ve{B}
&=\underbrace{\mu\ve{H}}_{\tx{local}}
+\underbrace{\frac{\chi}{\omega\gamma}
\left(\mu\nabla\times\ve{H}+i\chi\nabla\times\ve{E}\right)}_{\tx{nonlocal}}.
\end{align}
\end{subequations}
These relations, which seem to be presented here for the first time, indeed involve spatial derivatives of the excitations, which indicates that a chiral medium is \emph{nonlocal}, with the nonlocal and local parts indicated in~\eqref{eq:const_rel_nl}.

Given the assumed linearity, one may expand all the fields in terms of their spatial angular spectra -- or plane waves $\ves{\tilde{\Psi}}(\ve{k})e^{i\ve{k}\cdot\ve{r}}$ ($\ves{\Psi}=\ve{D},\ve{B},\ve{E},\ve{H}$) --~\cite{Clemmow_2013}, which transforms~\eqref{eq:const_rel_nl} into
\begin{subequations}\label{eq:const_rel_nl_k}
\begin{align}\label{eq:const_rel_nl_D_k}
\ve{\tilde{D}}
&=\epsilon\ve{\tilde{E}}
+\frac{\chi}{\omega\gamma}\left(i\epsilon\ve{k}\times\ve{\tilde{E}}
+\chi\ve{k}\times\ve{\tilde{H}}\right), \\
\label{eq:const_rel_nl_B_k}
\ve{\tilde{B}}
&=\mu\ve{\tilde{H}}
+\frac{\chi}{\omega\gamma}
\left(i\mu\ve{k}\times\ve{\tilde{H}}-\chi\ve{k}\times\ve{\tilde{E}}\right).
\end{align}
\end{subequations}
Projecting these equations onto the $x$, $y$ and $z$ directions leads to the matrix relation
\begin{subequations}\label{eq:chir_nl}
\begin{equation}\label{eq:chir_nl_expl}
\begin{pmatrix}
\tilde{D}_x \\ \tilde{D}_y \\ \tilde{D}_z \\ \tilde{B}_x \\ \tilde{B}_y \\ \tilde{B}_z
\end{pmatrix}
=
\begingroup
\setlength\arraycolsep{2pt}
\begin{pmatrix}
\epsilon & -a_\epsilon k_z & a_\epsilon k_y & 0 & -a_\chi k_z & a_\chi k_y \\
a_\epsilon k_z & \epsilon & -a_\epsilon k_x & a_\chi k_z & 0 & -a_\chi k_x \\
-a_\epsilon k_y & a_\epsilon k_x & \epsilon & -a_\chi k_y & a_\chi k_x & 0 \\
0 & a_\chi k_z & -a_\chi k_y & \mu & -a_\mu k_z & a_\mu k_y \\
-a_\chi k_z & 0 & a_\chi k_x & a_\mu k_z & \mu & -a_\mu k_x \\
a_\chi k_y & -a_\chi k_x & 0 & -a_\mu k_y & a_\mu k_x & \mu
\end{pmatrix}
\endgroup
\begin{pmatrix}
\tilde{E}_x \\ \tilde{E}_y \\ \tilde{E}_z \\ \tilde{H}_x \\ \tilde{H}_y \\ \tilde{H}_z,
\end{pmatrix},
\end{equation}
with
\begin{equation}
\begin{pmatrix}
a_\epsilon & a_\chi \\
-a_\chi & a_\mu
\end{pmatrix}
=\frac{\chi}{\omega\gamma}
\begin{pmatrix}
i\epsilon & \chi \\
-\chi & i\mu
\end{pmatrix}.
\end{equation}
Equation~\eqref{eq:chir_nl_expl} may be more compactly written as
\begin{equation}
\begin{pmatrix}
\ve{\tilde{D}} \\
\ve{\tilde{B}}
\end{pmatrix}
=
\begin{pmatrix}
\tilde{\te{\epsilon}}(\ve{k}) & \tilde{\te{\xi}}(\ve{k}) \\
\tilde{\te{\zeta}}(\ve{k}) & \tilde{\te{\mu}}(\ve{k})
\end{pmatrix}
\cdot
\begin{pmatrix}
\ve{\tilde{E}} \\
\ve{\tilde{H}}
\end{pmatrix},
\end{equation}
\noindent where
\begin{equation}
\begin{pmatrix}
\tilde{\te{\epsilon}}(\ve{k}) & \tilde{\te{\xi}}(\ve{k}) \\
\tilde{\te{\zeta}}(\ve{k}) & \tilde{\te{\mu}}(\ve{k})
\end{pmatrix}
=
\begin{pmatrix}
\epsilon\te{I}+a_\epsilon\ve{k}\times\te{I}
& a_\chi\ve{k}\times\te{I} \\
 -a_\chi\ve{k}\times\te{I}
& \mu\te{I}+a_\mu\ve{k}\times\te{I}
\end{pmatrix},
\end{equation}
\noindent which involves the skewon (or antisymmetric) dyadic $\ve{k}\times\te{I}$, which characterizes gyrotropic media~\cite{Hehl_Obukhov_2003}.
\end{subequations}

The tensorial formulation~\eqref{eq:chir_nl} of chirality, where the tildes may be unambiguously removed if the wave excitation is restricted to a single spatial spectral component $\ve{k}$, is the spatial-dispersion counterpart of~\eqref{eq:chir_const_rel}~\cite{Landau_1984}. Compared to~\eqref{eq:chir_const_rel}, it has the advantage of providing explicit relations between the components of the response fields and the components of the excitation fields. It is interesting to note that the chiral constitutive relations are either scalar and non-explicit~\footnote{The non-explicit nature of~\eqref{eq:chir_const_rel} can be easily realized by trying to write a response-field component as a function of the excitation-field components. For instance, Eq.~\eqref{eq:chir_const_rel_D} implies $D_x=\epsilon E_x+i\chi H_x$. But, this relation is incomplete per se, because it is coupled to~\eqref{eq:chir_const_rel_B}, where $H_x$ may be written as $H_x=(B_x+i\chi E_x)/\mu$. Substituting this relation into the previous one yields $D_x=(\epsilon+i\chi/\mu)E_x+B_x/\mu$. Now coupling is accounted for, but the electric-response component $D_x$ does not depend only on the electric excitation ($E_x$) but also on the magnetic-excitation component $B_x$, and it is not possible to go any further.}, as is the case in the direct spatial domain ($\ve{r}$) [Eq.~\eqref{eq:chir_const_rel}], or tensorial and explicit, as is the case in the spectral spatial domain ($\ve{k}$) [Eq.~\eqref{eq:chir_nl}], but they cannot be written in a simultaneously scalar and explicit form.

In the general bianisotropic case, the relations~\eqref{eq:const_rel_nl} and~\eqref{eq:const_rel_nl_k} generalize to (see Appendix~\ref{sec:bianis_nonloc})
\begin{subequations}\label{eq:bianis_const_rel_alt}
\begin{align}\label{eq:bianis_const_rel_D_alt}
\ve{D}=&\te{\epsilon}\cdot\ve{E}-\frac{i}{\omega}\te{\xi}\cdot
\left(\te{\mu}-\te{\zeta}\cdot\te{\epsilon}^{-1}\cdot\te{\xi}\right)^{-1}
\cdot\left(\nabla\times\ve{E}+\te{\zeta}\cdot\te{\epsilon}^{-1}\cdot\nabla\times\ve{H}\right), \\
\label{eq:bianis_const_rel_B_alt}
\ve{B}=&
\frac{i}{\omega}\te{\zeta}\cdot
\left(\te{\epsilon}-\te{\xi}\cdot\te{\mu}^{-1}\cdot\te{\zeta}\right)^{-1}
\cdot\left(\nabla\times\ve{H}+\te{\xi}\cdot\te{\mu}^{-1}\cdot\nabla\times\ve{E}\right)
+\te{\mu}\cdot\ve{H},
\end{align}
\end{subequations}
and
\begin{subequations}\label{eq:bianis_const_rel_alt_nl}
\begin{align}\label{eq:bianis_const_rel_D_alt_nl}
\tilde{\ve{D}}=&\left(\te{\epsilon}+\te{T}_D\cdot\ve{k}\times\te{I}\right)\cdot\tilde{\ve{E}}
+\left(\te{T}_D\cdot\te{\zeta}\cdot\te{\epsilon}^{-1}\cdot\ve{k}\times\te{I}\right)\cdot\tilde{\ve{H}},\\
&\qquad\tx{with}\quad
\te{T}_D=\frac{1}{\omega}\te{\xi}\cdot
\left(\te{\mu}-\te{\zeta}\cdot\te{\epsilon}^{-1}\cdot\te{\xi}\right)^{-1} \\
\label{eq:bianis_const_rel_B_alt_nl}
\tilde{\ve{B}}=&
\left(\te{T}_B\cdot\te{\xi}\cdot\te{\mu}^{-1}\cdot\ve{k}\times\te{I}\right)\cdot\tilde{\ve{E}}
+\left(\te{\mu}+\te{T}_B\cdot\ve{k}\times\ve{I}\right)\cdot\tilde{\ve{H}} \\
&\qquad\tx{with}\quad
\te{T}_B=-\frac{1}{\omega}\te{\zeta}\cdot
\left(\te{\epsilon}-\te{\xi}\cdot\te{\mu}^{-1}\cdot\te{\zeta}\right)^{-1},
\end{align}
\end{subequations}
respectively.

\subsection{Application to the Omega Particles}\label{sec:nonlocal_omega}
This section applies the nonlocal explicit formulation of chirality presented in Sec.~\ref{sec:speat_spec_const_rel} to better relate the microscopic behavior of Omega-type particles (Sec.~\ref{sec:two_part}) to their macroscopic response (Secs.~\ref{sec:macr_const_rel}).

Let us start with the planar Omega particle, shown in Fig.~\ref{fig:two_omega_part}(a) and studied in Fig.~\ref{fig:straight_omega}. Applying~\eqref{eq:bianis_const_rel_alt_nl} to the tensorial structure~\eqref{eq:planar_Omeg_tens} found for this particle in Sec.~\ref{sec:MaxAmp_straight_omega} yields, after translating~\eqref{eq:planar_Omeg_tens} into its macroscopic form with~\eqref{eq:exzm_from_xhi} and setting~$\tilde{E}_y=\tilde{E}_z=\tilde{H}_x=\tilde{H}_z=k_x=k_y=0$ for the $(\tilde{E}_x,\tilde{H}_y,k_z)$ polarization in Fig.~\ref{fig:straight_omega},
\begin{subequations}\label{eq:planar_Omega_spat_disp}
\begin{align}
\tilde{D}_x&=\epsilon_\tx{r}^{xx}\left(\epsilon_0+\frac{k_z\chi_\tx{em}^{xy}}{\eta_0\omega\varphi}\right)\tilde{E}_x
-\frac{k_z\left(\chi_\tx{em}^{xy}\right)^2}{\omega\varphi}\tilde{H}_y, \\
\tilde{D}_y&=\tilde{D}_z=0, \\
\tilde{B}_y&=-\frac{k_z\left(\chi_\tx{em}^{xy}\right)^2}{\omega\varphi}\tilde{E}_x
+\mu_\tx{r}^{yy}\left(\mu_0-\frac{k_z\eta_0\chi_\tx{em}^{xx}}{\omega\varphi}\right)\tilde{H}_y, \\
\tilde{B}_x&=\tilde{B}_z=0,
\end{align}
with
\begin{equation}
\varphi=1+\chi_\tx{ee}^{xx}+\left(\chi_\tx{em}^{xy}\right)^2+\chi_\tx{mm}^{yy}+\chi_\tx{ee}^{xx}\chi_\tx{mm}^{yy}
\end{equation}
\end{subequations}
where the term $\left(\chi_\tx{em}^{xy}\right)^2$ corresponds to the explicit term $-\chi_\tx{em}^{xy}\chi_\tx{me}^{yx}$ with the result $\chi_\tx{me}^{yx}=-\chi_\tx{em}^{xy}$ following from~\eqref{eq:alpha_meyx_emxy},
and
\begin{equation}\label{eq:er_mr_eta0}
\epsilon_\tx{r}^{xx}=1+\chi_\tx{ee}^{xx},
\quad\mu_\tx{r}^{yy}=1+\chi_\tx{mm}^{yy},\quad\eta_0=\sqrt{\frac{\mu_0}{\epsilon_0}}.
\end{equation}

The relations~\eqref{eq:planar_Omega_spat_disp} are in full agreement with the observations made in Sec.~\ref{sec:MaxAmp_straight_omega} in conjunction with Fig.~\ref{fig:straight_omega}, and provide in addition the exact quantitative response of the related medium.

Particularly, applying~\eqref{eq:bianis_const_rel_alt_nl} to the perpendicular polarization $(\tilde{E}_y,-\tilde{H}_x,k_z)$ yields the trivial free-space solution $\tilde{D}_y=\epsilon_0\tilde{E}_y$ and $\tilde{B}_x=-\mu_0\tilde{H}_x$ (with $\tilde{D}_x=\tilde{D}_z=\tilde{B}_y=\tilde{B}_z=0$), already noted in Sec.~\ref{sec:MaxAmp_straight_omega}. In the case of the oblique incident polarization $[(\tilde{E}_x,\tilde{E}_y),(\tilde{H}_y,-\tilde{H}_x),k_z]$, applying~\eqref{eq:bianis_const_rel_alt_nl}
naturally adds, by superposition, $\tilde{D}_y=\epsilon_0\tilde{E}_y$ and $\tilde{B}_x=-\mu_0\tilde{H}_x$ to the response in~\eqref{eq:planar_Omega_spat_disp}, leading to a response with polarization angle
\begin{subequations}
\begin{equation}
\theta_\ve{D}=\tan^{-1}\left(\frac{\tilde{D}_y}{\tilde{D}_x}\right)
=\tan^{-1}\left(\frac{\epsilon_0\tilde{E}_y}{\epsilon^{xx}\tilde{E}_x+\xi^{xy}\tilde{H}_y}\right),
\end{equation}
with
\begin{equation}
\epsilon^{xx}=\epsilon_\tx{r}^{xx}\left(\epsilon_0+\frac{k_z\chi_\tx{em}^{xy}}{\eta_0\omega\varphi}\right)
\quad\tx{and}\quad
\xi^{xy}=-\frac{k_z\left(\chi_\tx{em}^{xy}\right)^2}{\omega\varphi}.
\end{equation}
\end{subequations}
Clearly, the angle $\theta_\ve{D}$ generally differs from the excitation angle  $\theta_\ve{E}=\tan^{-1}\left(\tilde{E}_y/\tilde{E}_x\right)$, i.e., $\theta_\ve{D}\neq\theta_\ve{E}$, which indicates polarization rotation. So, while it preserves the  polarizations~$(\tilde{E}_x,\tilde{H}_y,k_z)$ and $(\tilde{E}_y,-\tilde{H}_x,k_z)$, the medium actually rotates an incident wave with oblique polarization, as a result of birefringence, where different eigenstates are `presented' to the $x$- and $y$-polarized waves. However, this is a \emph{linear birefringence rotation} phenomenon, as commonly occurring in natural crystals, and this rotation phenomenon is fundamentally distinct from chiral (circular-birefringence) rotation, since it strongly depends on the incident polarization, whereas chiral polarization rotation is independent of the incident polarization (additional proof of~\ding{195} in Fig.~\ref{fig:trilogy}).

Let us now analyze the twisted Omega or helix particle, shown in Fig.~\ref{fig:two_omega_part}(b) and studied in Fig.~\ref{fig:twisted_omega}. Analogously to the case of the straight Omega particle, we shall apply~\eqref{eq:bianis_const_rel_alt_nl} to the tensorial structure~\eqref{eq:twisted_Omeg_tens} found for this particle in Sec.~\ref{sec:MaxAmp_twisted_omega}, after translating~\eqref{eq:twisted_Omeg_tens} into its macroscopic form with~\eqref{eq:exzm_from_xhi} and for the appropriate polarization. However, since this structure induces polarization rotation (Sec.~\ref{sec:MaxAmp_twisted_omega}) and since its eigenstates are consequently CP modes (Sec.~\ref{sec:chir_eig_states}), the medium resolves LP incident waves into their two CP states (RH and LH) (Sec.~\ref{sec:pol_rot}), given by~\eqref{eq:eigen_states} and~\eqref{eq:eigen_states_m}. This means that the $\ve{k}$ in~\eqref{eq:bianis_const_rel_alt_nl} corresponds actually to the eigenstates $\ve{k}^\pm_\pm$ in reference to~\eqref{eq:eigen_states} and~\eqref{eq:eigen_states_m}. Let us consider, as in Fig.~\ref{fig:twisted_omega}, a forward LH-CP wave (bottom configuration $\equiv$ top rotation multiplied by $e^{-i\pi/2}=-i$) propagating along the $+z$ direction, which corresponds to [superscript $+$ and subscript $-$ in~\eqref{eq:eigen_states}] as
\begin{subequations}
\begin{equation}
\ve{k}^+_-
=\beta^+_-\hatv{z}
=\omega\left(\sqrt{\epsilon\mu}-\chi\right)\hatv{z},
\end{equation}
\begin{equation}
\ve{E}_-^+(z)
=E_0\left(\hatv{x}-i\hatv{y}\right)e^{i\beta_-^+ z},
\end{equation}
\begin{equation}
\ve{H}_-^+(z)
=i\frac{E_0}{\eta}\left(\hatv{x}-i\hatv{y}\right)e^{i\beta_-^+ z}.
\end{equation}
\end{subequations}
For this wave, we have to set $\ve{k}=\ve{k}^+_-=\beta^+_-\hatv{z}$ ($k_x=k_y=0$), $\tilde{\ve{E}}=\tilde{\ve{E}}^+_-=E_0\left(\hatv{x}-i\hatv{y}\right)$ and $\tilde{\ve{H}}=\tilde{\ve{H}}^+_-=i(E_0/\eta)\left(\hatv{x}-i\hatv{y}\right)$ in~\eqref{eq:bianis_const_rel_alt_nl}. The result is
\begin{subequations}\label{eq:helical_part_spat_disp}
\begin{align}
\tilde{D}_x&=\left[\epsilon_0\epsilon_\tx{r}^{xx}+i\frac{k_z\chi_\tx{em}^{xx}}{\eta_0\omega\varphi}
\left(1+\chi_\tx{ee}^{xx}+\chi_\tx{me}^{xx}\right)\right]E_0, \\
\tilde{D}_y&=-i\epsilon_0E_0 \\
\tilde{D}_z&=0, \\
\tilde{B}_x&=\left[\mu_0\mu_\tx{r}^{xx}-i\frac{k_z\eta_0\chi_\tx{me}^{xx}}{\omega\varphi}\left(\chi_\tx{em}^{xx}+\mu_\tx{r}^{xx}\right)
\right]\frac{E_0}{\eta_0}, \\
\tilde{B}_y&=-i\frac{\mu_0}{\eta_0}E_0, \\
\tilde{B}_z&=0,
\end{align}
with
\begin{equation}
\varphi=1+\chi_\tx{ee}^{xx}+\left(\chi_\tx{em}^{xx}\right)^2+\chi_\tx{mm}^{xx}+\chi_\tx{ee}^{xx}\chi_\tx{mm}^{xx}.
\end{equation}
\end{subequations}
where the term $\left(\chi_\tx{em}^{xx}\right)^2$ corresponds to $-\chi_\tx{em}^{xx}\chi_\tx{me}^{xx}$ with the result $\chi_\tx{me}^{xx}=-\chi_\tx{em}^{xx}$ following from~\eqref{eq:alphamexxemxx}. These relations are harder to decipher than their planar-Omega counterparts, but certainly include precious information on the related bianisotropic chiral medium. If the monoatomic helical particle corresponding this medium would be transformed into the triatomic particle in Fig.~\ref{fig:twisted_omegas_3_orientations} to form the biisotropic chiral metamaterial shown in Fig.~\eqref{fig:chiral_metamaterial}, the complicated bianisotropic relations~\eqref{eq:helical_part_spat_disp} would naturally reduce to more fundamental and practical relations~\eqref{eq:chir_nl}.

\section{Temporal Dispersion (or Temporal Nonlocality)}\label{sec:temp_disp_design}
\subsection{Lorentz Response}
The medium susceptibilities must exhibit a frequency dependence that obeys causality or, mathematically, that follows the Kramers-Kronig relations~\cite{Jackson_1998}. In general, the response at a given time depends not only on the excitation at the same time but also at previous times. This corresponds to a temporal form of nonlocality~\cite{Jackson_1998}, but a form which is quite simpler than spatial nonlocality because of the unidimensionality of time.

For composites made of relatively low-loss particles, such as the straight and looped sections of the Omega-type particles considered in this paper (Fig.~\ref{fig:two_omega_part}), dispersion follows the Lorentz model~\cite{Jackson_1998,Ishimaru_1990}
\begin{equation}\label{eq:disp_em}
\chi_{ab}(\omega)
=-\frac{\omega_{\tx{p},ab}^2}{\omega^2-\omega_{0,ab}^2+i\omega\nu_{ab}},
\end{equation}
where $ab$ stands as usual for ee, em, me or em, and where $\omega_{0,ab}$, $\omega_{\tx{p},ab}$ and $\nu_{ab}$ denote the resonance frequencies, plasma frequencies and damping factors, respectively, while $\omega$ is the frequency of the wave impinging on the particle.

According to~\eqref{eq:exzm_from_xhi_chir}, the dispersive susceptibility relations~\eqref{eq:disp_em} correspond to the relative chiral medium parameters
\begin{subequations}\label{eq:disp_param}
\begin{equation}
\epsilon_\tx{r}(\omega)
=\frac{\epsilon(\omega)}{\epsilon_0}
=1+\chi_\tx{ee}(\omega)
=\frac{\omega^2-\omega_\tx{p,$\epsilon$}^2-\omega_\tx{0,$\epsilon$}^2+i\omega\nu_\epsilon}
{\omega^2-\omega_\tx{0,$\epsilon$}^2+i\omega\nu_\epsilon},
\end{equation}
\begin{equation}
\chi_\tx{r}(\omega)
=\frac{\chi(\omega)}{\sqrt{\epsilon_0\mu_0}}
=-i\chi_\tx{em}(\omega)
=-\frac{\omega_\tx{p,$\chi$}^2}{\omega^2-\omega_\tx{0,$\chi$}^2+i\omega\nu_\chi},
\end{equation}
\begin{equation}
\mu_\tx{r}(\omega)
=\frac{\mu(\omega)}{\mu_0}
=1+\chi_\tx{mm}(\omega)
=\frac{\omega^2-\omega_\tx{p,$\mu$}^2-\omega_\tx{0,$\mu$}^2+i\omega\nu_\mu}
{\omega^2-\omega_\tx{0,$\mu$}^2+i\omega\nu_\mu},
\end{equation}
\end{subequations}
where we have replaced wherever appropriate the subscripts ee, mm and em by the subscripts $\epsilon$, $\mu$ and $\chi$, respectively, for notational convenience.

Using low-loss materials, we typically have $\omega\nu_{\epsilon,\mu,\chi}\ll\omega^2$, and the damping terms $\omega\nu_{\epsilon,\mu,\chi}$ can then be neglected to the first order. Moreover, the resonance frequency, is often considerably lower than the plasma and operating frequencies, i.e., $\omega_0\ll\omega_\tx{p,$\epsilon,\mu,\chi$},\omega$. Under such conditions, the functions~\eqref{eq:disp_em} reduce to the Drude lossless form $\chi_{\epsilon,\mu,\chi}(\omega)=-\left(\omega_\tx{p,$\epsilon,\mu,\chi$}/\omega\right)^2$, and Eqs.~\eqref{eq:disp_param} can be approximated as
\begin{subequations}\label{eq:disp_param_red}
\begin{equation}
\epsilon_\tx{r}(\omega)
=1+\chi_\tx{ee}(\omega)
\approx\frac{\omega^2-\omega_\tx{p,$\epsilon$}^2}{\omega^2},
\end{equation}
\begin{equation}
\chi_\tx{r}(\omega)
=\frac{\chi(\omega)}{\sqrt{\epsilon_0\mu_0}}
\approx-\frac{\omega_\tx{p,$\chi$}^2}{\omega^2},
\end{equation}
\begin{equation}
\mu_\tx{r}(\omega)
=1+\chi_\tx{mm}(\omega)
\approx\frac{\omega^2-\omega_\tx{p,$\mu$}^2}{\omega^2}.
\end{equation}
\end{subequations}
These relations provide precious insight into the dispersion response of a low-loss chiral medium. They most importantly reveal that in such media $\epsilon(\omega)$ and $\mu(\omega)$ tend to be respectively positive and negative above and below their plasma frequency while $\chi(\omega)$ is always negative.

\subsection{Metamaterial Resonance and Plasma Frequencies}

A metamaterial maximally interacts with electromagnetic energy near its lowest resonance (Sec.~\ref{sec:dim_constr}), which occurs at the frequency where the incoming wave best matches the boundary conditions of the metaparticles forming the metamaterial, while fulfilling the subwavelength dimensional constraint~\eqref{eq:subl_latt}. In the case of single-block particles, as the Omega-type particles considered in Sec.~\ref{sec:chir_mat_meta} (Fig.~\ref{fig:two_omega_part}), this corresponds to the situation where the size of the unfolded structure, $\ell$, is half the wavelength at resonance ($\ell=\lambda_\tx{res}/2$) if it is highly conductive, or somewhat less ($\lambda_\tx{res}/10<\ell<\lambda_\tx{res}/2$) if it is dielectric or plasmonic, due to field penetration, and not to the separate sizes of the straight-pair section and looped section. There is thus a unique and common resonance frequency for the three parameters in~\eqref{eq:disp_param}, i.e.,
\begin{equation}
\omega_0=\omega_{0,\epsilon}=\omega_{0,\mu}=\omega_{0,\chi}.
\end{equation}

Contrarily to the resonance frequency, the plasma frequencies may differ from each other. Indeed, the plasma frequencies in~\eqref{eq:disp_param} are proportional to the density of the corresponding dipole moments~\footnote{This is well-known to be the case in regular plasmas, where the plasma frequency of the conduction electrons is given by $\omega_\tx{p}=\sqrt{Ne^2/(m^*\epsilon_0)}$, where $N$, $e$ and $m^*$ are the density, charge and effective mass of these electron~\cite{Jackson_1998}, but this is also true in metamaterials, as shown in~\cite{Pendry_1999} for split-ring resonators upon safely defining the effective permeability as the ratio of the $\ve{B}$ and $\ve{D}$ fields.}, which are generally different for the electric ($\ve{p}$) and magnetic ($\ve{m}$) responses. For instance, in the case of the Omega particles in Fig.~\ref{fig:two_omega_part}, we have seen in Sec.~\ref{sec:chir_mat_meta} that most of the $\ve{p}$ response is related to the straight section while most of the $\ve{m}$ response is related to the looped section, which leads to particularizing~\eqref{eq:pm_dip_mmt} as
\begin{subequations}
\begin{equation}
\ve{p}_\tx{e}=\int_{V_\tx{straight}}\ve{r}'\rho(\ve{r}')d\ve{r}',\label{eq:m_dip_mm}
\end{equation}
\begin{equation}
\ve{p}_\tx{m}=\frac{\mu_0}{2}\int_{V_\tx{looped}}\ve{r}'\times\ve{J}(\ve{r}')d\ve{r}',\label{eq:m_dip_mmt}
\end{equation}
\end{subequations}
where the relevant integration volumes have been explicited. Omega particles with a much larger (resp. smaller) straight section and a much smaller (resp. larger) looped section, following the antagonistic dimensional constraint~\eqref{eq:antag_constr}, will then be characterized by larger densities of electric (resp. magnetic) dipole moments, and hence by larger electric (resp. magnetic) plasma frequency, while the cross-coupling frequency, involving the same geometrical parts, will be intermediate, i.e.,
\begin{subequations}\label{eq:plasma_seq}
\begin{equation}\label{eq:plasma_seq_straight}
2d\gg s\quad\Longleftrightarrow\quad
\omega_{\tx{p},\epsilon}>\omega_{\tx{p},\chi}>\omega_{\tx{p},\mu}
\end{equation}
and
\begin{equation}\label{eq:plasma_seq_loop}
2d\ll s\quad\Longleftrightarrow\quad
\omega_{\tx{p},\epsilon}<\omega_{\tx{p},\chi}<\omega_{\tx{p},\mu}.
\end{equation}
\end{subequations}

\subsection{Parametric Study}\label{sec:param_stud}
\begin{figure}[h]
    \centering
        \includegraphics[width=\linewidth]{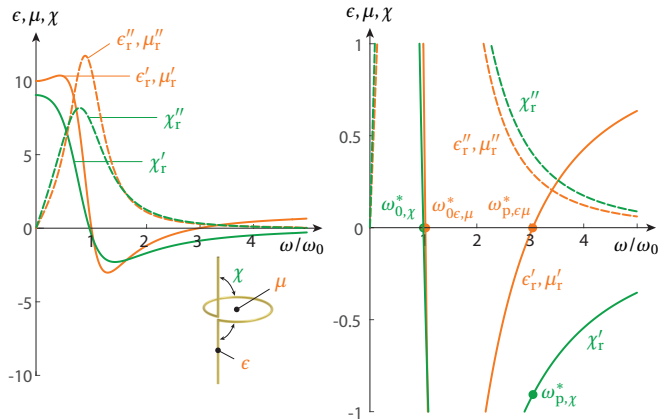}{
        \psfrag{x}[c][c][0.8]{$\omega/\omega_0$}
        \psfrag{y}[c][c][0.8]{$\epsilon,\mu,\chi$}
        \psfrag{e}[l][l][0.8]{\textcolor{orangeg}{$\epsilon_\tx{r}',\mu_\tx{r}'$}}
        \psfrag{i}[l][l][0.8]{\textcolor{orangeg}{$\epsilon_\tx{r}'',\mu_\tx{r}''$}}
        \psfrag{k}[r][r][0.8]{\textcolor{orangeg}{$\epsilon_\tx{r}'',\mu_\tx{r}''$}}
        \psfrag{c}[l][l][0.8]{\textcolor{greeng}{$\chi_\tx{r}'$}}
        \psfrag{m}[l][l][0.8]{\textcolor{greeng}{$\chi_\tx{r}''$}}
        \psfrag{a}[c][c][0.8]{\textcolor{orangeg}{$\omega_{0\epsilon,\mu}^*$}}
        \psfrag{b}[c][c][0.8]{\textcolor{orangeg}{$\omega_{\tx{p},\epsilon\mu}^*$}}
        \psfrag{j}[c][c][0.8]{\textcolor{greeng}{$\omega_{\tx{p},\chi}^*$}}
        \psfrag{d}[c][c][0.8]{\textcolor{greeng}{$\omega_{0,\chi}^*$}}
        \psfrag{s}[c][c][0.8]{\textcolor{orangeg}{$\epsilon$}}
        \psfrag{t}[c][c][0.8]{\textcolor{orangeg}{$\mu$}}
        \psfrag{z}[c][c][0.8]{\textcolor{greeng}{$\chi$}}
        }
        \vspace{-6mm}
        \caption{Dispersion responses~\eqref{eq:disp_param} for the parameters $\omega_{\tx{p},\epsilon}=\omega_{\tx{p},\mu}=\omega_{\tx{p},\chi}=\omega_\tx{p}=3\omega_0$, corresponding to helix particle with similar straight-section and looped-section dimensions, $\nu_\epsilon=\nu_\mu=0.8\omega_0$ and $\nu_\chi=1.2\omega_0$. (a)~Full scale. (b)~Zoom about the abscissa. The exact poles are $\omega_{0\chi}^*/\omega_0\approx 1.000$ and $\omega_{0\epsilon}^*/\omega_0=\omega_{0\mu}^*/\omega_0\approx 1.038$, and the exact zero is $\omega_{\tx{p}\epsilon}^*/\omega_0=\omega_{\tx{p}\mu}^*/\omega_0\approx 3.047$.}
   \label{fig:disp_resp}
\end{figure}

Figure~\ref{fig:disp_resp} plots the dispersive responses~\eqref{eq:disp_param} for a chiral medium with equal permittivity, permeability and chiral factor plasma frequencies, $\omega_{\tx{p},\epsilon}=\omega_{\tx{p},\mu}=\omega_{\tx{p},\chi}=\omega_\tx{p}$, occurring when the straight and looped sections of the helix particle have comparable dimensions. As predicted by~\eqref{eq:disp_param_red}, the real parts of the permittivity and permeability are positive (positive refractive index) and negative (negative refractive index) above and below that frequency, respectively~\cite{Caloz_2005}, while
$\chi$ is always negative in that frequency range. Moreover, the imaginary parts of $\epsilon$, $\mu$ and $\chi$ are always positive, consistently with the implicit assumption of passivity (Sec.~\ref{sec:loss} and Appendix~\ref{sec:loss_imag_sign}).

Interestingly, at the common -- or \emph{balanced} -- plasma frequency, we have $\epsilon=\mu=0$ -- and hence zero refractive index~\cite{Caloz_2005} -- and $\chi\neq 0$. Equations~\eqref{eq:chir_const_rel} reduce then to
\begin{subequations}\label{eq:pure_chir_const_rel}
\begin{align}\label{eq:pure_chir_const_rel_D}
\ve{D}(\omega_\tx{p})&=i\chi(\omega_\tx{p})\ve{H}(\omega_\tx{p}), \\
\label{eq:pure_chir_const_rel_B}
\ve{B}(\omega_\tx{p})&=-i\chi(\omega_\tx{p})\ve{E}(\omega_\tx{p}),
\end{align}
\end{subequations}
which correspond to a \emph{purely chiral} medium, also previously called \emph{chiral nihility}~\cite{Tretyakov_2003}. In this case, an LP wave is still rotated by the angle $\theta_\tx{C}(z)=\mp\omega\chi z$ [Eq.~\eqref{eq:chiral_rot_ang}, where $\theta_\tx{C}(z)$ does not depend on $(\epsilon,\mu)$], but it does not undergo any phase shift, since $\phi_\tx{C}(z)=\pm\omega\sqrt{\epsilon\mu}z=0$ [Eq.~\eqref{eq:chiral_rot_pha}].

At the resonance frequency, all the parameters change sign, which specifically means for $\chi(\omega)$ a change of the polarization-rotation direction, according to~\eqref{eq:chiral_rot_ang}, but this part of the spectrum is highly lossy and should therefore be avoided unless really needed.

\begin{figure}[h]
    \centering
        \includegraphics[width=\linewidth]{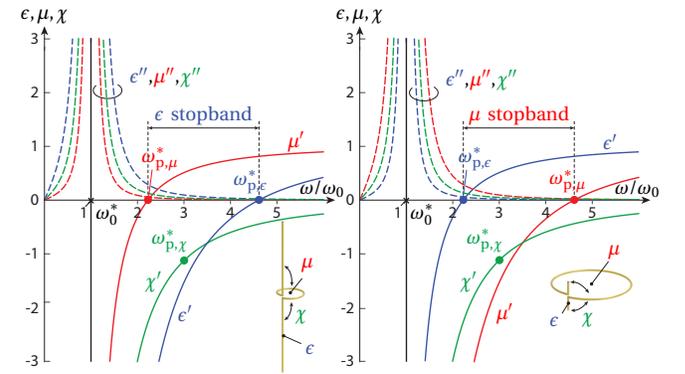}{
        \psfrag{x}[c][c][0.8]{$\omega/\omega_0$}
        \psfrag{y}[c][c][0.8]{$\epsilon,\mu,\chi$}
        \psfrag{p}[l][l][0.8]{\textcolor{blueg}{$\epsilon''$},\textcolor{redg}{$\mu''$},\textcolor{greeng}{$\chi''$}}
        \psfrag{e}[l][l][0.8]{\textcolor{redg}{$\mu'$}}
        \psfrag{m}[l][l][0.8]{\textcolor{blueg}{$\epsilon'$}}
        \psfrag{c}[l][l][0.8]{\textcolor{greeng}{$\chi'$}}
        \psfrag{a}[c][c][0.8]{$\omega_0$}
        \psfrag{b}[c][c][0.8]{$\omega_\tx{p}$}
        \psfrag{s}[c][c][0.8]{\textcolor{blueg}{$\epsilon$}}
        \psfrag{t}[c][c][0.8]{\textcolor{redg}{$\mu$}}
        \psfrag{z}[c][c][0.8]{\textcolor{greeng}{$\chi$}}
        \psfrag{k}[c][c][0.8]{\textcolor{redg}{$\omega_{\tx{p},\mu}^*$}}
        \psfrag{l}[c][c][0.8]{\textcolor{blueg}{$\omega_{\tx{p},\epsilon}^*$}}
        \psfrag{u}[c][c][0.8]{\textcolor{greeng}{$\omega_{\tx{p},\chi}^*$}}
        \psfrag{f}[c][c][0.8]{$\omega_0^*$}
        \psfrag{S}[c][c][0.8]{\textcolor{blueg}{$\epsilon$ stopband}}
        \psfrag{T}[c][c][0.8]{\textcolor{redg}{$\mu$ stopband}}
        }
        \vspace{-6mm}
        \caption{Dispersion responses~\eqref{eq:disp_param} for $\omega_{\tx{p},\chi}=3\omega_0$ and $\nu_\epsilon=\nu_\mu=\nu_\chi=0.1$ ($\omega_{0,\epsilon}^*/\omega_0\approx\omega_{0,\mu}^*/\omega_0\approx\omega_{0,\chi}^*/\omega_0\approx 1.00$). (a)~Helix with straight section much larger than looped section ($2d\gg s$), and $\omega_{\tx{p},\epsilon}=4.5\omega_0$ and $\omega_{\tx{p},\mu}=2\omega_0$ ($\omega_{\tx{p},\epsilon}^*/\omega_0=4.609$, $\omega_{\tx{p},\mu}^*/\omega_0=2.333$). (b)~Helix with looped section much larger than straight section ($s\gg 2d$), and $\omega_{\tx{p},\epsilon}=2\omega_0$ and $\omega_{\tx{p},\mu}=4.5\omega_0$ ($\omega_{\tx{p},\epsilon}^*/\omega_0=2.333$, $\omega_{\tx{p},\mu}^*/\omega_0=4.609$).}
   \label{fig:design}
\end{figure}

Figure~\ref{fig:design} investigates the situation of chiral media with different plasma frequencies, with Fig.~\ref{fig:design}(a) [resp. Fig.~\ref{fig:design}(b)] corresponding to a helix particle with a straight section that is much longer (resp. much shorter) than the looped section, whose plasma frequencies follow the sequence~\eqref{eq:plasma_seq_straight} [resp.~\eqref{eq:plasma_seq_loop}]. Here, the difference in the plasma frequencies leads to the opening of forbidden bands, or stopbands, corresponding imaginary phase shifts in~\eqref{eq:chiral_rot_pha}. These stopbands correspond to the frequency range extending between the double-negative (or negative refractive index) and double-positive (or positive refractive index) bands. The same comment as above applies for the common resonance frequency.

\section{Design Guidelines}\label{sec:des_princ}
A chiral metamaterial may be designed using the following procedure:
\begin{enumerate}
  \item Typically choose the balanced design of Fig.~\ref{fig:disp_resp}, rather than an unbalanced design as in Fig.~\ref{fig:design}, to avoid permittivity or permeability stopbands (Sec.~\ref{sec:param_stud}). We have then $\omega_{\tx{p},\epsilon}=\omega_{\tx{p},\mu}=\omega_{\tx{p},\chi}=\omega_\tx{p}$, corresponding to a particle with balanced straight-section and looped-section dimensions.
  \item If one targets a purely chiral response (polarization rotation without phase shift), corresponding to~\eqref{eq:pure_chir_const_rel}, then the medium should be operated at the common plasma frequency, $\omega_\tx{p}$; so, set the plasma frequency equal to the operation frequency, $\omega_\tx{op}$, i.e., $\omega_\tx{p}=\omega_\tx{op}$. Otherwise, set $\omega_\tx{p}=r\omega_\tx{op}$ with $r>0$ for negative-index response or $r<0$ for positive-index response; $|r|$ might initially be in the order of $1.1$ or $0.9$ for operation at $10\%$ off the plasma frequency.
  \item Select a particle with a 90$^\circ$ twist with between the straight and looped part of the structure, such as the helix in Fig.~\ref{fig:two_omega_part}(b).
  \item Since the operation frequency must be larger than the resonance frequency, $\omega_\tx{op}>\omega_\tx{0}$, and since Eq.~\eqref{eq:reson_part} indicates that the unfolded length of the particle is half the wavelength at resonance ($\ell=\lambda_\tx{res}/2$), that length much be somewhat greater at the operation frequency, i.e., $\ell>c/(2f_\tx{op})=\pi c/\omega_\tx{op}$. Set an initial guess for it, possibly using some analytical formulas~\cite{Tretyakov_1996}.
  \item Simulate the scattering parameters of the particle as indicated in Fig.~\ref{fig:helix_extrac}, and extract the (frequency-dependent) parameters $\chi$, $\epsilon$ and $\mu$ from the formulas (see Appendix~\ref{sec:param_extr_formulas})
\begin{subequations}\label{eq:extraction}
\begin{equation}
\chi
=-\frac{\tan^{-1}\left(\left|S_{21}^{yx}\right|/\left|S_{21}^{xx}\right|\right)}{\omega p},
\end{equation}
\begin{equation}
\epsilon=\frac{1}{Z_\tx{port}}\left(\frac{\angle S_{21}^{yx}}{\omega p}\right)\left(\frac{1-S_{11}^{xx}}{1+S_{11}^{xx}}\right),
\end{equation}
\begin{equation}
\mu=Z_\tx{port}\left(\frac{\angle S_{21}^{yx}}{\omega p}\right)\left(\frac{1+S_{11}^{xx}}{1-S_{11}^{xx}}\right),
\end{equation}
\end{subequations}
where $p$ is the period (see Fig.~\ref{fig:metam_ex}). Note that this technique automatically accounts for interparticle coupling in case the dilute-medium condition is not satisfied, thanks to the utilization of periodic boundary conditions in the $x$ and $y$ directions (infinite periodicity) and to the utilization of a sufficient number of periods in the $z$ direction for simulation convergence to the periodic regime.
\begin{figure}[h]
    \centering
        \includegraphics[width=0.8\linewidth]{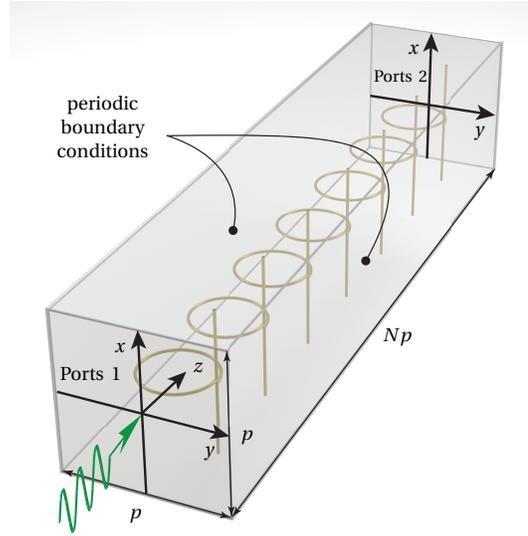}{
        \psfrag{1}[c][c][0.8]{Ports~1}
        \psfrag{2}[c][c][0.7]{Ports~2}
        \psfrag{p}[c][c][0.8]{$p$}
        \psfrag{N}[c][c][0.8]{$Np$}
        \psfrag{x}[c][c][0.8]{$y$}
        \psfrag{y}[c][c][0.8]{$x$}
        \psfrag{z}[c][c][0.8]{$z$}
        \psfrag{B}[c][c][0.8]{\begin{minipage}{2cm}\centering periodic \\ boundary \\ conditions \end{minipage}}
        }
        \vspace{-3mm}
        \caption{Full-wave simulation set up of the chiral particle for synthesis of the corresponding chiral metamaterial (Fig.~\ref{fig:metam_ex}) by iterative analysis. The computational box is a $z$-oriented cylinder
        with rectangular cross section, of dimension $p\times p$ and length $Np$ (i.e., $N$ periods, $N\in[5,10]$ -- here $N=7$ -- for convergence to a periodic response upon washing out the port-edge aperiodic effects), whose two sides perpendicular to $\hatv{z}$ are defined as input and output ports while the four parallel sides parallel to $\hatv{z}$ are defined as periodic boundary conditions (infinite periodicity in $x$ and $y$ directions). The ports~1 are P$_1^x$ and P$_1^y$ while the ports~2 are P$_2^x$ and P$_2^y$ (e.g. $S_{21}^{yx}$ represents the transmission from the port P$_1^x$ to the port P$_2^y$). The figure shows the monoatomic helix metaparticle, which might be sufficient to design the full chiral metamaterial from mapping with the spatial-dispersion explicit formulas~\eqref{eq:helical_part_spat_disp}. Otherwise, for simply using the formulas~\eqref{eq:extraction} and safely accounting for coupling between the particles in the different directions, one should rather replace the monoatomic metaparticle by its triatomic counterpart in Fig.~\ref{fig:twisted_omegas_3_orientations}.}
   \label{fig:helix_extrac}
\end{figure}
  \item Fine-tune the structure parameters following some appropriate optimization tool until satisfaction.
\end{enumerate}

\section{Conclusions}\label{sec:concl}

We have presented a first-principle and global perspective of electromagnetic chirality following a bottom-up construction from chiral particle or metaparticles (microscopic scale), through the electromagnetic theory of chiral media (macroscopic scale), to the establishment of advanced properties and design principles of chiral materials and metamaterials. The main conclusions and results may be summarized as follows:
\begin{enumerate}
  \item \label{list:chir_def} Chirality is a geometric property according to which an object is mirror asymmetric or, equivalently, different from its image in a mirror, irrespectively to orientation.
  \item As a consequence of~\ref{list:chir_def}), a chiral particle must have a volume; a purely planar particle is always mirror symmetric, and therefore never chiral. Moreover, the particle must include some `twisting' that breaks spatial symmetry. So, volume is a necessary, but not a sufficient condition for chirality.
  \item \label{list:three_obs} A chiral material or metamaterial is a medium constituted of chiral particles or metaparticles. Such a medium induces polarization rotation (irrespectively to the polarization of the incident wave) associated with magnetoelectric coupling (coupling between the electric and magnetic responses).
  \item According to~\ref{list:chir_def}) and \ref{list:three_obs}), chirality is intimately related to the concepts of mirror asymmetry, polarization rotation and magnetoelectric coupling. However, these concepts are not trivially interdependent:
      \begin{itemize}
        \item  According to~\eqref{list:chir_def}, mirror asymmetry is a necessary and sufficient condition for chirality.
        \item Mirror asymmetry implies polarization rotation and magnetoelectric coupling, but neither polarization rotation nor magnetoelectric coupling implies mirror asymmetry.
        \item Polarization rotation and magnetoelectric coupling do not imply each other.
      \end{itemize}
  \item Although nature includes many chiral substances, such substances suffer from chemical instability, high loss and restricted spectrum, and metamaterials are therefore a more promising technology for electromagnetic applications of chirality.
  \item In the dilute-medium regime (material with negligible interactions between its constitutive particles), the macroscopic properties of a medium can be inferred from their microscopic inspection: there is a one-to-one correspondence between the susceptibility tensors of the medium and the polarizability tensors of the particles that constitute it, the former being a density of averages of the latter. This fact is the foundation of the bottom-up development of this paper, where all the results have been inferred from the initial definition of chirality without any other a priori assumption. In the dense-medium regime, the coupling between particles may be accounted for by adding interaction tensors and using full-wave electromagnetic simulations, and fundamental findings presented in the paper can still be used as qualitative principles.
  \item In a magnetoelectric particle, the electric-to-magnetic and magnetic-to-electric responses are always opposite to each other.
  \item A planar -- and hence achiral -- particle involving magnetoelectric coupling, such as the planar Omega particle studied in the paper, cannot form the basis of an isotropic medium; the resulting media are always bianisotropic (and achiral). In contrast, a chiral particle, such as the helix particle studied in the paper, can, assuming a triatomic metaparticle with a copy of the basic chiral particle in each of the three directions of space (or a random arrangement of such copies). What is called a chiral medium in the physics community is generally an isotropic, or biisotropic, chiral medium, but a chiral medium (with mirror asymmetry, and hence magnetoelectric coupling and polarization rotation) can also be anisotropic, as a matter of fact.
  \item A biisotropic medium is not necessarily chiral. It is chiral if its electric and magnetic responses are in quadrature with their magnetic and electric counterpart excitations. In this case, the medium is indeed chiral, and is also called a Pasteur medium, and such a medium is reciprocal. If these responses and excitations are in phase, the biisotropic medium is called a Tellegen medium; such a medium is nonreciprocal, and therefore requires magnetization.
  \item Whereas the permittivity and permeability functions are even under space reversal, the chirality parameter function is odd under space reversal.
  \item The fundamental modes, or eigenstates, of a chiral medium are circularly polarized left-handed or right-handed waves, whose handedness changes with the direction of propagation. The chiral eigenstates `see' the chiral medium as monoisotropic.
  \item The polarization rotation in a chiral medium is a direct consequence of its circularly polarized eigenstates: the chiral medium resolves an incident linearly polarized wave in its right-handed and left-handed components and assigns them the corresponding different wavenumbers (or refractive indices), which results in a net polarization rotation, with rotation direction depending on the handedness of the chiral particles forming the medium. This is called circular birefringence, not to be confused with linear birefringence (e.g. in the planar Omega particle or standard birefringent crystals), where rotation occurs only for some incident polarizations.
  \item In contrast to Faraday rotation, which is nonreciprocal (and monoanisotropic), chiral or Pasteur polarization rotation, unwinds to its original state upon time reversal.
  \item Different right-handed and left-handed eigenstate absorptions in the presence of loss distort circular polarization into elliptic polarization, a phenomenon called circular dichroism.
  \item A chiral medium is spatially dispersive (or spatially nonlocal), i.e., its response at a given point of space depends also on the excitation in the vicinity of that point. Spatial dispersion leads to an alternative of the standard bianisotropic or biisotropic constitutive relations where the component of the response fields can be explicitly written in terms of the components of the excitation fields. This formulation of chirality is explicit but tensorial, whereas the standard formulation is scalar but implicit. The spatial dispersion perspective sheds on chirality physics a light that complements the standard perspective.
  \item A chiral medium is also temporally dispersive (or temporally nonlocal), due to causality, typically with Lorentz dispersion in the three constitutive parameters (permittivity, permeability and chirality factor). In the lossless case, the three parameters are purely real, and all are complex in the general, lossy case. The three parameters share the same resonance frequency, corresponding to the unfolded length of the particle, but have different and antagonistic electric and magnetic plasma frequencies, depending on the densities of the straight and folded parts of the chiral particle. In the balanced regime, where the electric and magnetic plasma frequencies are merged, the medium is purely chiral at the common plasma frequency, with double negative (or negative index) and double positive (or positive index) below and above that frequency, respectively. In the unbalanced regime, a stopband exists between the two plasma frequencies.
  \item A chiral metamaterial can be designed following a given procedure that leverages the concepts developed in the paper and uses full-wave simulation extraction in a setup involving periodic boundary conditions and orthogonal ports.
\end{enumerate}

Although it was discovered more then 200 years ago, chirality has been relatively little used in electromagnetic (microwave, terahertz and optical) applications to date, probably because of implementation difficulties and theoretical complexity. We expect that current developments of nanotechnologies and modern metamaterial/metasurface concepts will spur a diversity of developments in this area, and we hope that the present overview will contribute to such developments.

\section*{Acknowledgment}
The three-dimensional drawings of the particles in the figures have been realized by Amar Al-Bassam, using the 3D computer graphics software Blender.

\bibliography{Electromagnetic_Chirality_Caloz}

\appendix

\section*{Appendices}

\section{Units of the Polarizabilities and of the Constitutive Parameters}\label{app:units_const_par}

The units of the polarizabilities may be found by the definitions~\eqref{eq:polariz_tens} as
\begin{subequations}
\begin{equation}
\left[\te{\alpha}_\tx{ee}\right]
=\left[\frac{\ve{p}_\tx{ee}}{\ve{E}_\tx{loc}}\right]
=\frac{\tx{Asm}}{\tx{V/m}}
=\frac{\text{Asm$^2$}}{\text{V}},
\end{equation}
\begin{equation}
\left[\te{\alpha}_\tx{em}\right]
=\left[\frac{\ve{p}_\tx{em}}{\ve{H}_\tx{loc}}\right]
=\frac{\tx{Asm}}{\tx{A/m}}
=\tx{sm}^2,
\end{equation}
\begin{equation}
\left[\te{\alpha}_\tx{me}\right]
=\left[\frac{\ve{p}_\tx{me}}{\ve{E}_\tx{loc}}\right]
=\frac{\tx{Vsm}}{\tx{V/m}}
=\tx{sm}^2,
\end{equation}
\begin{equation}
\left[\te{\alpha}_\tx{mm}\right]
=\left[\frac{\ve{p}_\tx{mm}}{\ve{H}_\tx{loc}}\right]
=\frac{\tx{Vsm}}{\tx{A/m}}
=\frac{\text{Vsm$^2$}}{\text{A}}.
\end{equation}
\end{subequations}

The units of the medium parameters in the constitutive relations~\eqref{eq:bianis_const_rel} or~\eqref{eq:biis_const_rel},~\eqref{eq:biis_const_rel_split},~\eqref{eq:Tellegen_rel} or~\eqref{eq:chir_const_rel} -- specifically ($\te{\epsilon}$, $\epsilon$), ($\te{\mu}$, $\mu$), ($\te{\xi}$, $\xi$, $\tau$, $\chi$) and ($\te{\zeta}$, $\zeta$, $\tau$, $\chi$) -- may be found by first isolating the parameter of interest in the same relations, next replacing $\ve{D}$ and $\ve{B}$ by their respective expressions in Maxwell equations [Eqs.~\eqref{eq:Maxwell_curl}, and finally remembering that the units of $\nabla$, $\ve{E}$, $\ve{H}$ and $\omega$ are rad/m~\footnote{In Maxwell equations, the differential operator $\nabla$ refers to space, i.e., $\nabla=\nabla_\ve{r}$, and applies to electromagnetic waves, which, assuming linearity, decompose in plane waves, $\ves{\tilde{\Psi}}(\ve{k})e^{i\ve{k}\cdot\ve{r}}$~\cite{Clemmow_2013}. We have then that $\nabla_\ve{r}\times=i\ve{k}\times$ or, more generally, $\nabla_\ve{r}=i\ve{k}$, whose units is clearly rad/m.}, V/m, A/m and rad/s, respectively. This yields for instance for~\eqref{eq:biis_const_rel}
\begin{subequations}
\begin{equation}
\left[\epsilon\right]
=\left[\frac{\ve{D}}{\ve{E}}\right]
=\left[\frac{\frac{\nabla\times\ve{H}}{-i\omega}}{\ve{E}}\right]
=\frac{\frac{\text{(rad/m)(A/m)}}{\text{rad/s}}}{\text{V/m}}
=\frac{\text{As}}{\text{Vm}},
\end{equation}
\begin{equation}
\left[\mu\right]
=\left[\frac{\ve{B}}{\ve{H}}\right]
=\left[\frac{\frac{\nabla\times\ve{E}}{i\omega}}{\ve{H}}\right]
=\frac{\frac{\text{(rad/m)(V/m)}}{\text{rad/s}}}{\text{A/m}}
=\frac{\text{Vs}}{\text{Am}},
\end{equation}
\begin{equation}
\left[\xi\right]
=\left[\frac{\ve{D}}{\ve{H}}\right]
=\left[\frac{\frac{\nabla\times\ve{H}}{-i\omega}}{\ve{H}}\right]
=\frac{\frac{\text{(rad/m)(A/m)}}{\text{rad/s}}}{\text{A/m}}
=\frac{\text{s}}{\text{m}},
\end{equation}
or
\begin{equation}
\left[\zeta\right]
=\left[\frac{\ve{B}}{\ve{E}}\right]
=\left[\frac{\frac{\nabla\times\ve{E}}{i\omega}}{\ve{E}}\right]
=\frac{\frac{\text{(rad/m)(V/m)}}{\text{rad/s}}}{\text{V/m}}
=\frac{\text{s}}{\text{m}},
\end{equation}
\end{subequations}
where the last two results are identical, consistently with~\eqref{eq:biis_xi_ze_split}.

\section{Signs of the Imaginary Parts of the Chiral Constitutive Parameters in the Lossy Case}\label{sec:loss_imag_sign}
Let us first consider the case of~$\epsilon$. According to~\eqref{eq:exzm_from_xhi_chir},
\begin{subequations}
\begin{equation}
\epsilon_\tx{r}=\frac{\epsilon}{\epsilon_0}=\epsilon_\tx{r}'+i\epsilon_\tx{r}''
=1+\chi_\tx{ee}
=\left(1+\chi_\tx{ee}'\right)+i\chi_\tx{ee}'',
\end{equation}
i.e.,
\begin{equation}
\epsilon_\tx{r}'=1+\chi_\tx{ee}'
\quad\tx{and}\quad
\epsilon_\tx{r}''=\chi_\tx{ee}'',
\end{equation}
\end{subequations}
where the primed and double-primed quantities denote the real and imaginary parts, respectively.

Given the assumed plane-wave forward ($+r$-direction propagating) spacetime dependence $\psi\propto e^{ikr}=e^{i(k'+ik'')r}=e^{ik'r}e^{-k''r}$ [Eq.~\eqref{eq:PW_space}], one must have $k''>0$ for exponential decay in the $+r$ direction if the medium is lossy. Assuming first a nonmagnetic medium, we have
\begin{equation}
\begin{split}
k=k_0\sqrt{\epsilon_\tx{r}}
=k_0\sqrt{\epsilon_\tx{r}'+i\epsilon_\tx{r}''}
&=k_0\sqrt{\epsilon_\tx{r}'}\sqrt{1+i\frac{\epsilon''}{\epsilon'}} \\
&\overset{\epsilon''\ll\epsilon'}{\approx}k_0\sqrt{\epsilon_\tx{r}'}\left(1+i\frac{\epsilon''}{2\epsilon'}\right),
\end{split}
\end{equation}
where we have made the low-loss assumption $\epsilon''\ll\epsilon'$ in the last equality. We have thus found that $k''=k_0\epsilon''/(2\epsilon')$ and, assuming $\epsilon'>0$, the $k''$ positivity required for lossy absorption translates into $\epsilon''$ positivity, i.e., $\epsilon''>0$.

Exactly the same argument leads to the conclusion that we have also $\mu''>0$ in a lossless medium. The case of $\chi$ is slightly more subtle, but can be resolved by considering the circularly-polarized eigenstates of~\eqref{eq:E_4sols}, $\beta^+_+=\omega(\sqrt{\epsilon\mu}+\chi)$. At the electric plasma frequency or magnetic plasma frequency (Sec.~\ref{sec:param_stud}), we have $\sqrt{\epsilon\mu}=0$ and hence $\beta^+_+=\omega\chi=\omega(\chi'+i\chi'')=k'+ik''$, which indicates that we also have $\chi''>0$ if the medium is purely lossy.

\section{Derivation of the Chiral Eigenstates}\label{app:der_chir_eig}

Inserting the $z$-forward plane-wave fields~\eqref{eq:fwd_EH} into the chiral-explicit Maxwell equations~\eqref{eq:Maxwell_curl_mod}, using the $z$-propagating plane-wave conditions $\partial/\partial x=\partial/\partial y=0$, and projecting onto the $x,y,z$ axes, yields
\begin{subequations}\label{eq:Maxwell_proj}
\begin{equation}\label{eq:Maxwell_proja}
-i\beta E_y=\omega\chi E_x+i\omega\mu H_x,
\end{equation}
\begin{equation}\label{eq:Maxwell_projb}
i\beta E_x=\omega\chi E_y+i\omega\mu H_y,
\end{equation}
\begin{equation}\label{eq:Maxwell_projc}
-i\beta H_y=-i\omega\epsilon E_x+\omega\chi H_x,
\end{equation}
\begin{equation}\label{eq:Maxwell_projd}
i\beta H_x=-i\omega\epsilon E_y+\omega\chi H_y,
\end{equation}
\end{subequations}
which form a $4\times 4$ homogeneous linear system of equations in $E_x$, $E_y$, $H_x$ and $H_z$. Solving now~\eqref{eq:Maxwell_proja} and~\eqref{eq:Maxwell_projb} for $H_x$ and $H_y$, respectively, yields
\begin{subequations}\label{eq:HxHx_vs_ExEy}
\begin{equation}
H_x
=i\frac{\chi}{\mu}E_x-\frac{\beta}{\omega\mu}E_y,
\end{equation}
\begin{equation}
H_y
=\frac{\beta}{\omega\mu}E_x+i\frac{\chi}{\mu}E_y,
\end{equation}
\end{subequations}
and inserting these results into~\eqref{eq:Maxwell_projc} and~\eqref{eq:Maxwell_projd} reduces~\eqref{eq:Maxwell_proj} to the $2\times 2$ system
\begin{subequations}\label{eq:Ex_Ey_syst}
\begin{equation}\label{eq:Ex_Ey_systa}
\left[\beta^2-\omega^2\left(\epsilon\mu-\chi^2\right)\right]E_x
+2i\beta\chi\omega E_y
=0,
\end{equation}
\begin{equation}\label{eq:Ex_Ey_systb}
-2i\beta\chi\omega E_x
+\left[\beta^2-\omega^2\left(\epsilon\mu-\chi^2\right)\right]E_y
=0.
\end{equation}
\end{subequations}

Finally, nullifying the determinant of this system for a nontrivial solution leads to the quartic equation
\begin{equation}\label{eq:quartic_eq}
\left[\beta^2+\omega^2\left(\chi^2-\epsilon\mu\right)\right]^2
-4\beta^2\omega^2\chi^2=0.
\end{equation}
The resolution of this equation provides the four modal wavenumbers
\begin{equation}
\beta^\pm_\pm
=\omega\left(\pm\sqrt{\epsilon\mu}\pm\chi\right),
\end{equation}
while inserting these wavenumbers into~\eqref{eq:Ex_Ey_systa} or~\eqref{eq:Ex_Ey_systb}, and subsequently comparing the coefficients of $E_x$ and $E_y$, provides the four modal electric fields
\begin{subequations}\label{eq:Epm_der}
\begin{align}
\ve{E}^+_+&=E_0\left(\hatv{x}+i\hatv{y}\right)e^{i\beta^+_+z},\\
\ve{E}^+_-&=E_0\left(\hatv{x}-i\hatv{y}\right)e^{i\beta^+_-z},\\
\ve{E}^-_+&=E_0\left(\hatv{x}+i\hatv{y}\right)e^{i\beta^-_+},\\
\ve{E}^-_-&=E_0\left(\hatv{x}-i\hatv{y}\right)e^{i\beta^-_-z}.
\end{align}
\end{subequations}

The corresponding four modal magnetic field are then easily found by substituting~\eqref{eq:Epm_der} into~\eqref{eq:HxHx_vs_ExEy}, or by considering that, for plane-wave excitation, the simple relation \mbox{$\ve{H}=\hatv{z}\times\ve{E}/\eta$} ($\eta=\sqrt{\mu/\epsilon}$) hold.

\section{Bianisotropic Spatial Nonlocality}\label{sec:bianis_nonloc}
Inserting~\eqref{eq:bianis_const_rel} into~\eqref{eq:Maxwell_curl} yields
\begin{subequations}\label{eq:curl_EH_bianis}
\begin{align}
\label{eq:curl_E_bianis}
\nabla\times\ve{E}
&=i\omega\te{\zeta}\cdot\ve{E}+i\omega\te{\mu}\cdot\ve{H}, \\
\label{eq:curl_H_bianis}
\nabla\times\ve{H}
&=-i\omega\te{\epsilon}\cdot\ve{E}-i\omega\te{\xi}\cdot\ve{H}.
\end{align}
\end{subequations}
This system of equations can be solved for $\ve{E}$ in terms of \mbox{$\nabla\times\ve{E}$} and $\nabla\times\ve{H}$ by pre-dotmultiplying~\eqref{eq:curl_E_bianis} by $\te{\xi}\cdot\te{\mu}^{-1}$ and summing the resulting equation with~\eqref{eq:curl_H_bianis} so as to eliminate $\ve{H}$, and finally isolating $\ve{E}$. Similarly, it can be solved for $\ve{H}$ in terms of $\nabla\times\ve{E}$ and $\nabla\times\ve{H}$ by pre-dotmultiplying~\eqref{eq:curl_H_bianis} by $\te{\zeta}\cdot\te{\epsilon}^{-1}$ and summing the resulting equation with~\eqref{eq:curl_E_bianis} so as to eliminate $\ve{E}$, and finally isolating $\ve{H}$, where we have avoided to involve the inverses of $\te{\xi}$ and $\te{\zeta}$ because, in contrast to $\te{\epsilon}$ and $\te{\mu}$, these tensors may be non invertible. The result is
\begin{subequations}
\begin{align}
\label{eq:E_vs_curls_bian_A}
\ve{E}&=\frac{i}{\omega}
\left(\te{\epsilon}-\te{\xi}\cdot\te{\mu}^{-1}\cdot\te{\zeta}\right)^{-1}
\cdot\left(\nabla\times\ve{H}+\te{\xi}\cdot\te{\mu}^{-1}\cdot\nabla\times\ve{E}\right),
\\
\label{eq:H_vs_curls_bian_A}
\ve{H}&=-\frac{i}{\omega}
\left(\te{\mu}-\te{\zeta}\cdot\te{\epsilon}^{-1}\cdot\te{\xi}\right)^{-1}
\cdot\left(\nabla\times\ve{E}+\te{\zeta}\cdot\te{\epsilon}^{-1}\cdot\nabla\times\ve{H}\right).
\end{align}
\end{subequations}
We next substitute these relations into the expressions of~\eqref{eq:bianis_const_rel} that are associated with bianisotropy, i.e., having the coefficients $\te{\xi}$ and $\te{\zeta}$. Specifically, inserting~\eqref{eq:H_vs_curls_bian_A} into~\eqref{eq:bianis_const_rel_D} and \eqref{eq:E_vs_curls_bian_A} into~\eqref{eq:bianis_const_rel_B} provides the alternative constitutive relations
\begin{subequations}\label{eq:bianis_const_rel_alt_A}
\begin{align}\label{eq:bianis_const_rel_D_alt_A}
\ve{D}=&\te{\epsilon}\cdot\ve{E}-\frac{i}{\omega}\te{\xi}\cdot
\left(\te{\mu}-\te{\zeta}\cdot\te{\epsilon}^{-1}\cdot\te{\xi}\right)^{-1}
\cdot\left(\nabla\times\ve{E}+\te{\zeta}\cdot\te{\epsilon}^{-1}\cdot\nabla\times\ve{H}\right), \\
\label{eq:bianis_const_rel_B_alt_A}
\ve{B}=&
\frac{i}{\omega}\te{\zeta}\cdot
\left(\te{\epsilon}-\te{\xi}\cdot\te{\mu}^{-1}\cdot\te{\zeta}\right)^{-1}
\cdot\left(\nabla\times\ve{H}+\te{\xi}\cdot\te{\mu}^{-1}\cdot\nabla\times\ve{E}\right)
+\te{\mu}\cdot\ve{H},
\end{align}
\end{subequations}
which upon angular spectrum decomposition reduce to
\begin{subequations}\label{eq:bianis_const_rel_alt_as}
\begin{align}\label{eq:bianis_const_rel_D_alt_as}
\tilde{\ve{D}}=&\te{\epsilon}\cdot\tilde{\ve{E}}+\frac{1}{\omega}\te{\xi}\cdot
\left(\te{\mu}-\te{\zeta}\cdot\te{\epsilon}^{-1}\cdot\te{\xi}\right)^{-1}
\cdot\left(\ve{k}\times\tilde{\ve{E}}+\te{\zeta}\cdot\te{\epsilon}^{-1}\cdot\ve{k}\times\tilde{\ve{H}}\right), \\
\label{eq:bianis_const_rel_B_alt_as}
\tilde{\ve{B}}=&
-\frac{1}{\omega}\te{\zeta}\cdot
\left(\te{\epsilon}-\te{\xi}\cdot\te{\mu}^{-1}\cdot\te{\zeta}\right)^{-1}
\cdot\left(\ve{k}\times\tilde{\ve{H}}+\te{\xi}\cdot\te{\mu}^{-1}\cdot\ve{k}\times\tilde{\ve{E}}\right)
+\te{\mu}\cdot\tilde{\ve{H}},
\end{align}
\end{subequations}
or, using the identity $\ve{k}\times\ve{a}=\left(\ve{k}\times\te{I}\right)\cdot\ve{a}$,
\begin{subequations}\label{eq:bianis_const_rel_alt_nl_A}
\begin{align}\label{eq:bianis_const_rel_D_alt_nl_A}
\tilde{\ve{D}}=&\left(\te{\epsilon}+\te{T}_D\cdot\ve{k}\times\te{I}\right)\cdot\tilde{\ve{E}}
+\left(\te{T}_D\cdot\te{\zeta}\cdot\te{\epsilon}^{-1}\cdot\ve{k}\times\te{I}\right)\cdot\tilde{\ve{H}},\\
&\qquad\tx{with}\quad
\te{T}_D=\frac{1}{\omega}\te{\xi}\cdot
\left(\te{\mu}-\te{\zeta}\cdot\te{\epsilon}^{-1}\cdot\te{\xi}\right)^{-1} \\
\label{eq:bianis_const_rel_B_alt_nl_A}
\tilde{\ve{B}}=&
\left(\te{T}_B\cdot\te{\xi}\cdot\te{\mu}^{-1}\cdot\ve{k}\times\te{I}\right)\cdot\tilde{\ve{E}}
+\left(\te{\mu}+\te{T}_B\cdot\ve{k}\times\ve{I}\right)\cdot\tilde{\ve{H}} \\
&\qquad\tx{with}\quad
\te{T}_B=-\frac{1}{\omega}\te{\zeta}\cdot
\left(\te{\epsilon}-\te{\xi}\cdot\te{\mu}^{-1}\cdot\te{\zeta}\right)^{-1}.
\end{align}
\end{subequations}

For a biisotropic medium (Sec.~\ref{sec:gen_biis_med}), transforming the tensors into scalars reduces~\eqref{eq:bianis_const_rel_alt} to
\begin{subequations}\label{eq:bianis_const_rel_nl}
\begin{align}\label{eq:bianis_const_rel_D_nl}
\nonumber
\ve{D}&=\epsilon\ve{E}-\frac{i}{\omega}\left(\frac{\xi}{\mu-\zeta\xi/\epsilon}\right)
\left(\nabla\times\ve{E}+\frac{\zeta\nabla\times\ve{H}}{\epsilon}\right) \\
&=\epsilon\ve{E}-\frac{i}{\omega}\frac{\xi}{\gamma}\left(\epsilon\nabla\times\ve{E}+\zeta\nabla\times\ve{H}\right), \\
\label{eq:bianis_const_rel_B_nl}\nonumber
\ve{B}&=\frac{i}{\omega}\left(\frac{\zeta}{\epsilon-\xi\zeta/\mu}\right)
\left(\nabla\times\ve{H}+\frac{\xi\nabla\times\ve{E}}{\mu}\right) +\mu\ve{H}\\
&=\frac{i}{\omega}\frac{\zeta}{\gamma}\left(\xi\nabla\times\ve{E}+\mu\nabla\times\ve{H}\right)
+\mu\ve{H},
\end{align}
where
\begin{equation}
\gamma=\epsilon\mu-\xi\zeta,
\end{equation}
\end{subequations}
which further reduce to~\eqref{eq:const_rel_nl} in the chiral case ($\xi=i\chi$ and $\zeta=-i\chi$).

\section{Parameter Extraction Formulas}\label{sec:param_extr_formulas}
We consider a linearly polarized (LP) incident wave. Solving Eqs.~\eqref{eq:chiral_rot} respectively for $\chi$ and $\sqrt{\epsilon\mu}$, with $z=p$ representing the lattice period (Fig.~\ref{fig:metam_ex}), yields
\begin{subequations}\label{eq:chiral_extr}
\begin{equation}\label{eq:chiral_rot_ang_extr}
\chi
=-\frac{\theta_\tx{C}}{\omega p}
=-\frac{\tan^{-1}\left(E_y/E_x\right)}{\omega p}
=-\frac{\tan^{-1}\left(\left|S_{21}^{yx}\right|/\left|S_{21}^{xx}\right|\right)}{\omega p},
\end{equation}
\begin{equation}\label{eq:chiral_rot_pha}
\sqrt{\epsilon\mu}
=\frac{\phi_\tx{C}}{\omega p}
=\frac{\angle S_{21}^{xx}}{\omega p}=\frac{\angle S_{21}^{yx}}{\omega p},
\end{equation}
\end{subequations}
where the (frequency-dependent) scattering transmission parameters $S_{21}^{uv}$ ($u,v=x,y$) correspond to the simulation setup in Fig.~\ref{fig:helix_extrac}.

The chiral parameter, $\chi$, is readily provided by~\eqref{eq:chiral_rot_ang_extr}, but more information is required to discriminate $\epsilon$ and $\mu$ in~\eqref{eq:chiral_rot_pha}. Such information may be accessed via the (frequency-dependent) scattering reflection parameters $S_{11}^{uv}$ ($u,v=x,y$). Due to reciprocity, where any rotation incurred to the wave in one direction must be undone in the opposite direction, the polarization of the wave reflected by the particle must be identical to the incident wave~\footnote{This may be demonstrated ad absurdum as follows~\cite{Sounas_2019}. If we had $S_{11}^{yx}\neq0$ (reflection rotation), then we would also have, from isotropy, $S_{11}^{xy}\neq0$ and $S_{11}^{xy}=-S_{11}^{yx}$. The latter result following from the fact that the wave would rotate in the same direction for the two experiments (given the same particle chirality), so that $S_{11}^{yx}=E_y^\tx{refl,a}/E_x^\tx{inc,a}$ and $S_{11}^{xy}=E_x^\tx{refl,b}/E_y^\tx{inc,b}=(-|E_x^\tx{refl,b}|)/E_y^\tx{inc,b}=-S_{11}^{yx}$. But reciprocity (or time-reversal symmetry) requires $S_{11}^{xy}=S_{11}^{yx}$. Therefore, one must have $S_{11}^{yx}=S_{11}^{xy}=0$, i.e., no reflection rotation.}. Therefore, one needs to consider only the incident polarization, where the reflection parameter is simply given by the normal-incidence Fresnel coefficient $S_{11}^{xx}=(\eta-Z_\tx{port})/(\eta+Z_\tx{port})$, with $\eta=\sqrt{\mu/\epsilon}$. Solving this relation for $\eta$ yields
\begin{equation}\label{eq:eta_fc_Z}
\eta=Z_\tx{port}\frac{1+S_{11}^{xx}}{1-S_{11}^{xx}}=\sqrt{\frac{\mu}{\epsilon}}.
\end{equation}

Finally, taking successfully the product and the ratio of Eqs.~\eqref{eq:chiral_rot_pha} and~\eqref{eq:eta_fc_Z} provides the missing parameters:
\begin{subequations}
\begin{equation}
\mu=Z_\tx{port}\left(\frac{\angle S_{21}^{yx}}{\omega p}\right)\left(\frac{1+S_{11}^{xx}}{1-S_{11}^{xx}}\right),
\end{equation}
\begin{equation}
\epsilon=\frac{1}{Z_\tx{port}}\left(\frac{\angle S_{21}^{yx}}{\omega p}\right)\left(\frac{1-S_{11}^{xx}}{1+S_{11}^{xx}}\right).
\end{equation}
\end{subequations}

\end{document}